\documentclass[a4paper,11pt]{article}
\usepackage{jheppub}
\usepackage{amsmath}
\usepackage{graphicx}
\usepackage{multirow}
\usepackage{bm}
\usepackage{bbm}
\usepackage{color}
\usepackage{slashed}
\usepackage{scalefnt}
\usepackage{grffile}
\usepackage{booktabs}
\allowdisplaybreaks

\title{Complete NNLO QCD Calculations of $B_c^{(*)}\to J/\psi(\eta_c)$ Form Factors and Semileptonic Decays}
\author[a,1]{Wei Tao,\note{Corresponding author.}}
\author[b]{Zhen-Jun Xiao,}
\author[b]{Ruilin Zhu}
\author[a]{and Qin Chang}%
\affiliation[a]{Institute of Particle and Nuclear Physics,	Henan Normal University, Xinxiang 453007, China}
\affiliation[b]{Department of Physics and Institute of Theoretical Physics, Nanjing Normal University, Nanjing, Jiangsu 210023, China}
\emailAdd{taowei@htu.edu.cn}
\emailAdd{xiaozhenjun@njnu.edu.cn}
\emailAdd{rlzhu@njnu.edu.cn}
\emailAdd{changqin@htu.edu.cn}

\abstract{
We present the first complete next-to-next-to-leading-order (NNLO) QCD calculation of all 34 vector, axial-vector, tensor, and axial-tensor form factors for the four $B_c^{(*)}\to J/\psi(\eta_c)$ transitions within nonrelativistic QCD (NRQCD) factorization.
Using automated multi-loop techniques based on \texttt{FeynCalc} and \texttt{AMFlow}, we provide a dataset with at least 40 significant digits at 35 exact rational points in the $(m_b,m_c,q^2)$ parameter space, covering $q^2\in[-10,10]~\mathrm{GeV}^2$.
In the spacelike region ($q^2<0$) and near $q^2=0$, the NLO corrections typically enhance the LO form factors by $\mathcal{O}(70\%)$, while the NNLO corrections amount to $\mathcal{O}(20\%)$ of the NLO predictions and substantially reduce the renormalization-scale dependence, demonstrating good perturbative convergence. In contrast, fixed-order NRQCD predictions suffer from threshold instabilities and become less reliable at higher timelike $q^2$. 
We therefore extrapolate the form factors to the full physical region via $z$-series fits to the NRQCD results at spacelike and near-zero $q^2$, finding broad agreement with the latest lattice QCD results for $B_c\to J/\psi$.
With the resulting NRQCD+$z$-series form factors, we predict the decay widths, branching fractions, and lepton-flavor-universality (LFU) ratios for $B_c^{(*)}\to J/\psi(\eta_c)\,\ell\nu_\ell$ ($\ell=e,\mu,\tau$) at LO, NLO, and NNLO.
Higher-order corrections substantially enhance the absolute rates but largely cancel in the ratios, whose uncertainties are dominated by the choice of fitting points.
In particular, our NNLO central prediction $\mathcal{R}(B_c\to J/\psi)=0.2836$ lies between the latest lattice QCD determination and the most recent LHCb measurement.
}

\preprint{~}

\begin{document}

\maketitle

\section{Introduction}

The beauty-charmed mesons $B_c^{(*)}$ occupy a unique position in the Standard Model (SM) as bound states composed of two distinct heavy quark flavors. Their open heavy flavor forbids strong and electromagnetic annihilation, and the ground-state pseudoscalar meson $B_c$ decays solely through the weak interaction. The $B_c$ meson was first observed by the CDF Collaboration at the Tevatron through the semileptonic decays $B_c\to J/\psi\,\ell\nu_\ell$ ($\ell=e,\mu$)~\cite{CDF:1998ihx}. These decays provide a laboratory for studying the interplay between weak interactions and heavy-quark bound-state dynamics~\cite{Richman:1995wm}. In particular, the transitions $B_c\to J/\psi(\eta_c)\,\ell\nu_\ell$ ($\ell\in\{e,\mu,\tau\}$) proceed through the underlying $\bar b\to\bar c\ell\nu_\ell$ interaction, with the spectator charm quark becoming part of the final charmonium state, and complement $B\to D^{(*)}\ell\nu_\ell$ in tests of lepton flavor universality (LFU).

At the LHC, the LHCb Collaboration reported the first measurement of the LFU ratio $\mathcal{R}(B_c\to J/\psi)\equiv\mathcal{B}(B_c\to J/\psi\,\tau\nu_\tau)/\mathcal{B}(B_c\to J/\psi\,\mu\nu_\mu)$, obtaining $0.71\pm0.17\pm0.18$ with Run~1 data~\cite{LHCb:2017vlu}. Subsequent determinations include the CMS measurement of $0.17\pm0.33$~\cite{CMS:2024seh}, the updated CMS combination of $0.49\pm0.26$~\cite{CMS:2025jfx}, and the LHCb Run~2 result of $0.51\pm0.12\pm0.08$~\cite{LHCb:2026dev}. The latest LHCb central value has shifted downward relative to Run~1 but remains somewhat above current SM expectations, including the latest HPQCD prediction $0.2597\pm0.0027$~\cite{Harrison:2025yan}, while the experimental uncertainties are still sizable. These measurements motivate improved theoretical determinations of the hadronic inputs entering the ratio.

The ground-state vector partner $B_c^*$ provides a complementary setting for studying the same underlying $\bar b\to\bar c$ weak transition with a vector initial state. It has recently been observed by the ATLAS Collaboration through the radiative decay $B_c^*\to B_c\gamma$~\cite{ATLAS:2026ubk}. This electromagnetic M1 transition is expected to dominate its total width~\cite{Fulcher:1998ka,Yang:2021crs,Martin-Gonzalez:2022qwd}, so that the branching fractions of its weak semileptonic decays $B_c^*\to J/\psi(\eta_c)\,\ell\nu_\ell$ are strongly suppressed and depend on the assumed total width~\cite{Chang:2025rna,Tao:2025kqf}. Studying all four $B_c^{(*)}\to J/\psi(\eta_c)$ transitions within a common theoretical framework covers different meson-spin configurations and quark-current structures. Such calculations characterize the weak transition dynamics and determine the expected sizes of the rare $B_c^*$ semileptonic modes, providing quantitative benchmarks for future experimental studies of these weak decays.

The hadronic dynamics of these transitions is encoded in Lorentz-invariant form factors. While the vector and axial-vector form factors govern the SM semileptonic decay rates, the tensor and axial-tensor form factors provide inputs for studies of possible nonstandard semileptonic interactions. The LFU ratios benefit from the cancellation of the common CKM factor and many hadronic uncertainties, but remain sensitive to the momentum-transfer dependence of the form factors, because the $\tau$ and muon modes weight the physical phase space differently, with the $\tau$ mode restricted to $q^2\ge m_\tau^2$. Reliable predictions for the absolute rates and LFU ratios therefore require both the magnitudes and the $q^2$ dependence of the form factors over the full physical region.

Over the past decades, the $B_c^{(*)}\to J/\psi(\eta_c)$ form factors and semileptonic observables have been investigated using QCD sum rules~\cite{Kiselev:2000hs,Kiselev:2002vz,Wang:2012hu} and various quark models~\cite{Ebert:2003cn,Ivanov:2006ni,Hernandez:2006gt,Wang:2008xt,Chang:2019obq,Chang:2020xvu,Sheng:2021iss,Sheng:2022peg,Wang:2024cyi,Wang:2018ryc,Chang:2026ctr}.
In lattice QCD, the HPQCD Collaboration has determined the $B_c\to J/\psi$ vector, axial-vector, tensor, and axial-tensor form factors across the full physical $q^2$ range~\cite{Harrison:2020gvo,Harrison:2020nrv,Harrison:2025yan}, and has presented preliminary results for the $B_c\to\eta_c$ form factors~\cite{Colquhoun:2016osw}, whereas analogous lattice determinations for the vector $B_c^*$ transitions are not yet available.

Since both the initial $B_c^{(*)}$ and final $J/\psi(\eta_c)$ mesons are heavy-quark bound states with small internal relative velocities, nonrelativistic QCD (NRQCD) factorization~\cite{Bodwin:1994jh} provides a systematic framework for calculating their transition form factors. At lowest order in the velocity expansion, each form factor is given by the product of a short-distance coefficient, calculable perturbatively in the strong coupling $\alpha_s$, and two nonperturbative long-distance matrix elements proportional to the wave functions at the origin of the initial and final mesons, respectively. 
A hard-gluon exchange with the spectator charm quark is already required at leading order (LO), so that the nonvanishing LO form factors start at $\mathcal{O}(\alpha_s)$ and the LO predictions depend strongly on the renormalization scale.

Next-to-leading-order (NLO) QCD corrections were first studied for the $B_c\to\eta_c$ vector and tensor form factors~\cite{Bell:2005gw,Bell:2006tz}, followed by calculations of the $B_c\to J/\psi(\eta_c)$ vector and axial-vector~\cite{Qiao:2011yz,Qiao:2012vt} as well as tensor and axial-tensor~\cite{Tao:2022yur} form factors. Relativistic corrections to these form factors have also been investigated~\cite{Zhu:2017lqu,Shen:2021dat}. More recently, following a tree-level analysis of the $B_c^*\to J/\psi$ vector and axial-vector form factors~\cite{Geng:2023ffc}, we obtained analytical NLO results for all vector, axial-vector, tensor, and axial-tensor form factors of $B_c^*\to J/\psi$~\cite{Chang:2025rna} and $B_c^*\to\eta_c$~\cite{Tao:2025kqf}. 
Complementary to fixed-order calculations, studies in the hierarchical limit $m_b\gg m_c\gg\Lambda_{\rm QCD}$ have derived the all-order structure of the leading double logarithms in the soft-overlap contribution to the $B_c\to\eta_c$ form factors at large recoil~\cite{Boer:2023tcs,Bell:2024bxg} and verified it through three loops using soft-collinear effective theory~\cite{Bell:2026ktc}.

Previous NLO studies~\cite{Bell:2005gw,Bell:2006tz,Qiao:2011yz,Qiao:2012vt,Tao:2022yur,Chang:2025rna,Tao:2025kqf} found sizable one-loop corrections and noticeable residual renormalization-scale dependence.  
An explicit next-to-next-to-leading-order (NNLO) calculation is
therefore needed to assess the convergence of the $\alpha_s$
expansion and the extent to which the scale dependence is reduced
at the next order, with the aim of improving the precision of
SM predictions.
A related issue concerns the use of these perturbative results
in phenomenology: integrated semileptonic observables require
form factors over the entire physical $q^2$ range, whereas the
small relative velocity inside each bound state does not by
itself ensure that the hard-scattering calculation remains
valid throughout this region.
Earlier analyses found that fixed-order NRQCD predictions
become less reliable at larger timelike $q^2$, toward zero
recoil~\cite{Qiao:2011yz,Chang:2025rna,Tao:2025kqf}. 
It is thus important to examine the momentum dependence of the corrections and their scale sensitivity in order to identify the kinematic region in which fixed-order NRQCD calculations provide reliable inputs for a full-$q^2$ parameterization.

In this work, we perform the first complete NNLO ($\mathcal{O}(\alpha_s^3)$) QCD calculation of all 34 vector, axial-vector, tensor, and axial-tensor form factors for $B_c^{(*)}\to J/\psi(\eta_c)$ within NRQCD, at leading order in the heavy-quark velocity expansion.
The two-loop calculation involves multiscale amplitudes depending
on $m_b$, $m_c$, and $q^2$, including closed heavy-quark loops
and flavor-singlet topologies. It also requires a careful
treatment of $\gamma_5$ in dimensional regularization~\cite{Larin:1991tj,Larin:1993tq,Moch:2015usa,Chang:2025rna,Tao:2025kqf}, together with
the renormalization of the QCD currents and NRQCD matrix elements. Building upon our automated multi-loop framework for heavy flavor-changing currents and $B_c^{(*)}$ decay constants~\cite{Tao:2022hos,Tao:2022qxa,Tao:2023mtw,Tao:2023vvf,Tao:2023pzv}, we express the two-loop amplitudes in terms of a minimal set of integral families, and evaluate them using the auxiliary-mass-flow method~\cite{Liu:2017jxz,Liu:2021wks,Liu:2022mfb,Liu:2020kpc,Liu:2022tji} implemented in \texttt{AMFlow}~\cite{Liu:2022chg,Huang:2026rjb} together with integration-by-parts (IBP) reductions~\cite{Chetyrkin:1981qh,Peraro:2019svx,Guan:2024byi,Lange:2025fba}.
We examine the perturbative convergence and scale dependence
in both spacelike and timelike regions over
$q^2\in[-10,10]~\mathrm{GeV}^2$, and study heavy-quark-mass
variations at $q^2=0$.

While earlier NLO analyses~\cite{Tao:2022yur,Chang:2025rna,Tao:2025kqf} combined NRQCD form-factor ratios with low-$q^2$ lattice data, we use the NRQCD form factors directly as inputs to the $z$-series fits. The wave functions at the origin are treated as independent nonperturbative inputs, with their ranges informed by decay constants and potential-model estimates. Based on the perturbative analysis, we select inputs in the spacelike and near-zero $q^2$ region and fit a truncated Bourrely--Caprini--Lellouch (BCL)-type $z$-series parameterization~\cite{Bourrely:2008za,Khodjamirian:2011ub,Chang:2019obq} to extrapolate the form factors over the full physical region $0\le q^2\le q_{\max}^2$. 
Using the resulting vector and axial-vector NRQCD+$z$-series form factors at LO, NLO, and NNLO, we predict the decay widths, branching fractions, and LFU ratios for $B_c^{(*)}\to J/\psi(\eta_c)\,\ell\nu_\ell$ at the corresponding perturbative orders and compare our results with other theoretical determinations, including lattice QCD, and with experimental measurements from LHCb and CMS. 
We examine the different sensitivities of the absolute rates and LFU ratios to higher-order corrections and the choice of fitting points, and assess the uncertainties associated with the renormalization scale, factorization scale, wave functions at the origin, and fitting-point selection.  

The remainder of this paper is organized as follows. Section~\ref{sec:ffdef} defines the transition form factors and discusses their relations at $q^2=0$. Section~\ref{sec:calc} describes the NRQCD factorization framework, $\gamma_5$ prescription, multi-loop calculation, renormalization, and the decoupling and running of $\alpha_s$. Section~\ref{sec:ffnr} presents the fixed-order NRQCD results for the form factors and analyzes their perturbative behavior. Section~\ref{sec:ffq2nrz} provides the NRQCD+$z$-series form factors over the physical region, followed by the semileptonic decay predictions in section~\ref{sec:GmBrR}. We summarize in section~\ref{Summary}, and collect the formulas for the helicity amplitudes and differential decay widths in appendix~\ref{app:dGmdq2ffs}.  

\section{Definition of form factors}
\label{sec:ffdef}

In the Standard Model, the semileptonic decays $B_c^{(*)}\to J/\psi(\eta_c)\, \ell {\nu}_{\ell}$ proceed through
the weak transition $\bar{b}\to \bar{c}\ell {\nu}_{\ell}$. The decay amplitude is determined by the contraction of the leptonic current with the hadronic
transition matrix element of the corresponding $b$-$c$ quark current,
which can be expressed as~\cite{Richman:1995wm,Chang:2016cdi,Chang:2020xvu,Tran:2024phq,Kiselev:2000hs}
\begin{align}
\mathcal{M}
=
\frac{G_F}{\sqrt{2}}V_{cb}^*\,L^\mu\,
\langle H'|\bar b\,\Gamma_\mu \,c\,|H\rangle ,
\end{align}
where $H \in \{B_c, B_c^*\}$ and $H' \in \{J/\psi, \eta_c\}$ denote the initial- and final-state mesons, respectively. $L^\mu$ and $\Gamma_\mu$ represent the leptonic current and the quark-current vertex, respectively. For the Standard Model $V-A$ interaction, they take the forms:
$L^\mu = \bar{\nu}_\ell \gamma^\mu (1-\gamma_5) \ell$ and $\Gamma_\mu = \gamma_\mu(1-\gamma_5)$.

More generally, the hadronic transition matrix elements of various $b$-$c$ quark currents can be constructed by sandwiching the corresponding currents between the initial and final meson states. In particular, the vector, axial-vector, tensor, and axial-tensor currents lead to different Lorentz-covariant decompositions, whose coefficients are parameterized in terms of the corresponding form factors. The vector and axial-vector form factors enter the Standard Model semileptonic decays, while tensor and axial-tensor form factors are also relevant for probing possible new-physics effects.

We denote the momentum of the initial $B_c^{(*)}$ meson by $p$ and, for the vector $B_c^*$ state, its polarization vector by $\epsilon$. The momentum of the final-state $J/\psi(\eta_c)$ meson is denoted by $p'$, with $\epsilon'$ representing the polarization vector of the $J/\psi$.
Introducing the momentum transfer
$q=p-p'$ and the total momentum $P=p+p'$, 
the hadronic transition matrix elements can be decomposed into independent
Lorentz structures constructed from the polarization vectors
$\epsilon$ and $\epsilon'$, the momenta $q$ and $P$, and the
Levi-Civita tensor $\varepsilon_{\mu\nu\rho\sigma}$.
The $B_c^{(*)}\to J/\psi(\eta_c)$ vector, axial-vector, tensor,
and axial-tensor form factors are then defined through the hadronic transition
matrix elements of the corresponding $b$-$c$ quark currents.

For the $B_c\to J/\psi$ transition, 
the vector form factor $V^{B_c\to J/\psi}$, axial-vector form factors $A_{0,1,2}^{B_c\to J/\psi}$, tensor form factor $T_1^{B_c\to J/\psi}$, and axial-tensor form factors $T_{2,3}^{B_c\to J/\psi}$ are defined as follows~\cite{Tao:2022yur,Harrison:2020gvo,Harrison:2025yan},
\begin{align}
\left\langle J/\psi\left(\epsilon^{\prime }, p^{\prime }\right)\left|\bar{b} \gamma_\mu c\right| B_c\left(p\right)\right\rangle
=\,&  -\frac{i\varepsilon_{\mu \nu \alpha \beta} \epsilon^{ \prime * \nu}P^{\alpha} q^{ \beta}}{M+M'} V^{B_c\to J/\psi},\\
\left\langle J/\psi\left(\epsilon^{\prime }, p^{\prime }\right)\left|\bar{b} \gamma_\mu \gamma_5 c\right| B_c\left( p\right)\right\rangle
=\,& -
\frac{2\epsilon^{\prime *} \cdot q }{q^2} q_\mu  M' A_0^{B_c\to J/\psi} \nonumber\\
&
-\left(\epsilon_{\mu}^{ \prime * }-\frac{\epsilon^{\prime *} \cdot q}{q^2}q_\mu\right)(M+M')A_1^{B_c\to J/\psi}\nonumber\\
&
+\frac{\epsilon^{\prime *} \cdot q}{M+M'}\left(P_\mu-\frac{M^2-M^{ \prime 2}}{q^2}q_\mu\right)A_2^{B_c\to J/\psi},\\
\left\langle J/\psi\left(\epsilon^{\prime }, p^{\prime }\right)\left|\bar{b} \sigma_{\mu\nu} q^\nu c\right| B_c\left( p\right)\right\rangle=\,&  \varepsilon_{\mu \nu \alpha \beta} \epsilon^{ \prime * \nu}P^{\alpha} q^{ \beta} T_1^{B_c\to J/\psi},\\
\left\langle J/\psi\left(\epsilon^{\prime }, p^{\prime }\right)\left|\bar{b}  \sigma_{\mu\nu}\gamma_5 q^\nu  c\right| B_c\left( p\right)\right\rangle=\, &-
i\left[(M^2-M^{ \prime 2})\epsilon_\mu^{\prime *}-\epsilon^{\prime *} \cdot q P_\mu\right] T_2^{B_c\to J/\psi}\nonumber\\
&
-i \epsilon^{\prime *} \cdot q \left(q_\mu-\frac{q^2}{M^2-M^{ \prime 2}} P_\mu\right)T_3^{B_c\to J/\psi}.
\end{align}

For the $B_c\to \eta_c$ transition, the vector form factors $F_{p,0}^{B_c\to \eta_c}$ and  tensor form factor $F_{t}^{B_c\to \eta_c}$ are defined by~\cite{Bell:2005gw,Bell:2006tz,Colquhoun:2016osw}
\begin{align}
\left\langle \eta_c\left( p^{\prime }\right)\left|\bar{b} \gamma_\mu c\right| B_c\left( p\right)\right\rangle
=\, & -\left(P_\mu -\frac{M^2-M'^2}{q^2}q_\mu \right)F_p^{B_c\to \eta_c}\nonumber\\&
-\frac{M^2-M'^2}{q^2}q_\mu F_0^{B_c\to \eta_c},\\
\left\langle \eta_c\left( p^{\prime }\right)\left|\bar{b} \sigma_{\mu\nu} q^\nu c\right| B_c\left( p\right)\right\rangle
=\,& -i \frac{q^2 P_\mu -(M^2-M'^2) q_\mu}{M+M'}F_t^{B_c\to \eta_c},
\end{align}
while the axial-vector and axial-tensor matrix elements vanish due to parity conservation,
\begin{align}
\left\langle \eta_c\left( p^{\prime }\right)\left|\bar{b} \gamma_\mu \gamma_5 c\right| B_c\left( p\right)\right\rangle=\,&0,\\
\left\langle\eta_c\left( p^{\prime }\right)\left|\bar{b}  \sigma_{\mu\nu}\gamma_5 q^\nu  c\right| B_c\left( p\right)\right\rangle=\,& 0 .
\end{align}

For the $B_c^*\to J/\psi$ transition, the vector form factors $V_{1,2,3,4,5,6}^{B_c^*\to J/\psi}$, axial-vector form factors $A_{1,2,3,4}^{B_c^*\to J/\psi}$, tensor form factors $T_{2,3,4,6}^{B_c^*\to J/\psi}$, and axial-tensor form factors $T_{2,3,4}^{\prime B_c^*\to J/\psi}$ are parameterized as~\cite{Chang:2025rna}
\begin{align}
\left\langle J/\psi\left(\epsilon^{\prime }, p^{\prime }\right)\left|\bar{b} \gamma_\mu c\right| B_c^*\left(\epsilon, p\right)\right\rangle
=\,&  -\left(\epsilon \cdot \epsilon^{ \prime *}\right)\left[P_\mu V_1^{B_c^*\to J/\psi}-q_\mu V_2^{B_c^*\to J/\psi}\right]\nonumber\\& -\left(\epsilon\cdot q\right) \epsilon_\mu^{ \prime *} V_3^{B_c^*\to J/\psi}+\left(\epsilon^{\prime *} \cdot q\right) \epsilon_\mu V_4^{B_c^*\to J/\psi}\nonumber\\&+\left(\epsilon \cdot q\right)\left(\epsilon^{ \prime *} \cdot q\right)\bigg[\left(\frac{P_\mu }{M^{ 2}-M^{\prime 2}} -\frac{q_\mu}{q^2} \right) V_5^{B_c^*\to J/\psi}\nonumber\\&+\frac{q_\mu}{q^2}  V_6^{B_c^*\to J/\psi}\bigg],\\
\left\langle J/\psi\left(\epsilon^{\prime }, p^{\prime }\right)\left|\bar{b} \gamma_\mu \gamma_5 c\right| B_c^*\left(\epsilon, p\right)\right\rangle=\,& 
i \varepsilon_{\mu \nu \alpha \beta} \epsilon^{\alpha} \epsilon^{ \prime * \beta}\left[P^\nu A_1^{B_c^*\to J/\psi}- q^\nu A_2^{B_c^*\to J/\psi}\right] \nonumber\\
& +\frac{i \varepsilon_{\mu \nu \alpha \beta} P^\alpha q^\beta}{M^{ 2}-M^{ \prime 2}}\left[\epsilon^{\prime *} \cdot q \epsilon^{ \nu} A_3^{B_c^*\to J/\psi}\right.\nonumber\\&\left.-\epsilon \cdot q \epsilon^{ \prime * \nu} A_4^{B_c^*\to J/\psi}\right]
,\\
\langle J/\psi(\epsilon',p')\vert{}\bar{b}\sigma_{\mu\nu}q^{\nu}c\vert{}B_{c}^{*}(\epsilon,p)\rangle = \, & -i(M+M')(\epsilon\cdot\epsilon'^*)\left(\frac{q^2 P_\mu}{M^2-M'^2} - q_\mu\right) T_2^{B_c^*\to J/\psi} \nonumber\\ & 
+ i(M+M')(\epsilon\cdot q)\left[\frac{(\epsilon'^*\cdot q)P_\mu}{M^2-M'^2} - \epsilon'^*_\mu\right] T_3^{B_c^*\to J/\psi} \nonumber\\ & 
- i(M+M')(\epsilon'^*\cdot q)\left[\frac{(\epsilon\cdot q)P_\mu}{M^2-M'^2} - \epsilon_\mu\right] T_4^{B_c^*\to J/\psi} \nonumber\\ & 
- i\frac{(\epsilon\cdot q)(\epsilon'^*\cdot q)}{M+M'}\left(\frac{q^2 P_\mu}{M^2-M'^2} - q_\mu\right) T_6^{B_c^*\to J/\psi} ,\\
\left\langle J/\psi\left(\epsilon^{\prime }, p^{\prime }\right)\left|\bar{b}  \sigma_{\mu\nu}\gamma_5 q^\nu  c\right| B_c^*\left(\epsilon, p\right)\right\rangle=\, &
- \varepsilon_{\mu \nu \alpha \beta} \epsilon^{\alpha} \epsilon^{ \prime * \beta} q^\nu(M+M^{ \prime }) {T'_2}^{B_c^*\to J/\psi}  \nonumber\\
& +\frac{ \varepsilon_{\mu \nu \alpha \beta} P^\alpha q^\beta}{M+M^{ \prime }}\left[\epsilon^{\prime *} \cdot q \epsilon^{ \nu} {T'_3}^{B_c^*\to J/\psi}\right.\nonumber\\&\left.+\epsilon \cdot q \epsilon^{ \prime * \nu} {T'_4}^{B_c^*\to J/\psi}\right]
.
\end{align}

For the $B_c^*\to \eta_c$ transition, the vector form factor $V^{B_c^*\to \eta_c}$, axial-vector form factors $A_{0,1,2}^{B_c^*\to \eta_c}$, tensor form factor $T^{B_c^*\to \eta_c}$, and axial-tensor form factors $T_{1,2}^{\prime B_c^*\to \eta_c}$ are introduced as~\cite{Tao:2025kqf} 
\begin{align}
\left\langle \eta_c\left( p^{\prime }\right)\left|\bar{b} \gamma_\mu c\right| B_c^*\left(\epsilon, p\right)\right\rangle
&=
\frac{i \varepsilon_{\mu \nu \alpha \beta} \epsilon^{\nu}P^\alpha q^\beta}{M+M'}V^{B_c^*\to\eta_c},\\
\left\langle \eta_c\left( p^{\prime }\right)\left|\bar{b} \gamma_\mu \gamma_5 c\right| B_c^*\left(\epsilon, p\right)\right\rangle
&=\frac{ 2 M\epsilon \cdot q q_\mu}{q^2}A_0^{B_c^*\to\eta_c}
+ \left(\epsilon_\mu-\frac{\epsilon \cdot q}{q^2} q_\mu\right) (M+M')A_1^{B_c^*\to\eta_c}
\nonumber\\&\quad
+ \epsilon \cdot q \left(\frac{P_\mu}{M+M'}-\frac{M-M'}{q^2} q_\mu\right) A_2^{B_c^*\to\eta_c}
,\\
\left\langle \eta_c\left( p^{\prime }\right)\left|\bar{b} \sigma_{\mu\nu} q^\nu c\right| B_c^*\left(\epsilon, p\right)\right\rangle
&= -\varepsilon_{\mu \nu \alpha \beta} \epsilon^{\nu}P^\alpha q^\beta T^{B_c^*\to\eta_c},\\
\left\langle\eta_c\left( p^{\prime }\right)\left|\bar{b}  \sigma_{\mu\nu}\gamma_5 q^\nu  c\right| B_c^*\left(\epsilon, p\right)\right\rangle
&=i\left( \epsilon_\mu-\frac{\epsilon\cdot q}{q^2} q_\mu\right) (M^2-M'^2){T'_1}^{B_c^*\to\eta_c}\nonumber\\&\quad
-i \epsilon \cdot q \left(P_\mu-\frac{M^2-M'^2}{q^2} q_\mu\right) {T'_2}^{B_c^*\to\eta_c}
.%
\end{align}

Here, $\sigma_{\mu\nu} = \frac{i}{2} [\gamma_\mu, \gamma_\nu]$. $M=m_H \in \{m_{B_c}, m_{B_c^*}\}$ denotes the mass of the initial meson, and $M'=m_{H'} \in \{m_{J/\psi}, m_{\eta_c}\}$ denotes the mass of the final-state meson. 
(Note that in our subsequent NRQCD matching calculation, these meson masses are projected onto their free-quark values, $M \to m_b + m_c$ and $M' \to 2m_c$.)
For a semileptonic decay mode with lepton $\ell \in \{e, \mu, \tau\}$, the physical kinematic domain is bounded by $m_\ell^2 \le q^2 \le q_{\text{max}}^2= (M - M')^2$.  
At the maximum-recoil point ($q^2 = 0$, which corresponds to the lower boundary in the massless lepton limit), driven by kinematic constraints (to avoid $1/q^2$ singularities) and the four-dimensional Hodge/chiral duality relation between the tensor and axial-tensor currents ($\sigma_{\mu\nu} \gamma_5 = \frac{i}{2} \varepsilon_{\mu\nu\alpha\beta} \sigma^{\alpha\beta}$, with $\gamma_5=i\gamma^0\gamma^1\gamma^2\gamma^3$ and $\varepsilon^{0123}=-\varepsilon_{0123}=1$), the form factors satisfy the following exact relations:
\begin{align}
2M' A_0^{B_c\to J/\psi}(0)&=(M+M')A_1^{B_c\to J/\psi}(0)-(M-M')A_2^{B_c\to J/\psi}(0),\label{ff0relationbcjpsiav}\\
T_1^{B_c\to J/\psi}(0) &= T_2^{B_c\to J/\psi}(0),\\
F_p^{B_c\to \eta_c}(0)&=F_0^{B_c\to \eta_c}(0),\\
V_5^{B_c^*\to J/\psi}(0)&=V_6^{B_c^*\to J/\psi}(0),\\
(M+M')T_3^{B_c^*\to J/\psi}(0) &= (M+M'){T_2'}^{B_c^*\to J/\psi}(0) + (M-M'){T_4'}^{B_c^*\to J/\psi}(0),\label{ff0relationbcvjpsit3}\\
(M+M')T_4^{B_c^*\to J/\psi}(0) &= (M+M'){T_2'}^{B_c^*\to J/\psi}(0) - (M-M'){T_3'}^{B_c^*\to J/\psi}(0),\label{ff0relationbcvjpsit4}\\
2M A_0^{B_c^*\to\eta_c}(0)&=(M+M')A_1^{B_c^*\to\eta_c}(0)+(M-M')A_2^{B_c^*\to\eta_c}(0),\label{ff0relationbcvetacav}\\
T^{B_c^*\to\eta_c}(0)&={T_1'}^{B_c^*\to\eta_c}(0) = {T_2'}^{B_c^*\to\eta_c}(0) .
\end{align}

\section{Calculation procedures and details}
\label{sec:calc}

In this section, we outline the theoretical framework and computational procedures employed to evaluate the $B_c^{(*)} \to J/\psi(\eta_c)$ hadronic transition matrix elements and the corresponding form factors up to NNLO perturbative QCD corrections within the nonrelativistic QCD (NRQCD) factorization formalism~\cite{Bodwin:1994jh}.

\subsection{NRQCD factorization framework}

The $B_c^{(*)}\to J/\psi(\eta_c)$ transition involves two heavy quarkonium states, where both the initial and final mesons can be systematically described within the framework of NRQCD. 
Since both the initial $B_c^{(*)}$ and final-state $J/\psi(\eta_c)$ mesons are composed of heavy quark-antiquark pairs, the small relative velocities of the heavy quarks inside these bound states provide a natural expansion parameter.
Within the NRQCD framework, the hadronic transition matrix elements (form factors) in QCD are factorized into perturbatively calculable short-distance coefficients (SDCs) and nonperturbative NRQCD long-distance matrix elements (LDMEs) associated with the double-heavy-quark bound states.
The NRQCD LDMEs are organized according to powers of the heavy-quark relative velocity and, at leading order in the velocity expansion, the color-singlet $S$-wave LDMEs are proportional to the Schr\"odinger wave functions at the origin of the corresponding mesons.

Consequently, the $B_c^{(*)}\to J/\psi(\eta_c)$ transition form factors $F_i(q^2)$ can be schematically written as
\begin{align}
F_i(q^2)=C_i(\alpha_s,m_b,m_c,q^2,\mu,\mu_f)\,\psi_{B_c^{(*)}}(0)\psi_{J/\psi(\eta_c)}(0),
\end{align}
where $\mu$ and $\mu_f$ are the QCD renormalization scale and NRQCD factorization scale, respectively.
The meson wave functions at the origin, $\psi_{B_c^{(*)}}(0)$ and $\psi_{J/\psi(\eta_c)}(0)$, encode the nonperturbative contributions associated with the low-energy (soft) momentum regions of QCD and are generally regarded as universal nonperturbative quantities, which can be determined from potential-model calculations, lattice QCD, or phenomenological extractions from experimental data.
The short-distance coefficients $C_i$, on the other hand, capture the contributions from the hard momentum region of the QCD amplitudes and can be calculated perturbatively as a series in the strong coupling $\alpha_s$.

To determine the perturbative short-distance coefficient, we perform the matching between the on-shell amputated amplitude of the free heavy-quark states for the process $c\bar{b}\to c\bar{c}W^*$ in full QCD and the corresponding amplitude in NRQCD.
The covariant spin and color-singlet projectors are applied to project the free heavy-quark pairs onto the intended $S$-wave meson states~\cite{Petrelli:1997ge,Bodwin:2007ga,Bodwin:2002cfe}. 
By employing the expansion-by-regions method~\cite{Beneke:1997zp}, the short-distance coefficient can be extracted directly from the hard-region contribution to the full QCD amplitude, thereby avoiding the need to evaluate the full amplitude explicitly. 
At leading order in the velocity expansion, this amounts to setting the relative momentum between the heavy quark and antiquark inside each meson to zero~\cite{Tao:2023mtw,Tao:2023vvf}. Consequently, the heavy quark and antiquark carry fixed fractions of the total meson momentum, proportional to their respective masses.
Finally, the form factors can be extracted by projecting the calculated hard QCD amplitudes onto the independent Lorentz structures constructed from the external momenta and polarization vectors, as specified in the decomposition of the corresponding hadronic transition matrix elements.

\subsection{Amplitude generation and simplification}

\begin{figure}[htbp]
	\center{
		\includegraphics*[scale=0.6]{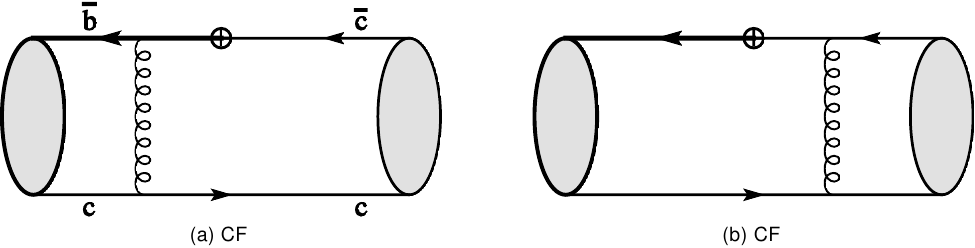}
		\caption {\label{fig:diagtree} Tree-level Feynman diagrams with the corresponding color structures indicated below for the $B_c^{(*)}\to J/\psi(\eta_c)$ transitions. The circled cross symbol “$\oplus$” denotes the insertion of a certain $b$-$c$ transition current vertex. 
				 }}
\end{figure}

\begin{figure}[htbp]
	\center{
		\includegraphics*[scale=0.6]{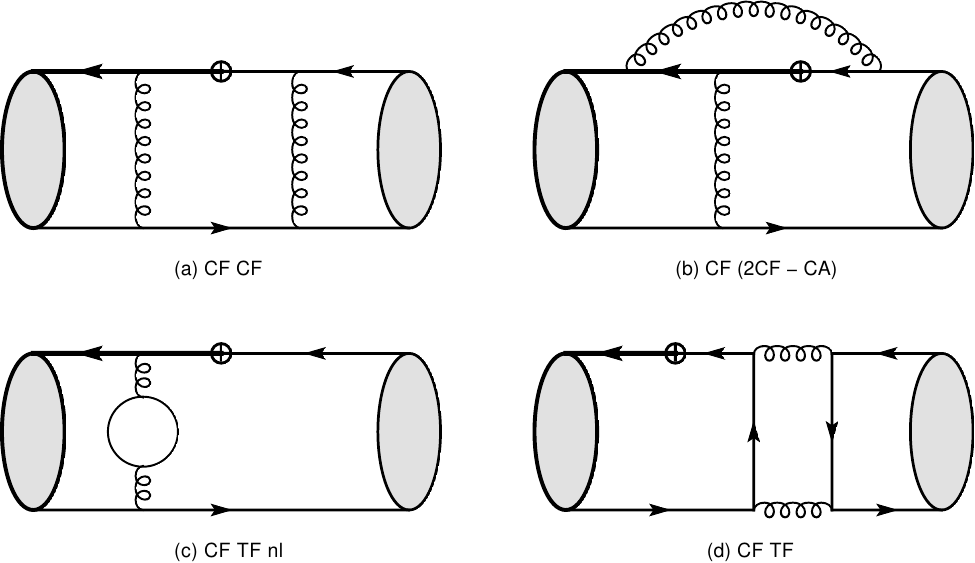}
		\caption {\label{fig:diagloop1} Representative one-loop Feynman diagrams with the corresponding color structures indicated below for the $B_c^{(*)}\to J/\psi(\eta_c)$ transitions. The circled cross symbol “$\oplus$” denotes the insertion of a certain $b$-$c$ transition current vertex. The very thin solid closed loop denotes a massless-quark loop with $n_l$ flavors.
		}}
\end{figure}

\begin{figure}[htbp]
	\center{
		\includegraphics*[scale=0.9]{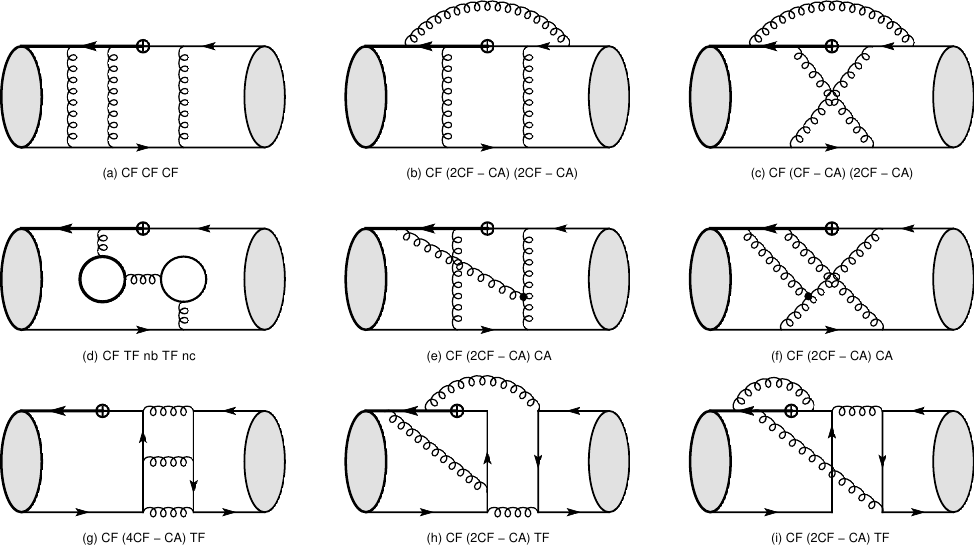}
		\caption {\label{fig:diagloop2} Representative two-loop Feynman diagrams with the corresponding color structures indicated below for the $B_c^{(*)}\to J/\psi(\eta_c)$ transitions. The circled cross symbol “$\oplus$” denotes the insertion of a certain $b$-$c$ transition current vertex. The thick and thin solid closed loops represent the bottom-quark loop with mass $m_b$ and $n_b$ flavors and the charm-quark loop with mass $m_c$ and $n_c$ flavors, respectively.
				}}
\end{figure}

In the actual calculation of QCD, we employ FeynCalc~\cite{Shtabovenko:2025lxq,Shtabovenko:2023idz} to generate the Feynman diagrams for the process $c\bar b\to c\bar c W^*$, taking into account all possible active quark flavors ($n_f = n_b + n_c + n_l$) running in the closed quark loop, including $n_b$ bottom-quark flavors of mass $m_b$, $n_c$ charm-quark flavors of mass $m_c$, and $n_l$ massless light-quark flavors.
After diagrammatic filtering, there are 2 tree-level, 37 one-loop, and 939 two-loop Feynman diagrams, with representative examples shown in figures~\ref{fig:diagtree},~\ref{fig:diagloop1}, and~\ref{fig:diagloop2}, respectively.

The Feynman diagrams can be classified into two categories: flavor-singlet diagrams (such as figure~\ref{fig:diagloop1}(d) and figures~\ref{fig:diagloop2}(g),~\ref{fig:diagloop2}(h), and~\ref{fig:diagloop2}(i)) and flavor-nonsinglet diagrams~\cite{Bell:2006tz,Tao:2022yur,Chang:2025rna,Tao:2025kqf}. 
The contribution from each flavor-singlet diagram to the form factors generally contains a nonzero imaginary part, while the contribution from each flavor-nonsinglet diagram is purely real; both contributions may contain divergent poles.
For the $B_c^{(*)}\to J/\psi$ form factors, we find that the sum of the flavor-singlet one-loop diagram contributions vanishes, whereas the sum of the flavor-singlet two-loop diagram contributions yields a finite complex value with a nonzero imaginary part.
For the $B_c^{(*)}\to \eta_c$ form factors, both the summed flavor-singlet one-loop and two-loop contributions are complex; the former are finite, whereas the latter require renormalization.
Further investigation indicates that the total contribution from the flavor-singlet diagrams is only about $\mathcal{O}(1\%)$ of that from the flavor-nonsinglet diagrams and is thus numerically negligible. 
These observations can be checked using the form factor dataset provided with this work.
For convenience, the small imaginary parts arising from flavor-singlet diagrams will be omitted in the following discussions.

To simplify the raw amplitudes corresponding to the Feynman diagrams, we mainly use the following functions implemented in the \texttt{FeynCalc} package~\cite{Shtabovenko:2025lxq}.
The $SU(N_c)$ color algebra is first carried out using \texttt{SUNSimplify}. The Dirac matrix manipulations, Lorentz index contractions, and Dirac trace calculations are then performed jointly with \texttt{Anti5}, \texttt{Contract}, \texttt{DiracSimplify}, and \texttt{DiracTrace}. 
As a result, the form factors are ultimately expressed in terms of scalar products involving external and loop momenta.

For the trace calculation of a fermion chain containing $\gamma_5$ matrices, we adopt the same $\gamma_5$ prescription as described in refs.~\cite{Chang:2025rna,Tao:2025kqf}.
This prescription can be equivalently formulated in terms of a fixed-reading-point $\gamma_5$ scheme, in which the cyclicity of the trace is not employed, while the anticommutation relation $\{\gamma_5,\gamma_\mu\}=0$ and $\gamma_5^2=1$ are preserved.
The essential step in this prescription is to determine a fixed reading point for each fermion chain. If the fermion chain contains the $b$-$c$ axial-current vertex $\Gamma_\mu = \gamma_\mu\gamma_5$ or $\sigma_{\mu\nu}\gamma_5 q^\nu$, the reading point is fixed according to the replacement rule
\begin{align}
\text{trace}\left(\Omega_1\cdot\Gamma_\mu\cdot \Omega_2\right)\to \text{trace}\left(\frac{\Gamma_\mu\cdot \Omega_2\cdot \Omega_1+\Omega_2\cdot \Omega_1\cdot\Gamma_\mu}{2}\right).
\end{align}
Otherwise, the reading point is fixed by
\begin{align}
\text{trace}\left(\Omega_1\cdot\gamma_5\cdot \Omega_2\right)\to \text{trace}\left(\Omega_2\cdot \Omega_1\cdot\gamma_5\right).
\end{align}
Here, $\Omega_1$ and $\Omega_2$ denote generic Dirac matrix structures.
Once the reading point has been fixed, we set \texttt{FCSetDiracGammaScheme["NDR"]} and use \texttt{Anti5} in FeynCalc to move $\gamma_5$ to the rightmost end of the fermion chain by repeatedly applying the anticommutation relation and $\gamma_5^2=1$.
After this manipulation, each fermion chain inside a trace either contains no $\gamma_5$ or contains a single $\gamma_5$ at the rightmost end of the chain.
The resulting traces can then be evaluated directly by setting \texttt{FCSetDiracGammaScheme["Larin"]} and applying \texttt{DiracSimplify} in FeynCalc.
Note, in the case of a fermion chain containing an even number of $\gamma_5$ matrices, the above fixed-reading-point $\gamma_5$ scheme is, in practice, equivalent to the naive dimensional regularization (NDR) $\gamma_5$ scheme.

\subsection{Multi-loop calculation}

With the aid of an in-house \texttt{Mathematica} package built upon \texttt{FeynCalc}~\cite{Shtabovenko:2025lxq}, the simplified amplitudes—consisting of scalar products of external and loop momenta—are expressed as linear combinations of Feynman integrals and the total number of independent integral families is minimized
by exploiting the symmetries among different integral families. In particular, the number of integral families is minimized from 169 to 19 at one loop and from 1240 to 234 at two loops.

The resulting 19 one-loop and 234 two-loop integral families can be directly evaluated numerically using the \texttt{AMFlow} package~\cite{Liu:2022chg,Huang:2026rjb}, which implements the auxiliary-mass-flow method~\cite{Liu:2017jxz,Liu:2021wks,Liu:2022mfb,Liu:2020kpc,Liu:2022tji}. 
In the internal workflow of \texttt{AMFlow}, the Feynman integrals are first reduced to master integrals, and the latter are subsequently determined by solving differential equations with respect to the auxiliary mass. 
Both steps involve integral reduction.
For independent cross-checks, we equip \texttt{AMFlow} with three reduction tools, including \texttt{Kira}~\cite{Lange:2025fba}, \texttt{Blade}~\cite{Guan:2024byi}, and \texttt{FiniteFlow}~\cite{Peraro:2019svx}, all of which are based on integration-by-parts (IBP) identities~\cite{Chetyrkin:1981qh}.

Besides evaluating these integral families directly with \texttt{AMFlow}, an alternative, more refined strategy is also possible (see refs.~\cite{Tao:2023mtw,Tao:2023vvf,Tao:2023pzv}).
In this approach, the integral families are first reduced to their corresponding master-integral families using dedicated reduction programs. The total number of independent master-integral families is then further minimized, and the resulting minimal set is passed to \texttt{AMFlow} for numerical evaluation.

In practice, we find that although the latter strategy reduces the number of integral families to be evaluated by \texttt{AMFlow} and thus lowers the computational cost of the master-integral calculation, 
the explicit IBP reduction and additional intermediate reduction steps will introduce exceedingly large and complicated rational coefficients when expanding the amplitude in terms of master integrals. As a result, substituting the reduction results and numerical master-integral results back into the amplitude and simplifying the final expressions becomes more time-consuming.

By contrast, directly feeding the 234 integral families into \texttt{AMFlow} maintains a much simpler algebraic structure, effectively avoiding massive coefficient inflation and leading to a simpler overall workflow and a shorter total computation time for the amplitude, despite \texttt{AMFlow} having to handle somewhat more integral families.
We therefore adopt the direct-family-evaluation strategy in our calculation.

\subsection{Renormalization}

To remove the divergences arising in the loop diagrams and obtain finite, physical results for the form factors, i.e., the renormalized form factors, we perform the renormalization procedure within dimensional regularization.

We first carry out the renormalization of the QCD amplitude. At NNLO, this requires summing the contributions from the two-loop diagrams, one-loop diagrams with one $\mathcal{O}(\alpha_s)$ counterterm vertex insertion, tree-level diagrams with one $\mathcal{O}(\alpha_s^2)$ counterterm vertex insertion, and tree-level diagrams with two $\mathcal{O}(\alpha_s)$ counterterm vertex insertions. These counterterm diagrams involve the QCD strong coupling renormalization constant $Z_{\alpha_s}$ in the modified minimal subtraction ($\overline{\mathrm{MS}}$) scheme~\cite{Mitov:2006xs,Chetyrkin:1997un,vanRitbergen:1997va}, as well as the QCD heavy-quark field ($Z_{2,b(c)}^{\mathrm{OS}}$) and mass ($Z_{m,b(c)}^{\mathrm{OS}}$) renormalization constants in the on-shell (OS) scheme~\cite{Fael:2020bgs}.

We also need to renormalize the QCD $b$-$c$ quark current vertices. This involves the QCD on-shell renormalization constants $Z_J^{\mathrm{OS}}$ for the vector ($v$), axial-vector ($a$), tensor ($t$), and axial-tensor ($t5$) $b$-$c$ quark currents. For the vector and axial-vector currents, one has
\begin{align}
Z_v^{\mathrm{OS}}=Z_a^{\mathrm{OS}}= 1.
\end{align}
For the tensor and axial-tensor currents, the detailed two-loop on-shell renormalization constants can be written as 
\begin{align}
Z_t^{\mathrm{OS}} = Z_{t5}^{\mathrm{OS}} = Z_t^{\overline{\mathrm{MS}}} z_t^g z_t^\mu, 
\end{align}
whose explicit analytical expressions can be found in our previous work~\cite{Tao:2023vvf,Tao:2023pzv}.

In addition, the odd-$\gamma_5$ trace prescription used in the QCD amplitudes requires the corresponding renormalization.
This may involve $\gamma_5$-scheme renormalization constants for the axial-vector and axial-tensor $b$-$c$ quark current vertices, as well as for the pseudoscalar meson projection operators. These renormalization constants may have both singlet and nonsinglet versions. 
Since the $b\text{-}c$ quark current is flavor-changing, its $\gamma_5$-scheme renormalization constant is non-singlet for all diagrams.
For the axial-vector form factors of the $B_c^* \to J/\psi$ transition, the $\gamma_5$-scheme renormalization constant is taken as
\begin{align}
Z_{5, \langle J/\psi | \bar b\gamma_\mu\gamma_5 c| B_c^* \rangle} = Z_{5,a}^{\text{ns}}  = Z_{5,a}^{\text{ns},\overline{\mathrm{MS}}}\, Z_{5,a}^{\text{ns,fin}},
\end{align}
where $Z_{5,a}^{\text{ns}}$ is the nonsinglet $\gamma_5$-scheme renormalization constant for the QCD axial-vector $b$-$c$ quark current vertex~\cite{Larin:1991tj,Larin:1993tq,Moch:2015usa},
factored into the $\overline{\rm MS}$ renormalization part $Z_{5,a}^{\rm ns,\overline{MS}}$ and the finite renormalization part $Z_{5,a}^{\rm ns,fin}$, which read
\begin{align}
Z_{5,a}^{\mathrm{ns},\overline{\mathrm{MS}}}&= 1 + \left( \frac{\alpha_s^{\left(n_{f}\right)}\left(\mu\right)}{4\pi} \right)^2 \frac{2  C_F}{\epsilon} \left(\frac{11}{3}C_A-\frac{4}{3} T_F n_f\right),\nonumber\\
Z_{5,a}^{\mathrm{ns,fin}}&=1 - \frac{\alpha_s^{\left(n_{f}\right)}\left(\mu\right)}{\pi} C_F   + \left(\frac{\alpha_s^{\left(n_{f}\right)}\left(\mu\right)}{4\pi}\right)^2 C_F\left( 22 C_F - \frac{107}{9}  C_A + \frac{4}{9} T_F   n_f\right).
\end{align}
For brevity in this work, the $\gamma_5$-scheme renormalization constants for all other form factors are uniformly set to unity, namely
\begin{equation}
Z_{5, \langle J/\psi(\eta_c) | \bar b\Gamma_\mu c| B_c^{(*)} \rangle} = 1 \quad \text{for } \langle J/\psi(\eta_c) | \bar b\Gamma_\mu c| B_c^{(*)} \rangle \neq \langle J/\psi | \bar b\gamma_\mu\gamma_5 c| B_c^* \rangle.
\end{equation}
This simplification induces only very small scheme-dependent finite-term deviations from the full treatment (e.g., see tables~\ref{tab:ff0numbcjpsi} and \ref{tab:ff0numbcvetac}), %
and has a negligible impact on the phenomenological results of this work.

Finally, the meson NRQCD long-distance matrix elements (or the meson wavefunctions at the origin) also require renormalization. This involves the NRQCD current $\overline{\rm MS}$ renormalization constants $\widetilde{Z}_{J,\mathcal{H}}$ associated with the initial (final) state mesons $\mathcal{H}$, which are given by~\cite{Tao:2022qxa,Tao:2023mtw,Tao:2023vvf,Tao:2023pzv}
\begin{align}
\widetilde{Z}_{p}&=1+\left(\frac{\alpha_{s}^{\left(n_{l}\right)}\left(\mu_f\right)}{\pi}\right)^{2}\pi^{2}C_{F}\frac{1}{\epsilon}\left(\frac{x^2+6 x+1}{8 (x+1)^2}C_{F}+\frac{1}{8}C_{A}\right),\\
\widetilde{Z}_{v}&=1+\left(\frac{\alpha_{s}^{\left(n_{l}\right)}\left(\mu_f\right)}{\pi}\right)^{2}\pi^{2}C_{F}\frac{1}{\epsilon}\left(\frac{3x^2+2x+3}{24\left(x+1\right)^2}C_{F}+\frac{1}{8}C_{A}\right),\\
\widetilde{Z}_{p,B_c}&= \left. \widetilde{Z}_p \right\vert{}_{x = \frac{m_c }{ m_b}},\\
\widetilde{Z}_{p,\eta_c}&=\left. \widetilde{Z}_p \right\vert{}_{x = 1},\\
\widetilde{Z}_{v,B_c^*}&=\left. \widetilde{Z}_v \right\vert{}_{x = \frac{m_c }{ m_b}},\\
\widetilde{Z}_{v,J/\psi}&=\left. \widetilde{Z}_v \right\vert{}_{x = 1}.
\end{align}

After renormalization, all divergent poles cancel exactly, and the NNLO results for all form factors are finite.
Furthermore, all renormalized NNLO form factors are invariant under renormalization-group (RG) scale variations~\cite{Tao:2022hos,Tao:2023mtw,Tao:2023vvf,Tao:2023pzv}, as verified numerically to a precision exceeding 40 significant digits.

To guarantee the reliability of our computational setup, extensive cross-checks have been performed at the NLO level. 
Under the same conventions and normalization prescriptions, our automated framework faithfully reproduces all existing NLO results available in the literature~\cite{Bell:2005gw,Bell:2006tz,Qiao:2011yz,Qiao:2012vt,Tao:2022yur,Chang:2025rna,Tao:2025kqf}.
Note that, to conform to the standard normalization convention for the meson wave functions at the origin adopted in the present work (which is consistent with refs.~\cite{Bell:2005gw,Bell:2006tz}), the analytic results of all form factors from our previous work~\cite{Tao:2022yur,Chang:2025rna,Tao:2025kqf} must be globally multiplied by a common overall factor of $1/\pi$. And, to account for the $\gamma_5$-scheme renormalization, the analytic results from our previous work~\cite{Chang:2025rna} for the $B_c^*\to J/\psi$ axial-vector form factors require a further multiplication by the finite renormalization factor $Z_{5,a}^{\rm ns,fin}$. 
After applying these corrections, the NLO analytic results from our previous work~\cite{Tao:2022yur,Chang:2025rna,Tao:2025kqf} agree with the present NLO numerical results for all form factors to at least 40 significant digits.

\subsection{Decoupling, running, and evaluation of $\alpha_s$}

Using the decoupling relation~\cite{Chetyrkin:2005ia,Kniehl:2006bg,Bernreuther:1981sg,Grozin:2011nk,Grozin:2012ic,Barnreuther:2013qvf,Grozin:2007fh,Gerlach:2019kfo,Ozcelik:2021zqt,Tao:2022hos,Tao:2023mtw,Tao:2022qxa,Tao:2023vvf,Tao:2023pzv} and the coupling running~\cite{Abreu:2022cco,Tao:2022hos,Tao:2023mtw,Tao:2022qxa,Tao:2023vvf,Tao:2023pzv} in $D=4-2\epsilon$ dimensions, the strong coupling constant can be systematically converted from $\alpha_s^{(n_f)}(\mu)$ and $\alpha_s^{(n_l)}(\mu_f)$ to $\alpha_s^{(n_l)}(\mu)$, thereby completing the matching between QCD and NRQCD. As a result, the form factors can be expressed as perturbative expansions in powers of $\alpha_s^{(n_l)}(\mu)$.

For the numerical evaluation and phenomenological analysis, the flavor-number parameters entering the form-factor expressions are set as
\begin{equation}
(n_b,n_c,n_l)=
\begin{cases}
(1,1,3), & \mu \geq m_b,\\
(0,1,3), & m_c \leq \mu < m_b,\\
(0,0,3), & \mu < m_c .
\end{cases}
\end{equation}
And the values of $\alpha_s^{(n_l=3)}(\mu)$ for $\mu \in [0.5, 10]~\text{GeV}$ are evaluated at three-loop accuracy using the \texttt{RunDec} function \texttt{AlphasExact}~\cite{Chetyrkin:2000yt,Schmidt:2012az,Deur:2016tte,Herren:2017osy} with the input value $\alpha_s^{(n_l=3)}(1.2729~\text{GeV})\approx 0.3825$, which is determined by running down from the initial value $\alpha_s^{(n_f=5)}(M_Z = 91.1876\allowbreak~\text{GeV}) = 0.1179$ and decoupling at the scale-invariant masses $\overline{m}_b(\overline{m}_b)=4.1859~\text{GeV}$ and $\overline{m}_c(\overline{m}_c)=1.2729~\text{GeV}$~\cite{ParticleDataGroup:2022pth,ParticleDataGroup:2024cfk,ParticleDataGroup:2026mpi,Chetyrkin:2017lif,Chetyrkin:2009fv,Egner:2021lxd} via \texttt{DecAsDownSI}.

\section{NRQCD results for form factors~\label{sec:ffnr}}

The primary outcome of our NRQCD calculation is a comprehensive, high-precision dataset for the NNLO QCD corrections to all 34  $B_c^{(*)}\to J/\psi(\eta_c)$ form factors at 35 exact rational points in the $(m_b,m_c,q^2)$ parameter space. At the central quark masses $(m_b,m_c)=(\frac{475}{10^2},\frac{150}{10^2})~\mathrm{GeV}$, %
31 different $q^2$ points are considered, covering both the spacelike ($q^2<0$) and timelike ($q^2>0$) regions as well as $q^2=0$:
\begin{align}\label{datasetq2all}
q^2 \in \Big\{ & -10, -9, -8, -7, -6, -5, -4, -3, -2, -1, \, 0, \, \frac{1}{10^2}, \, \frac{1}{4}, \, \frac{1}{2}, \, \frac{3}{4}, \, 1, \, \frac{3}{2}, \, 2, \nonumber \\
& \frac{2524921}{10^6}, \, \frac{273}{10^2}, \, 3, \, \frac{335}{10^2}, \, 4, \, 5, \, \frac{5049842}{10^6}, \, \frac{541}{10^2}, \, 6, \, 7, \, 8, \, 9, \, 10 \Big\}~\text{GeV}^2.
\end{align}
In addition, four mass-shifted points at $q^2=0$ are evaluated for the heavy-quark-mass uncertainty analysis:
\begin{align}
(m_b,m_c)\in
\left\{
\left(\frac{475}{10^2},\frac{100}{10^2}\right),
\left(\frac{475}{10^2},\frac{200}{10^2}\right),
\left(\frac{425}{10^2},\frac{150}{10^2}\right),
\left(\frac{525}{10^2},\frac{150}{10^2}\right)
\right\}~\mathrm{GeV}.
\end{align}
The complete results for the form factors, given as perturbative
series up to $\mathcal{O}(\alpha_s^3)$ with a numerical precision
of at least 40 significant digits, are provided in the ancillary
files accompanying this paper.

\subsection{Numerical setup and input parameters}

Before presenting the numerical results for the form factors and phenomenological observables, we summarize the relevant input parameters in table~\ref{tab:globalinputs}. Unless explicitly stated otherwise, these numerical values are adopted throughout the remainder of this work.
Note that we retain $m_{B_c^*}=6.331~\mathrm{GeV}$ in our numerical analysis, about 8~MeV ($0.13\%$) below the recent ATLAS central value of $6.3390~\mathrm{GeV}$~\cite{ATLAS:2026ubk}.

\begin{table}[htbp]
	\centering
	\caption{Summary of the numerical input parameters used throughout this work. Unless otherwise specified, default inputs are adopted from the Particle Data Group (PDG)~\cite{ParticleDataGroup:2024cfk,ParticleDataGroup:2022pth}.}
	\label{tab:globalinputs}
	\renewcommand{\arraystretch}{1.15}
	\begin{tabular}{lll}
		\toprule
		Parameter~ & Value & Source / Ref. \\
		\midrule
		\multicolumn{3}{l}{\textit{Decay constants}} \\
		$f_{B_c}$       & $0.434~\text{GeV}$   & Lattice QCD~\cite{Colquhoun:2015oha} \\
		$f_{B_c^*}$     & $0.422~\text{GeV}$   & Lattice QCD~\cite{Colquhoun:2015oha} \\
		$f_{J/\psi}$    & $0.4065~\text{GeV}$  & PDG \& Lattice~\cite{ParticleDataGroup:2024cfk,ParticleDataGroup:2022pth,Hatton:2020qhk,Donald:2012ga} \\
		$f_{\eta_c}$    & $0.3981~\text{GeV}$  & Lattice QCD~\cite{Hatton:2020qhk,Davies:2010ip} \\
		\midrule
		\multicolumn{3}{l}{\textit{Meson masses}} \\
		$m_{B_c}$       & $6.274~\text{GeV}$   & \\
		$m_{J/\psi}$    & $3.0969~\text{GeV}$  & \\
		$m_{\eta_c}$    & $2.9841~\text{GeV}$  & \\
		$m_{B_c^*}$     & $6.331~\text{GeV}$   & Lattice QCD~\cite{Mathur:2018epb,Gregory:2009hq,Dowdall:2012ab} \\
		$m_{B_{c0}^*(0^+)}$  & $6.710~\text{GeV}$   & Lattice \& Potential~\cite{Martin-Gonzalez:2022qwd,Wang:2012kw,Dowdall:2012ab,Boyd:1997kz,Leljak:2019eyw} \\
		$m_{B_{c1}(1^+)}$  & $6.740~\text{GeV}$   & Lattice \& Potential~\cite{Martin-Gonzalez:2022qwd,Wang:2012kw,Dowdall:2012ab,Boyd:1997kz,Leljak:2019eyw} \\
		\midrule
		\multicolumn{3}{l}{\textit{Electroweak, CKM, and lepton masses}} \\
		$G_F$           & $1.16638 \times 10^{-5}~\text{GeV}^{-2}$ & \\
		$|V_{cb}|$      & $0.0408$            & \\
		$m_e$           & $0.510999 \times 10^{-3}~\text{GeV}$     & \\
		$m_\mu$         & $0.10566~\text{GeV}$ & \\
		$m_\tau$        & $1.777~\text{GeV}$   & \\
		\midrule
		\multicolumn{3}{l}{\textit{Lifetimes and total widths}} \\
		$\tau_{B_c}$    & $0.51~\text{ps}$    & \\
		$\Gamma_{\text{tot}}({B_c^*})$ & $6.0 \times 10^{-8}~\text{GeV}$ & Potential models~\cite{Yang:2021crs,Fulcher:1998ka,Martin-Gonzalez:2022qwd} \\
		\bottomrule
	\end{tabular}
\end{table}

In principle, within the NRQCD factorization framework, the wave functions at the origin (or the corresponding NRQCD LDMEs)  depend on the factorization scale $\mu_f$, with this dependence  canceling against the $\mu_f$ dependence of the short-distance coefficients, order by order in $\alpha_s$, so that the physical form factors remain independent of $\mu_f$. Although the $\mu_f$-evolution of these LDMEs can be formally obtained by solving the corresponding RG equations, there is currently no well-established theoretical framework to rigorously fix their boundary conditions~\cite{Sang:2022kub,Sang:2022tnh,Tao:2023mtw}. We therefore treat the wave functions at the origin as independent nonperturbative inputs, varying their values within phenomenologically reasonable ranges and using the residual $\mu_f$ dependence of the short-distance coefficients to estimate the associated theoretical uncertainties~\cite{Tao:2022qxa}.

In our numerical calculations of the form factors, by invoking heavy-quark spin symmetry~\cite{Bodwin:1994jh,Tao:2023mtw,Tao:2022qxa}, the wave functions at the origin are taken as
\begin{align}
\psi_{B_c}(0)=\psi_{B_c^*}(0)&=0.325\pm 0.025~{\rm GeV}^{3/2},\nonumber\\
\psi_{J/\psi}(0)=\psi_{\eta_c}(0)&=0.225\pm 0.025~{\rm GeV}^{3/2},\label{wavefsval}
\end{align}
where the lower limits are based on the wave-function values extracted
from the PDG and lattice QCD results for the corresponding decay
constants in table~\ref{tab:globalinputs} via
\begin{align} 
\psi_\mathcal{H}(0) = f_\mathcal{H} \sqrt{\frac{m_\mathcal{H}}{4N_c}} \quad \text{for } \mathcal{H} \in \{B_c, B_c^*, J/\psi, \eta_c\}, 
\end{align} 
while the upper limits are taken from typical potential-model predictions~\cite{Patel:2008na,Braaten:2002fi,Biswas:2023bqz,Buchmuller:1980su,Eichten:1994gt,Eichten:1995ch,Ikhdair:2005xe,Tao:2022qxa}, with the central values chosen at the midpoint between the two.

The central values and variation ranges of the factorization scale, the renormalization scale, and the heavy-quark masses are chosen as~\cite{Egner:2021lxd,Chen:2025qgy,Zhang:2022nuf,Yang:2026kvh,Chen:2026maw,Shen:2014msa,Tao:2022qxa,Tao:2023vvf,Tao:2023pzv,Chang:2025rna,Tao:2025kqf,Tao:2023mtw,Bell:2005gw,Zhang:2026hgu}
\begin{align}
\mu_f &= 1.5^{+1.5}_{-1.0}~\text{GeV}, & \mu &= 4.75^{+5.25}_{-3.25}~\text{GeV}, \nonumber \\
m_b &= 4.75\pm 0.50~\text{GeV}, & m_c &= 1.50\pm 0.50~\text{GeV}.\label{mbmcmumufval}
\end{align}

\subsection{$F(0)$~\label{sec:ff0nr}}

\begin{figure}[htbp]%
	\centering
	\includegraphics[width=0.44\textwidth]{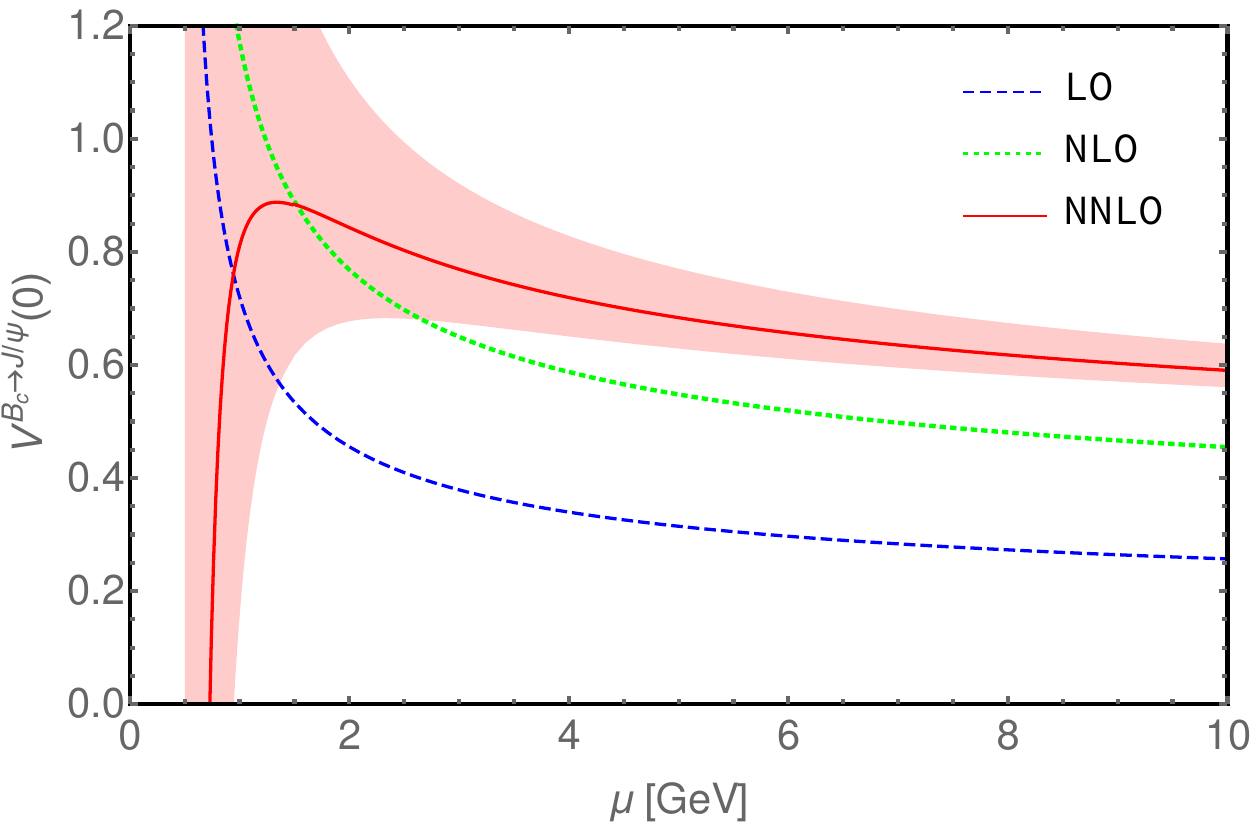}\qquad %
	\includegraphics[width=0.44\textwidth]{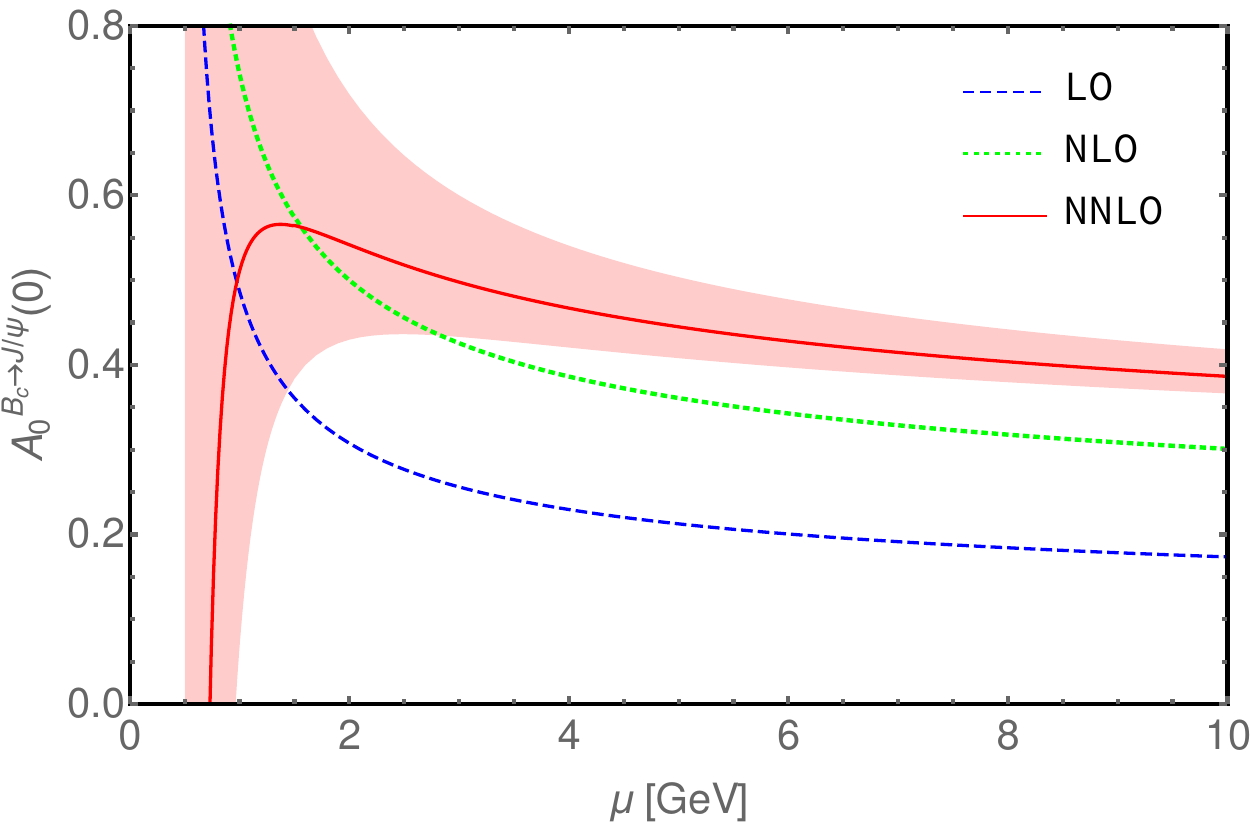}\qquad
	\includegraphics[width=0.44\textwidth]{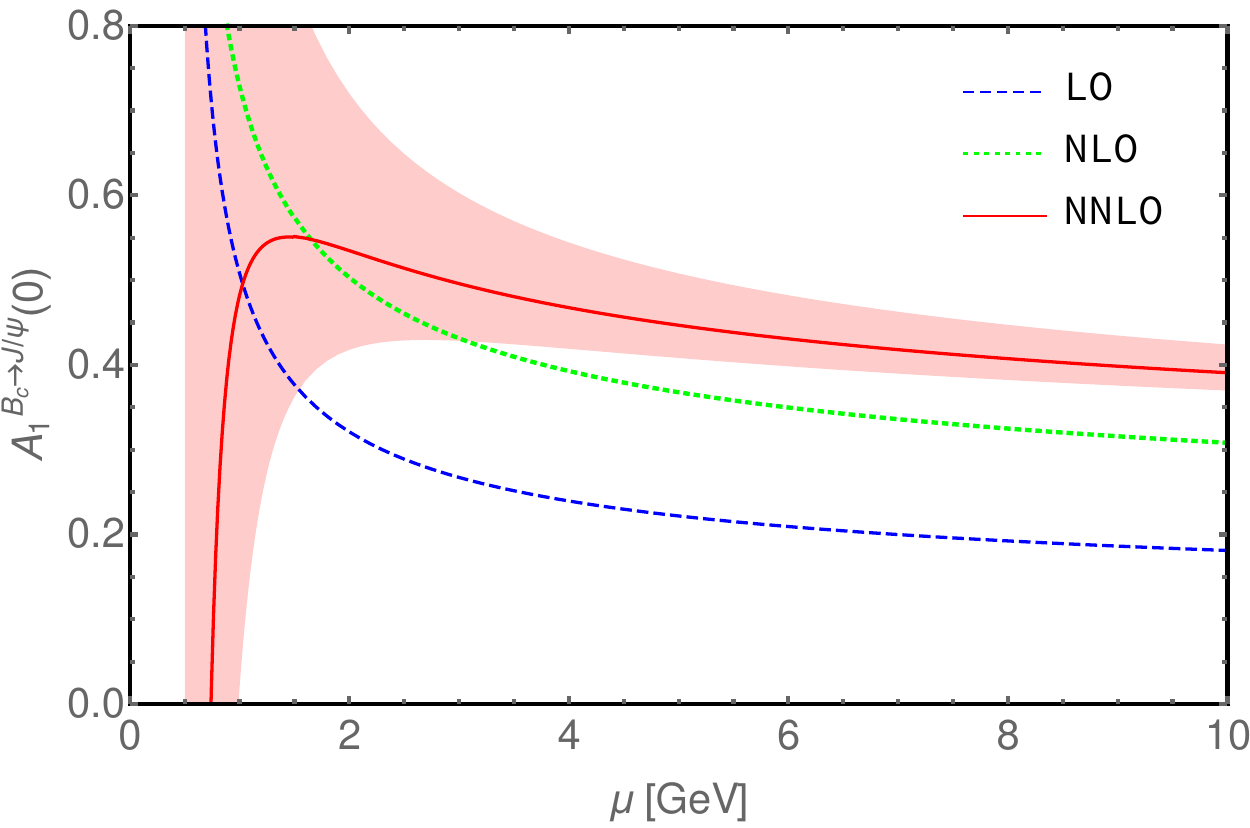}\qquad
	\includegraphics[width=0.44\textwidth]{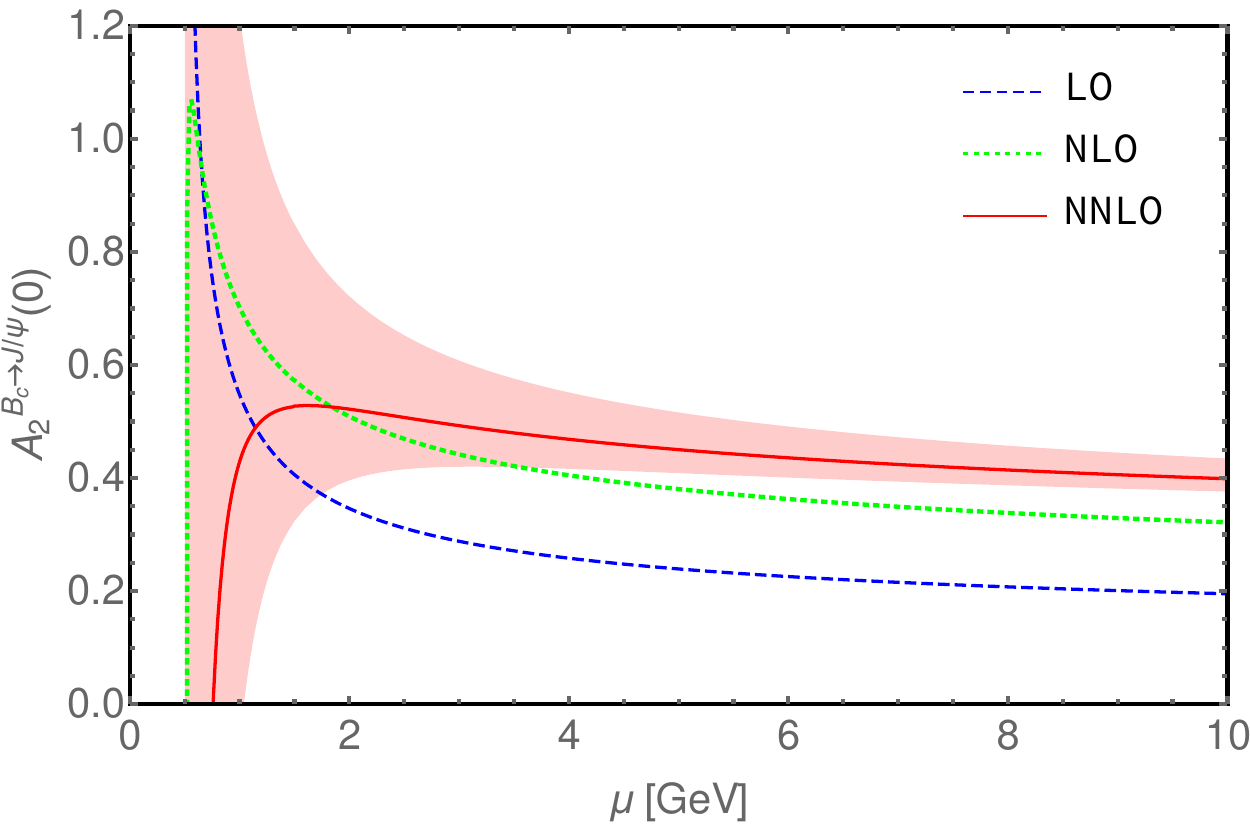}\qquad
	\includegraphics[width=0.44\textwidth]{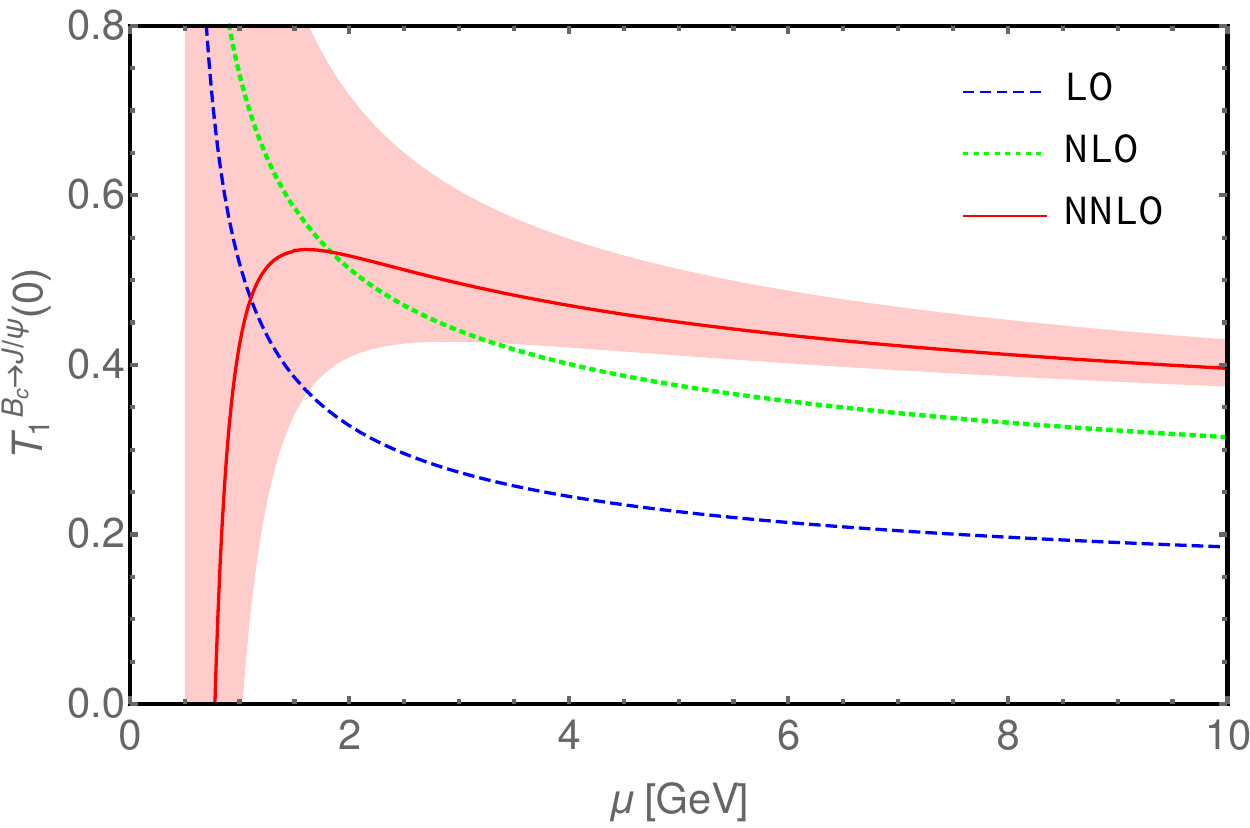}\qquad
	\includegraphics[width=0.44\textwidth]{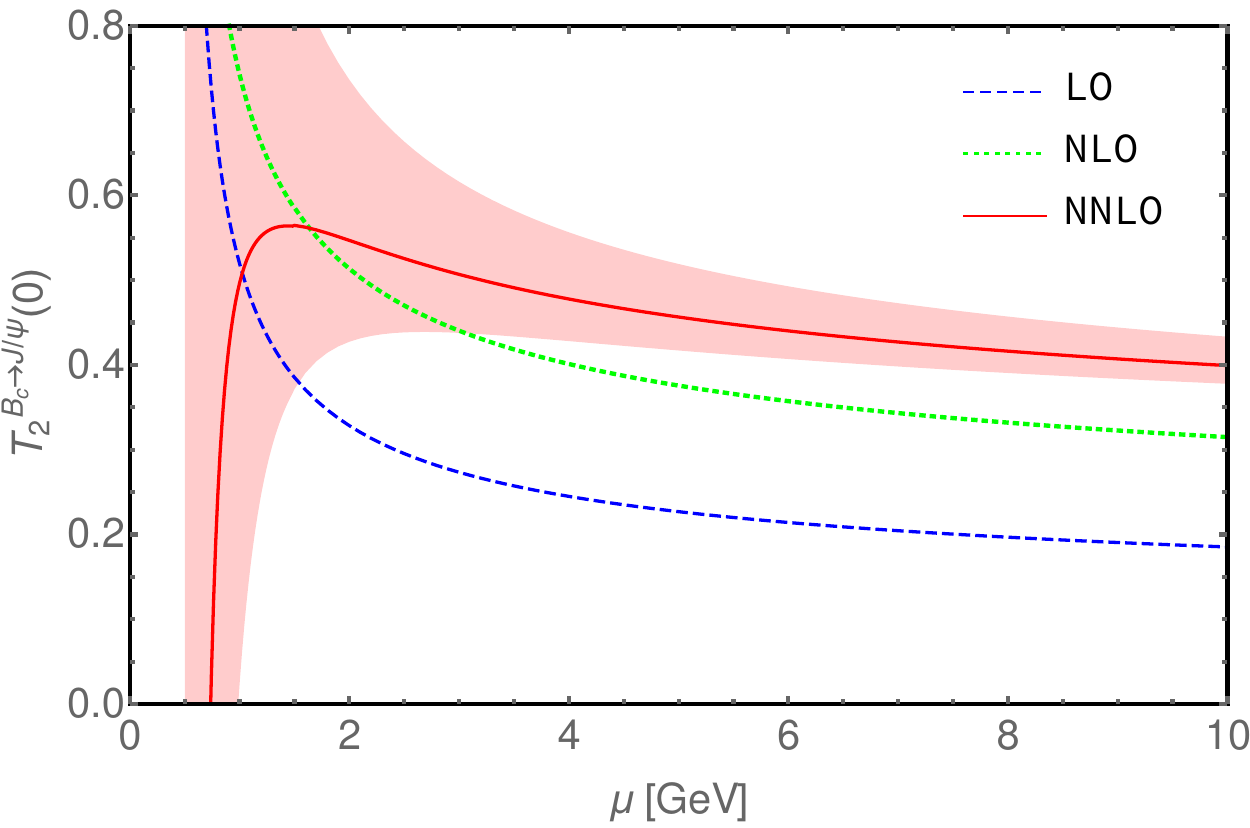}\qquad	
	\includegraphics[width=0.44\textwidth]{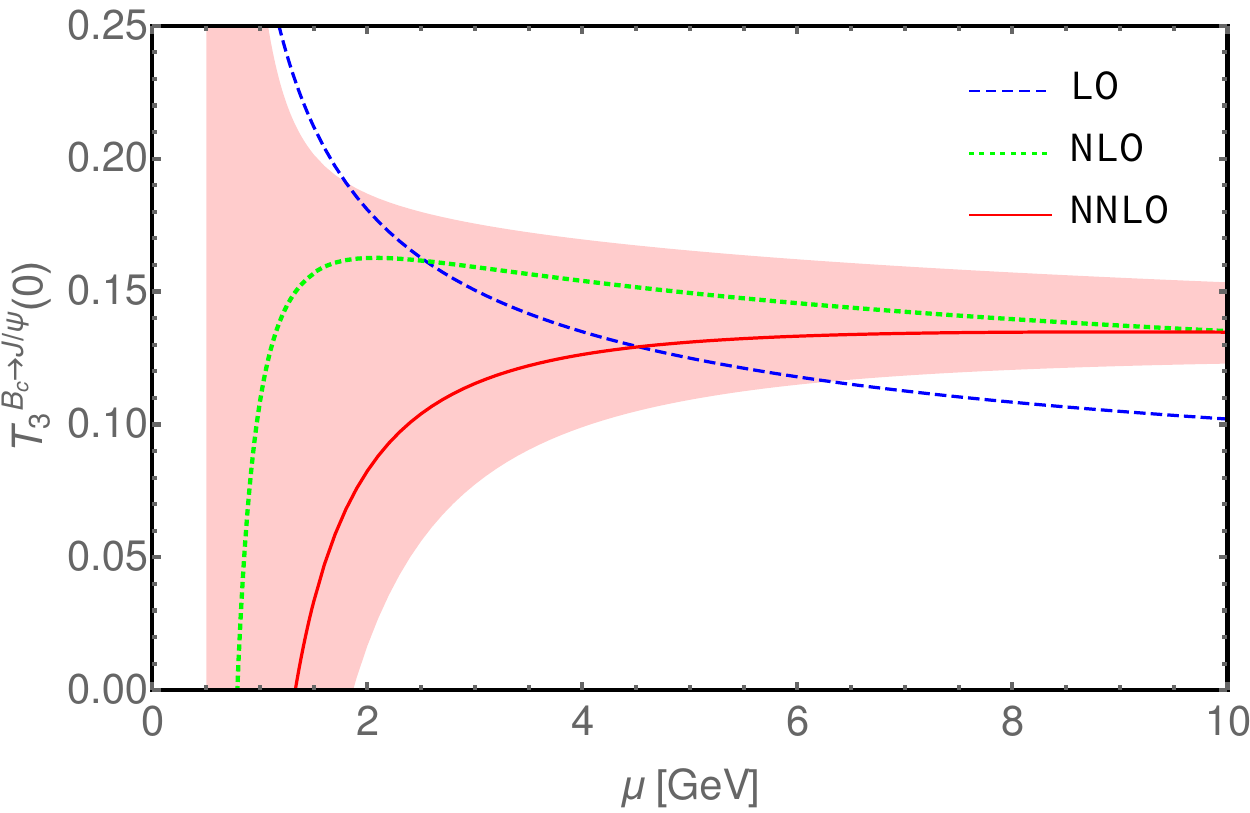}
	\caption{The renormalization-scale dependence of the $B_c\to J/\psi$ transition form factors at $q^2=0$ at LO, NLO, and NNLO accuracy.   
	The central values of  the form factors  are evaluated with   $\mu_f=1.5\,\mathrm{GeV}$,  $m_b=4.75\,\mathrm{GeV}$, $m_c=1.5\,\mathrm{GeV}$, $\psi_{B_c}(0)=0.325\,\mathrm{GeV}^{3/2}$, and $\psi_{J/\psi}(0)=0.225\,\mathrm{GeV}^{3/2}$. The error bands stem from varying  the NRQCD factorization scale $\mu_f$ between    0.5  and 3    $\mathrm{GeV}$,  corresponding to the upper and lower edges of the bands, respectively. The minor discrepancy between $T_1^{B_c\to J/\psi}(0)$ and $T_2^{B_c\to J/\psi}(0)$ at the NNLO level is due to the $\gamma_5$-trace prescription (see the discussion in the caption of table~\ref{tab:ff0numbcjpsi}).}
	\label{fig:ff0mubcjpsi}
\end{figure}

\begin{figure}[htbp]%
	\centering
	\includegraphics[width=0.44\textwidth]{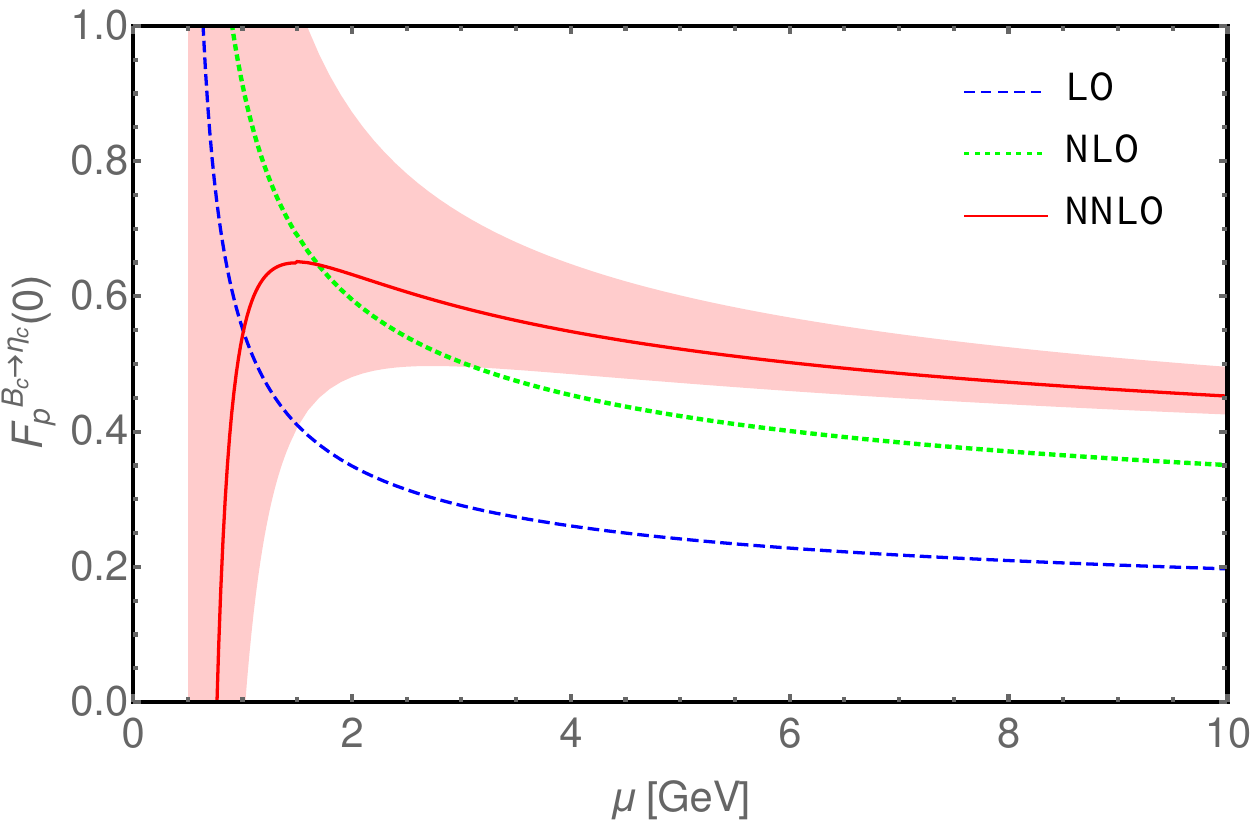}\qquad %
	\includegraphics[width=0.44\textwidth]{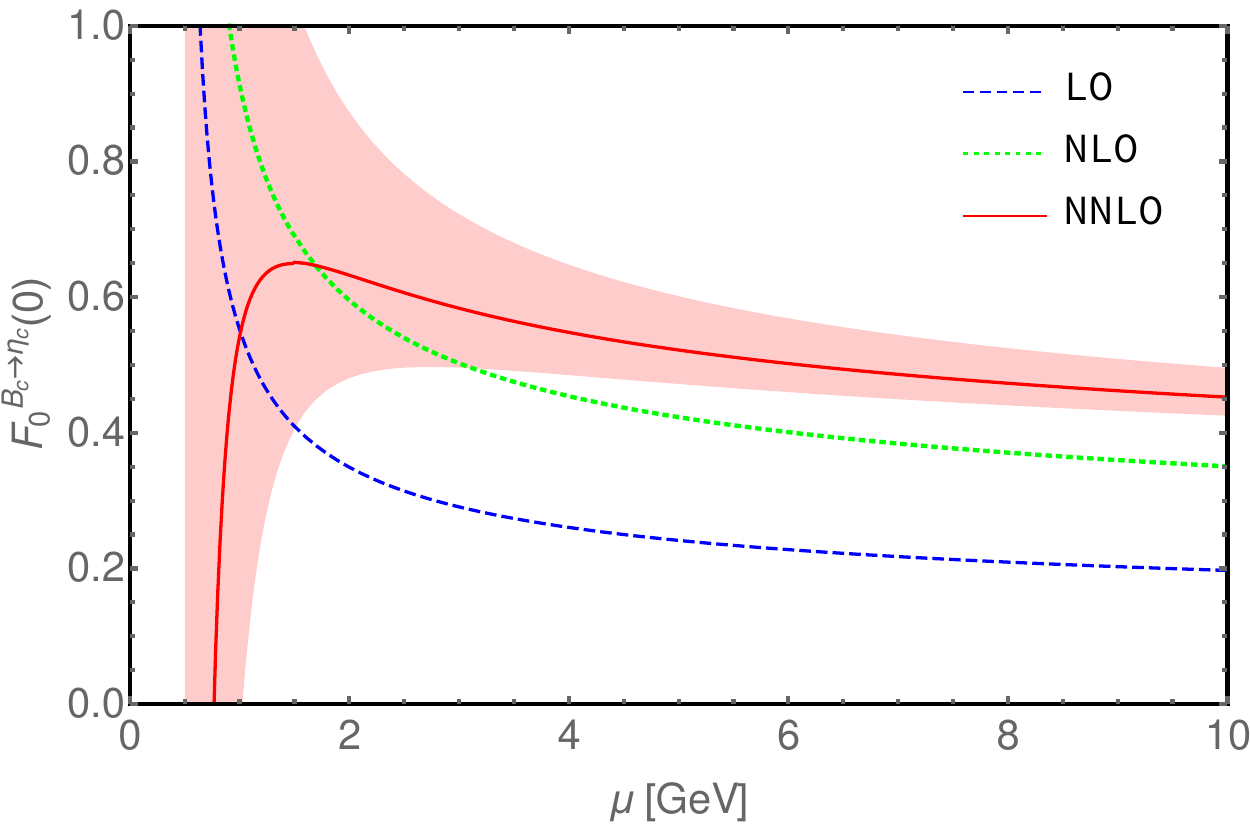}\qquad
	\includegraphics[width=0.44\textwidth]{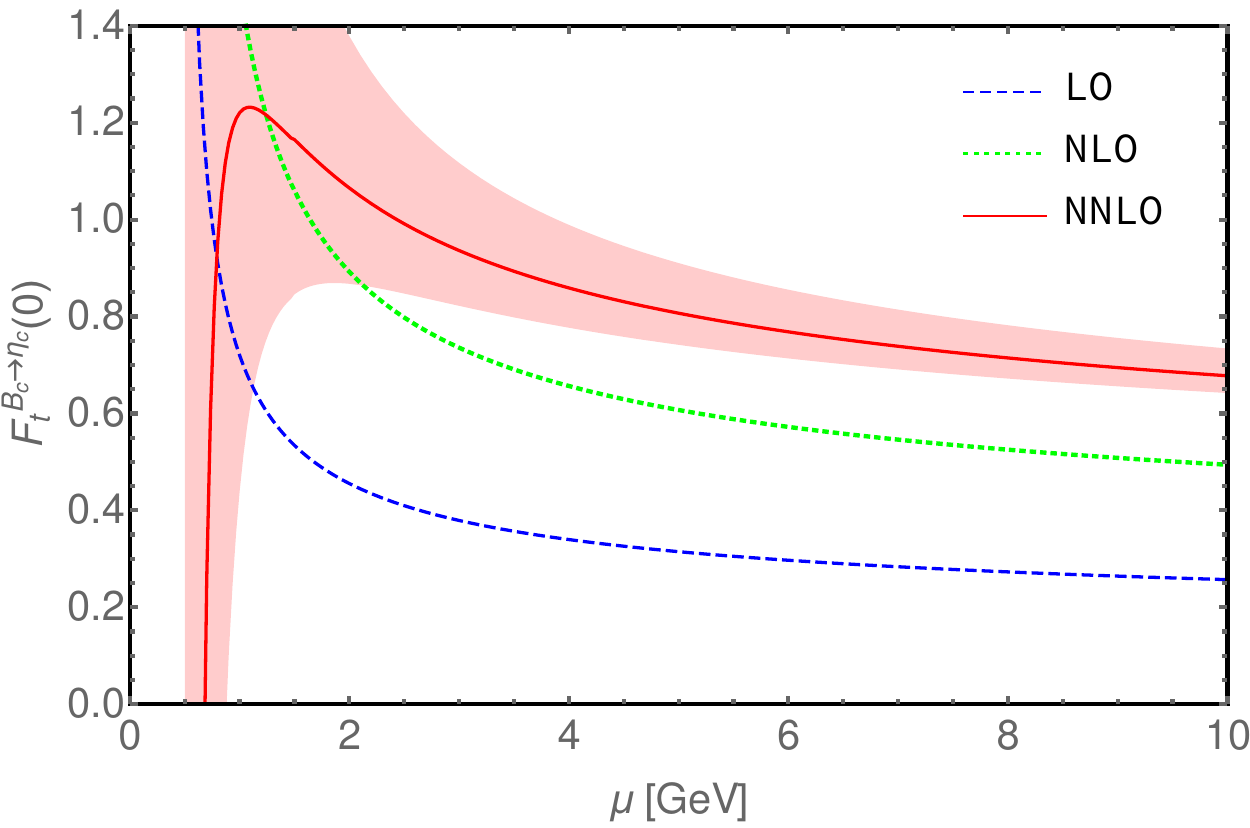}
	\caption{Same as figure~\ref{fig:ff0mubcjpsi}, but for $B_c\to\eta_c$, with $F_p^{B_c\to\eta_c}(0)=F_0^{B_c\to\eta_c}(0)$.
		 }
\end{figure}

\begin{figure}[htbp]%
	\centering
	\includegraphics[width=0.315\textwidth]{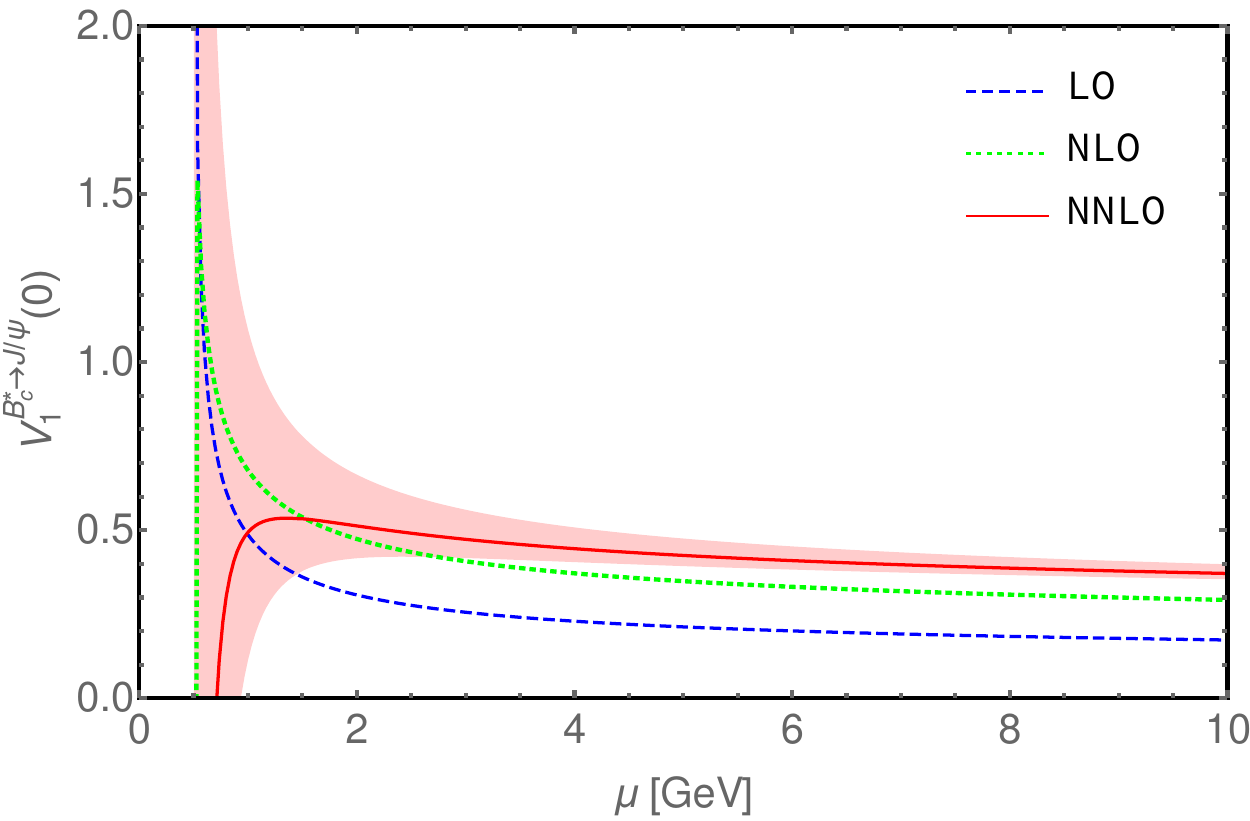}\quad %
	\includegraphics[width=0.315\textwidth]{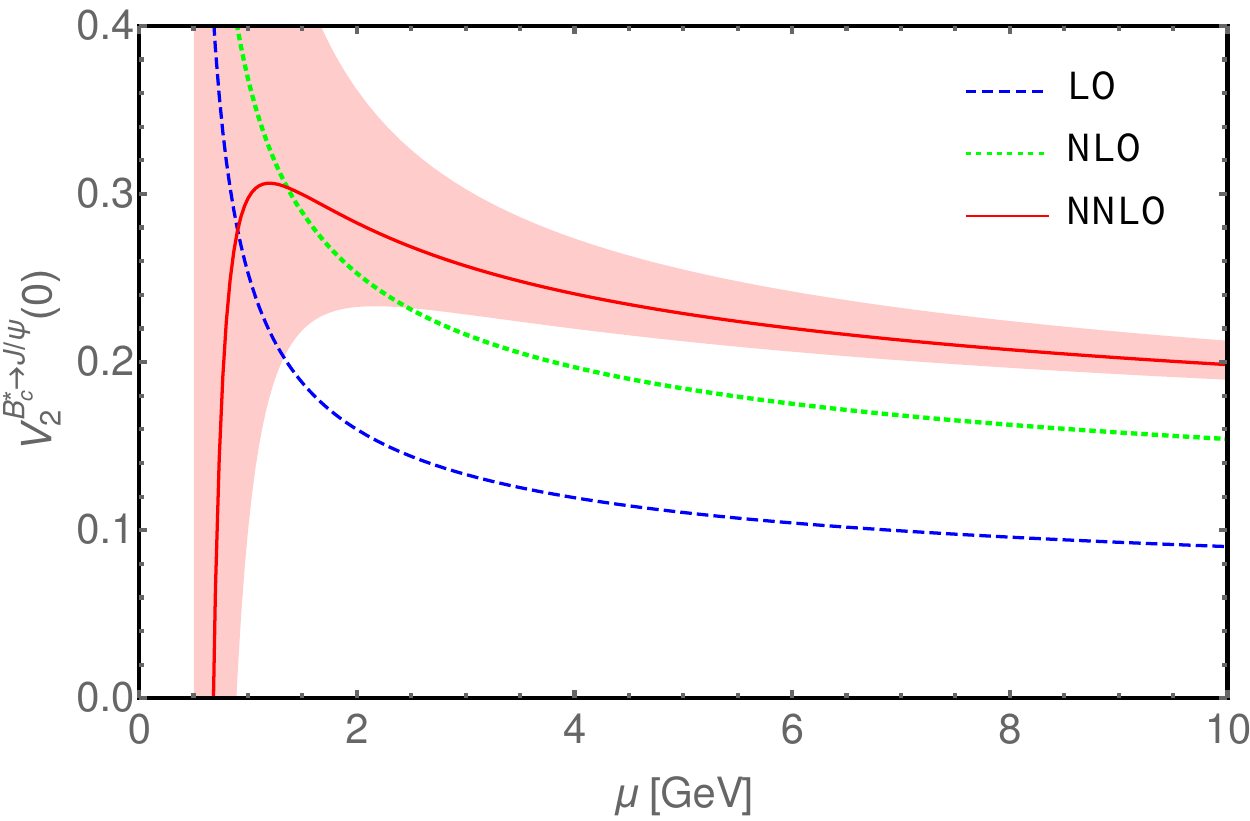}\quad
	\includegraphics[width=0.315\textwidth]{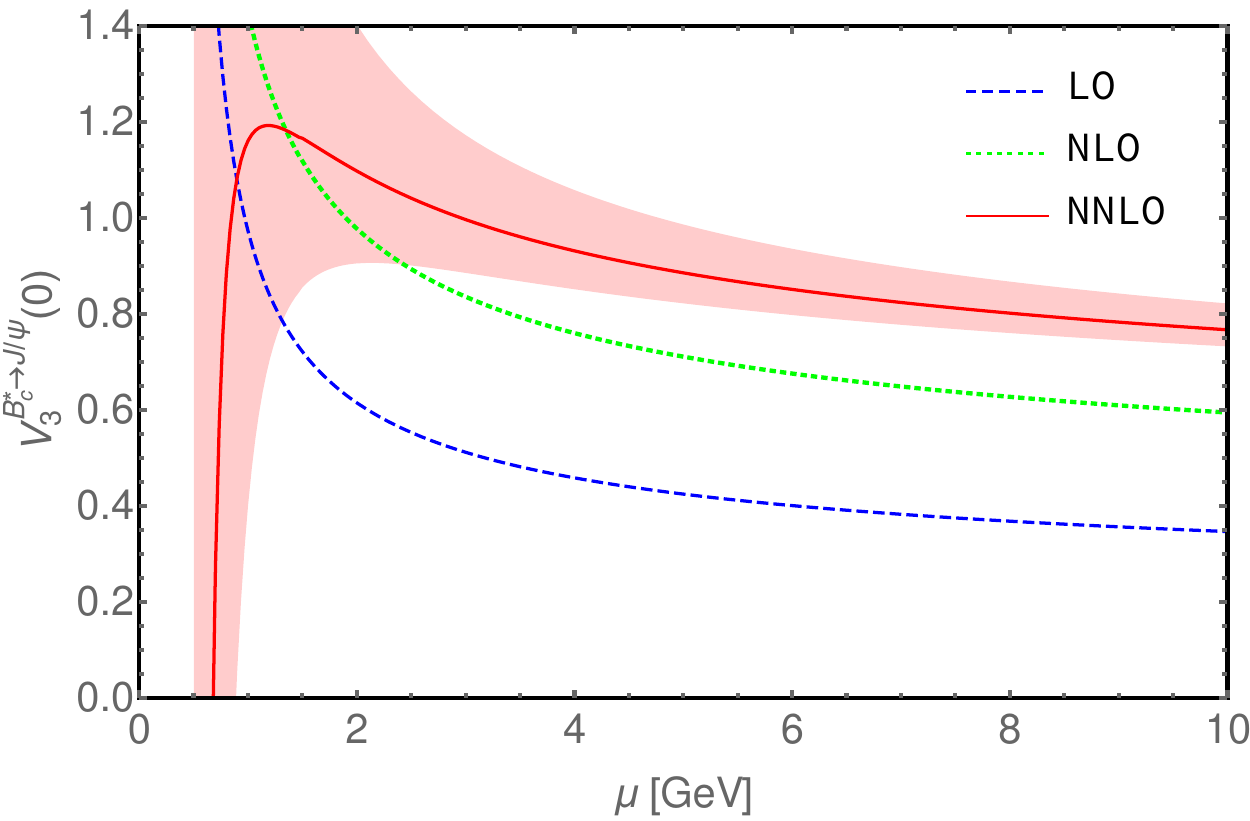}\quad
	\includegraphics[width=0.315\textwidth]{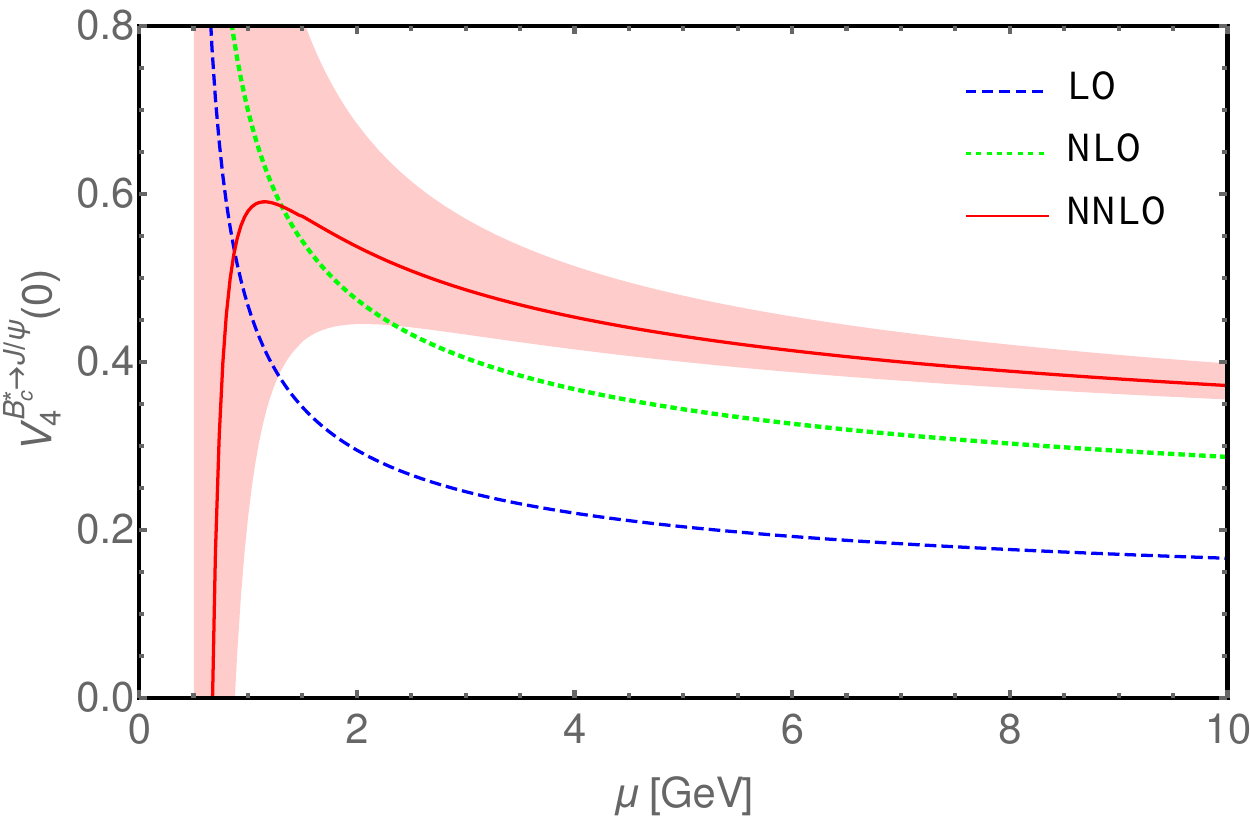}\quad
	\includegraphics[width=0.315\textwidth]{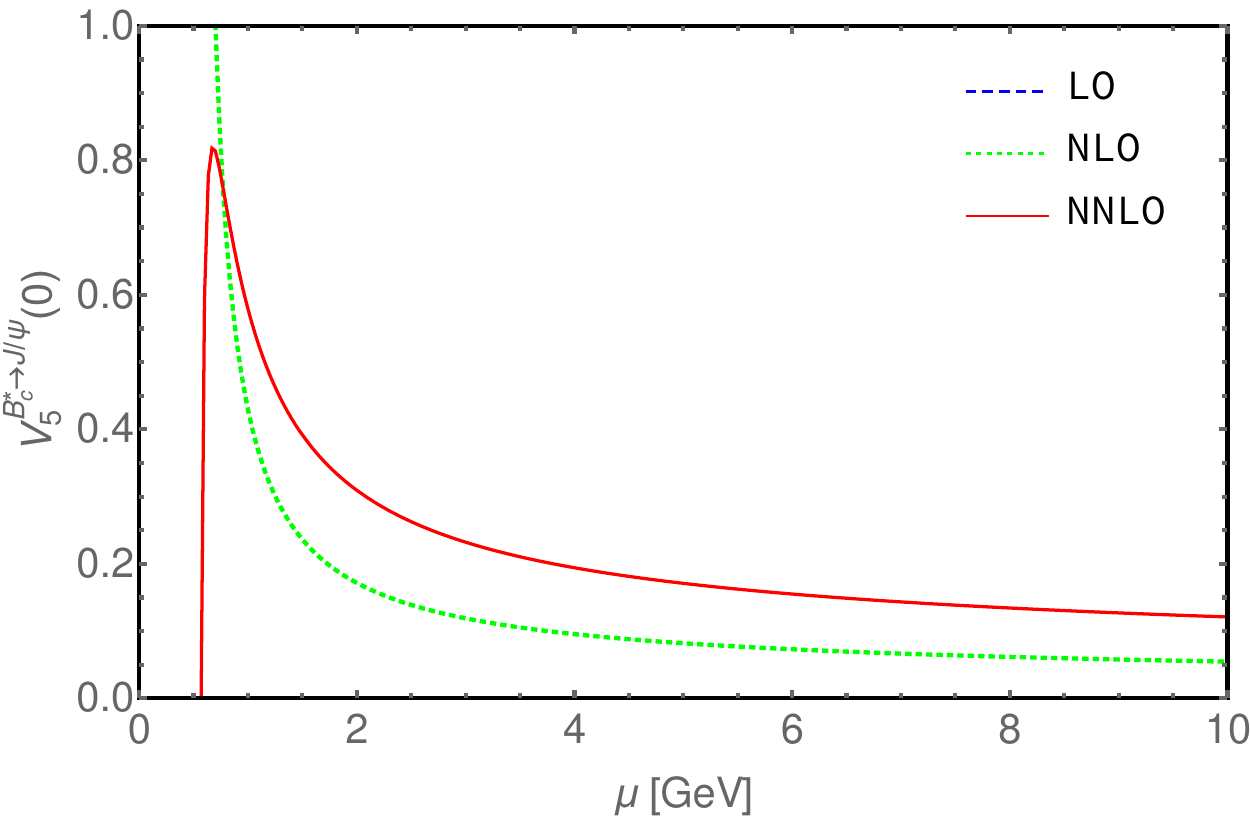}\quad
	\includegraphics[width=0.315\textwidth]{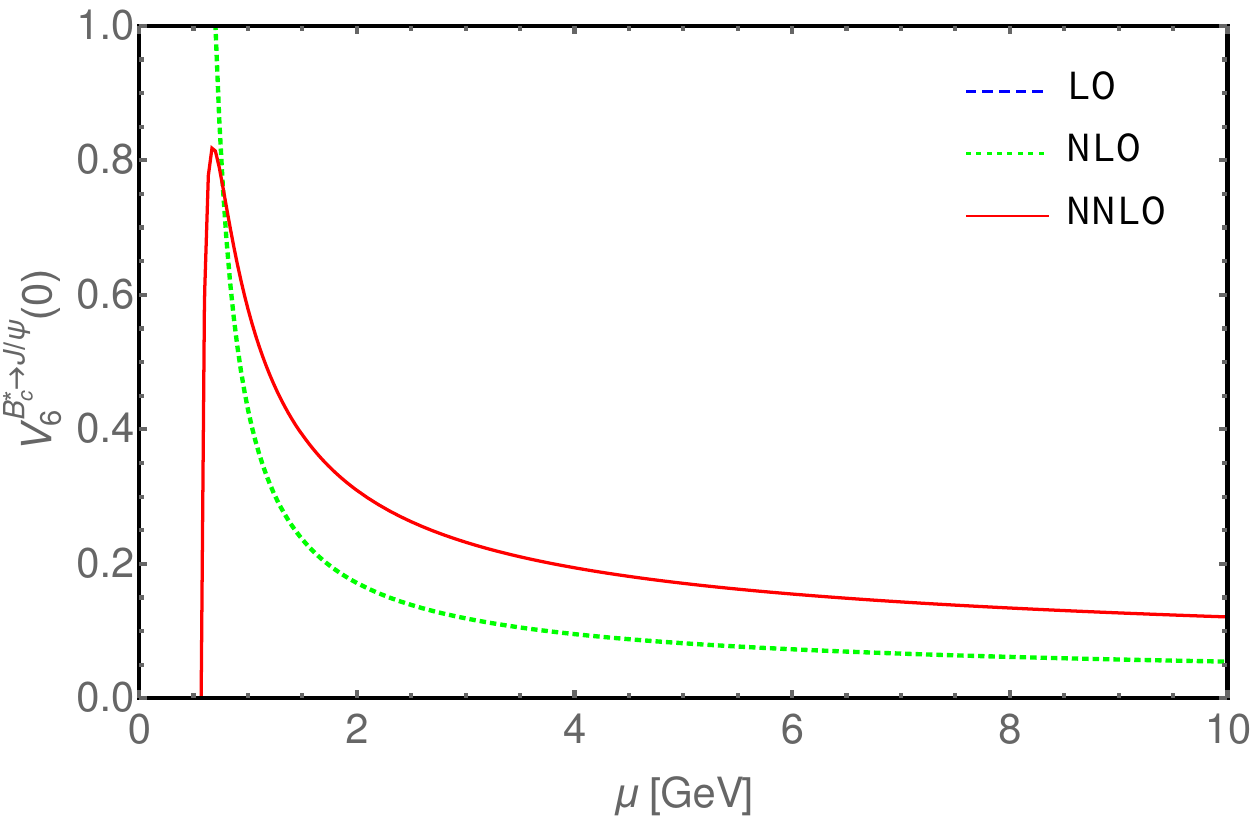}\quad
	\includegraphics[width=0.315\textwidth]{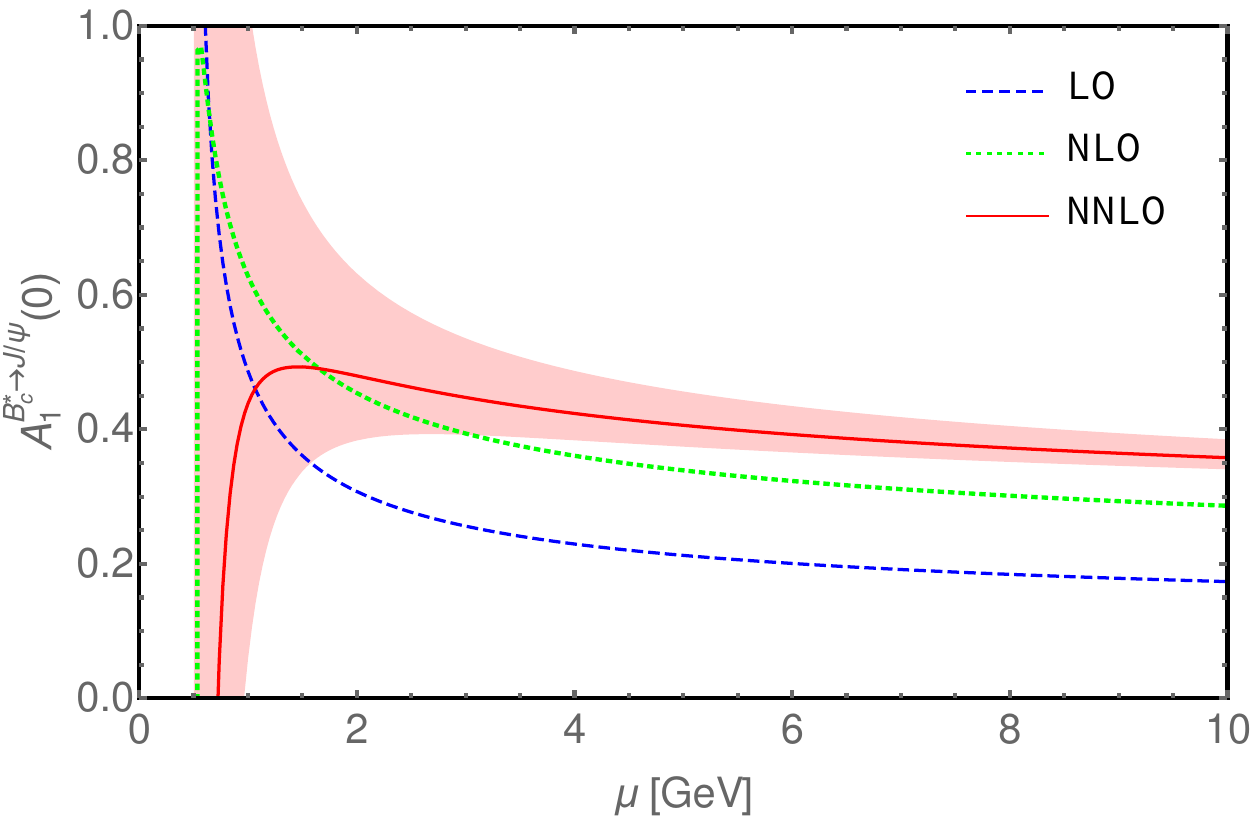}\quad	
	\includegraphics[width=0.315\textwidth]{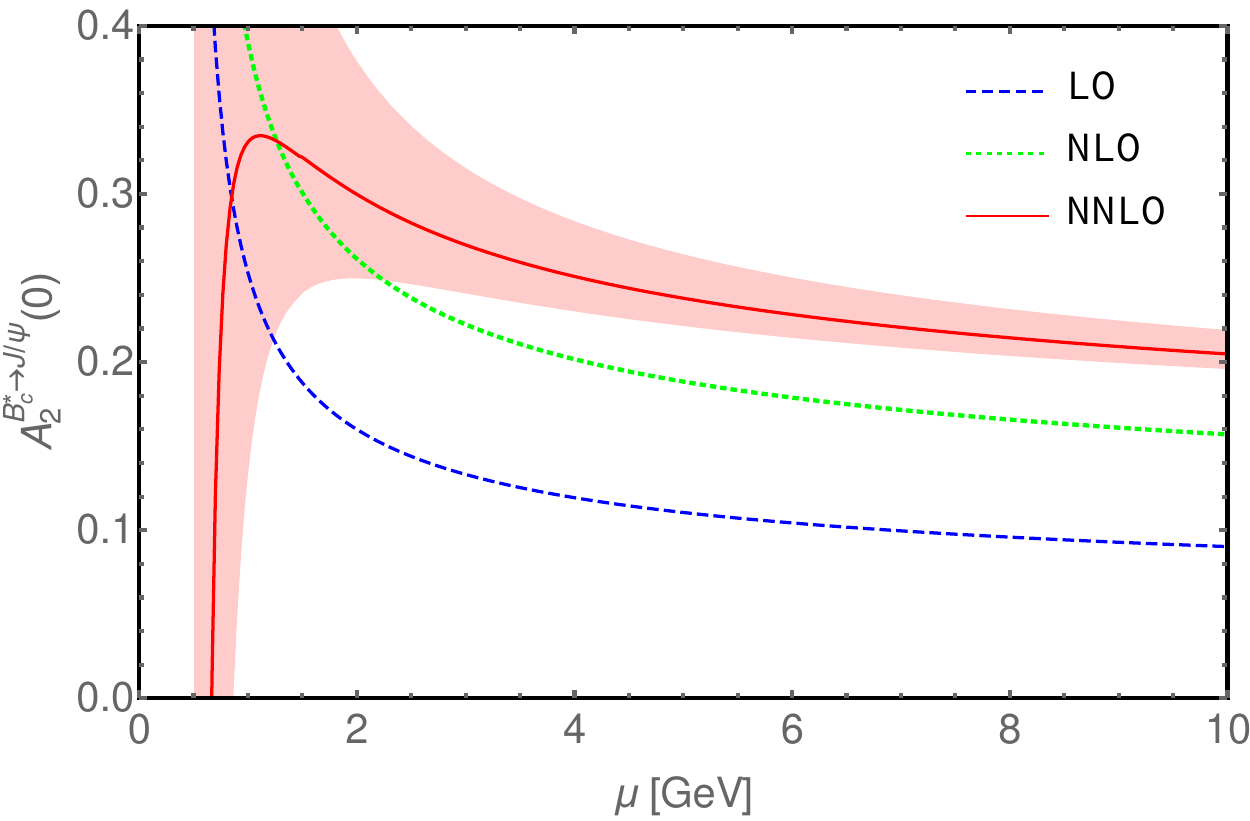}\quad
	\includegraphics[width=0.315\textwidth]{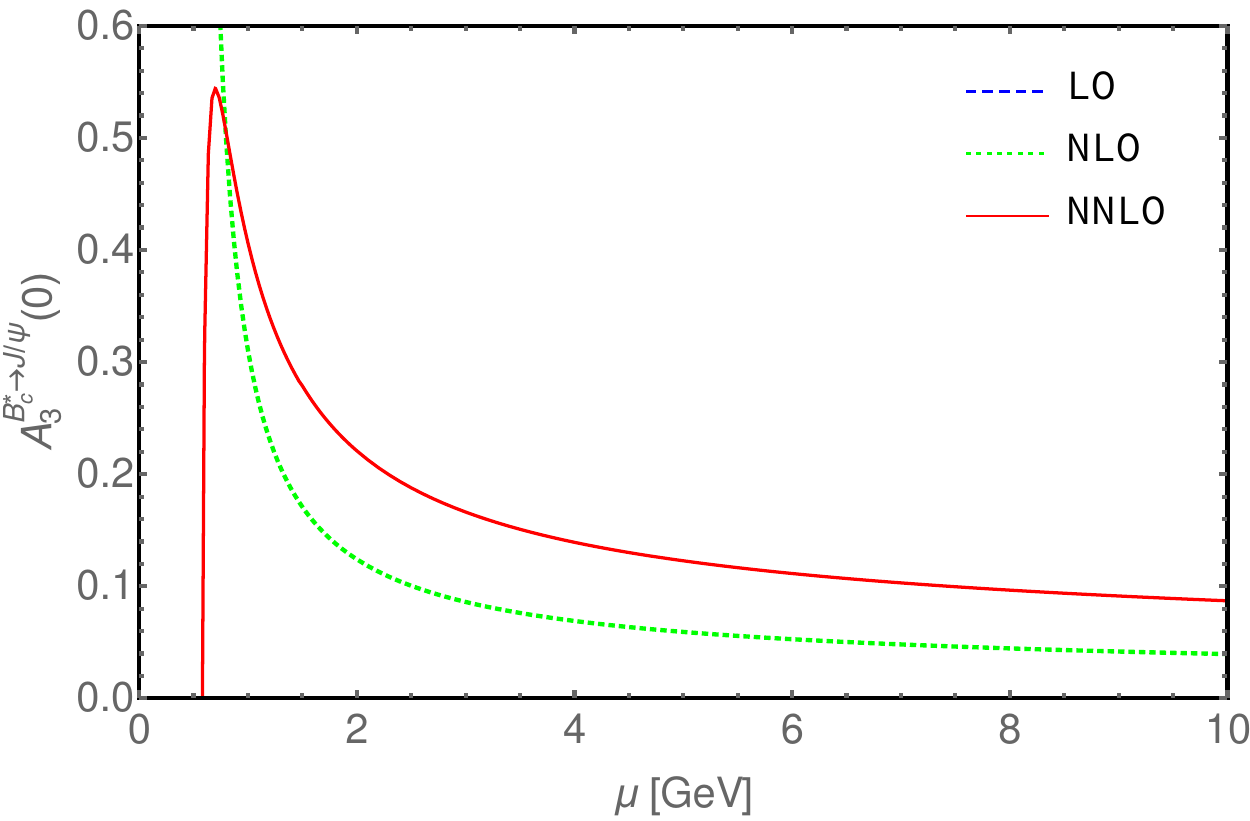}\quad
	\includegraphics[width=0.315\textwidth]{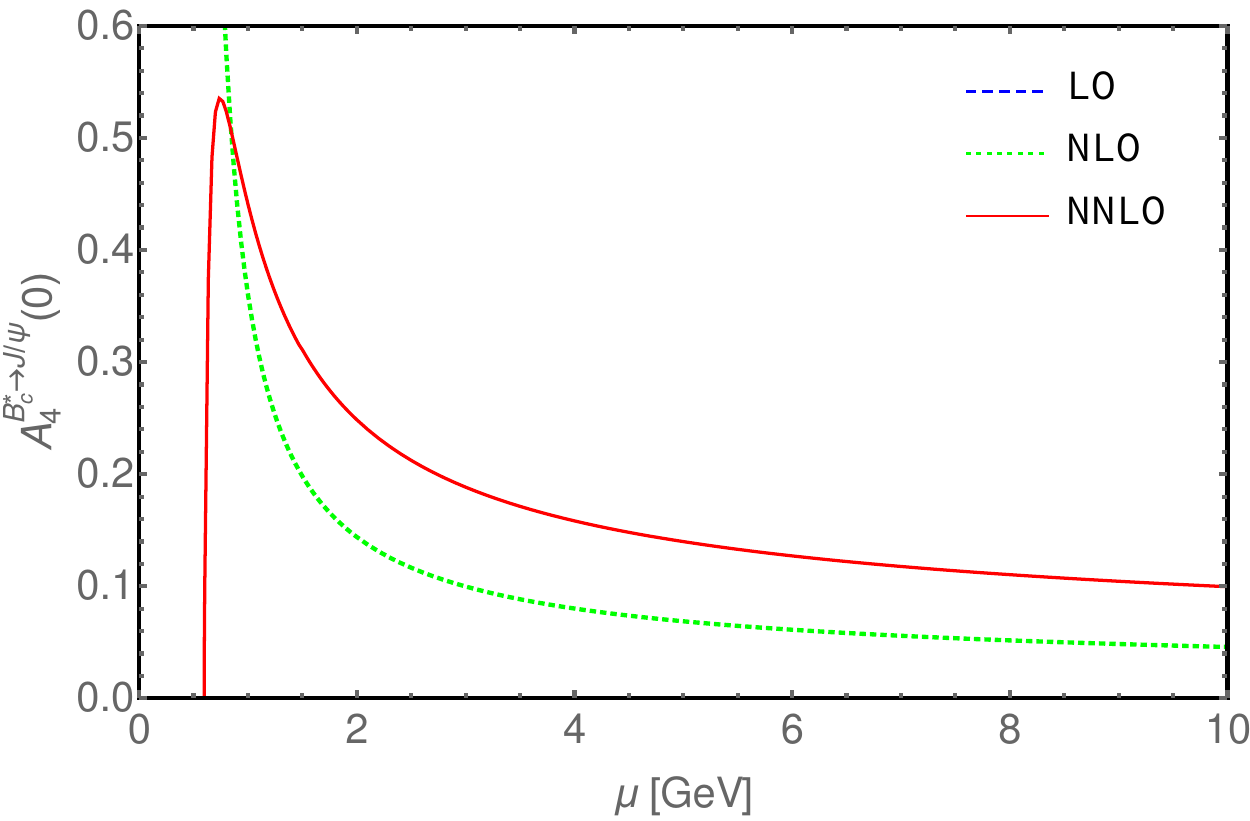}\quad		
	\includegraphics[width=0.315\textwidth]{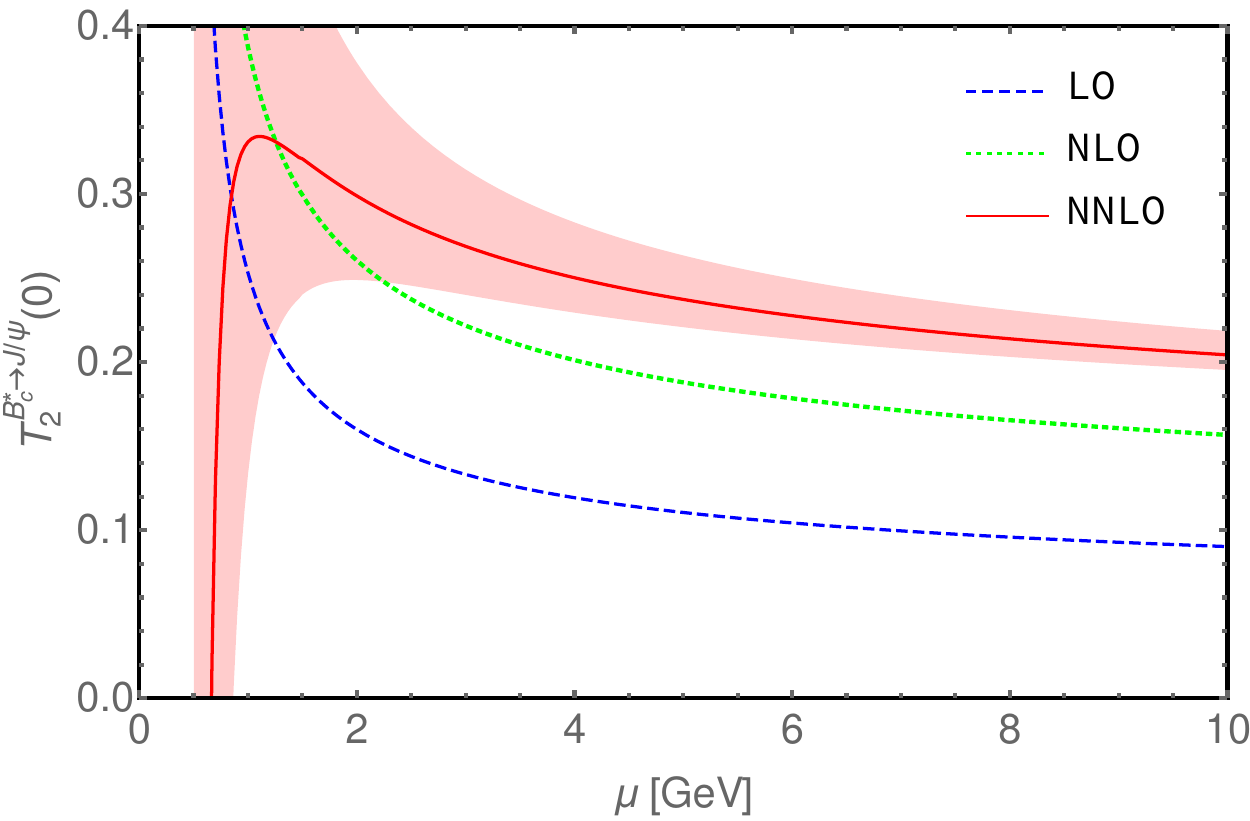}\quad
	\includegraphics[width=0.315\textwidth]{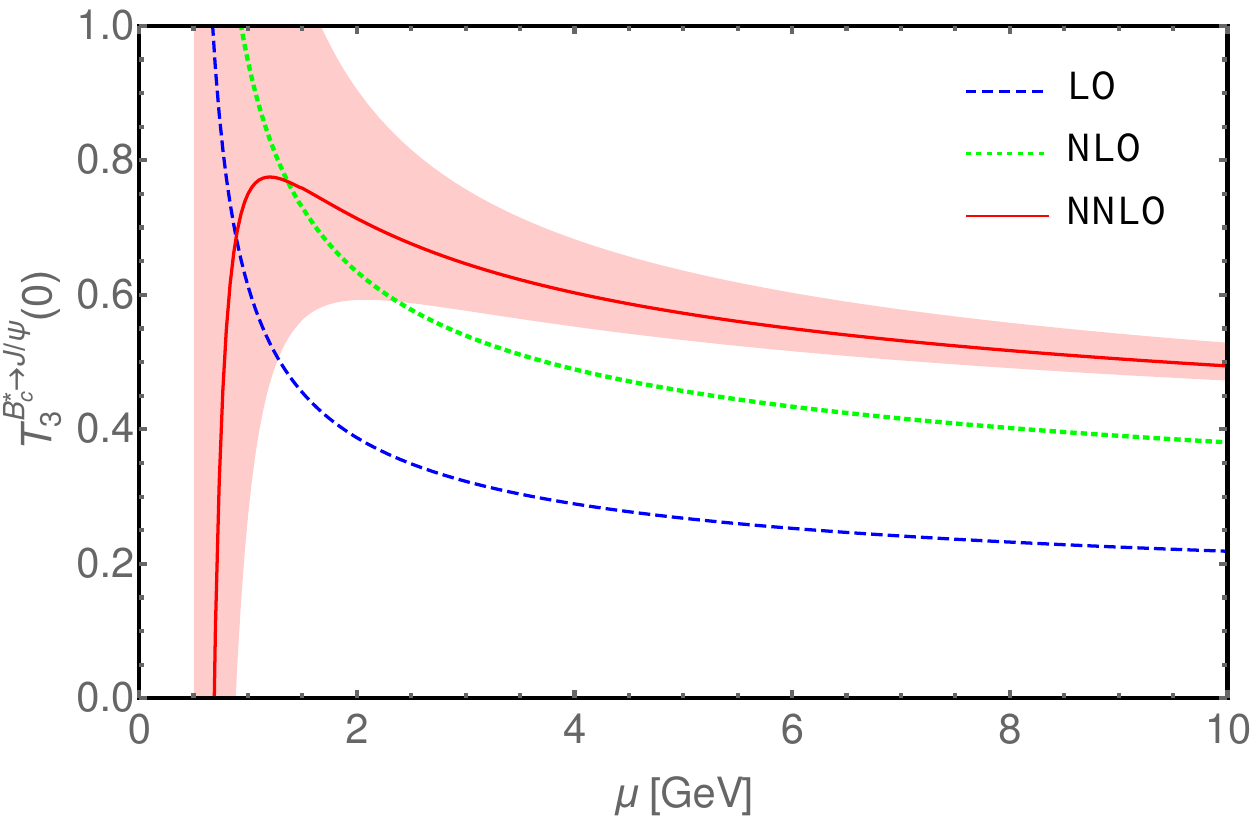}\quad
	\includegraphics[width=0.315\textwidth]{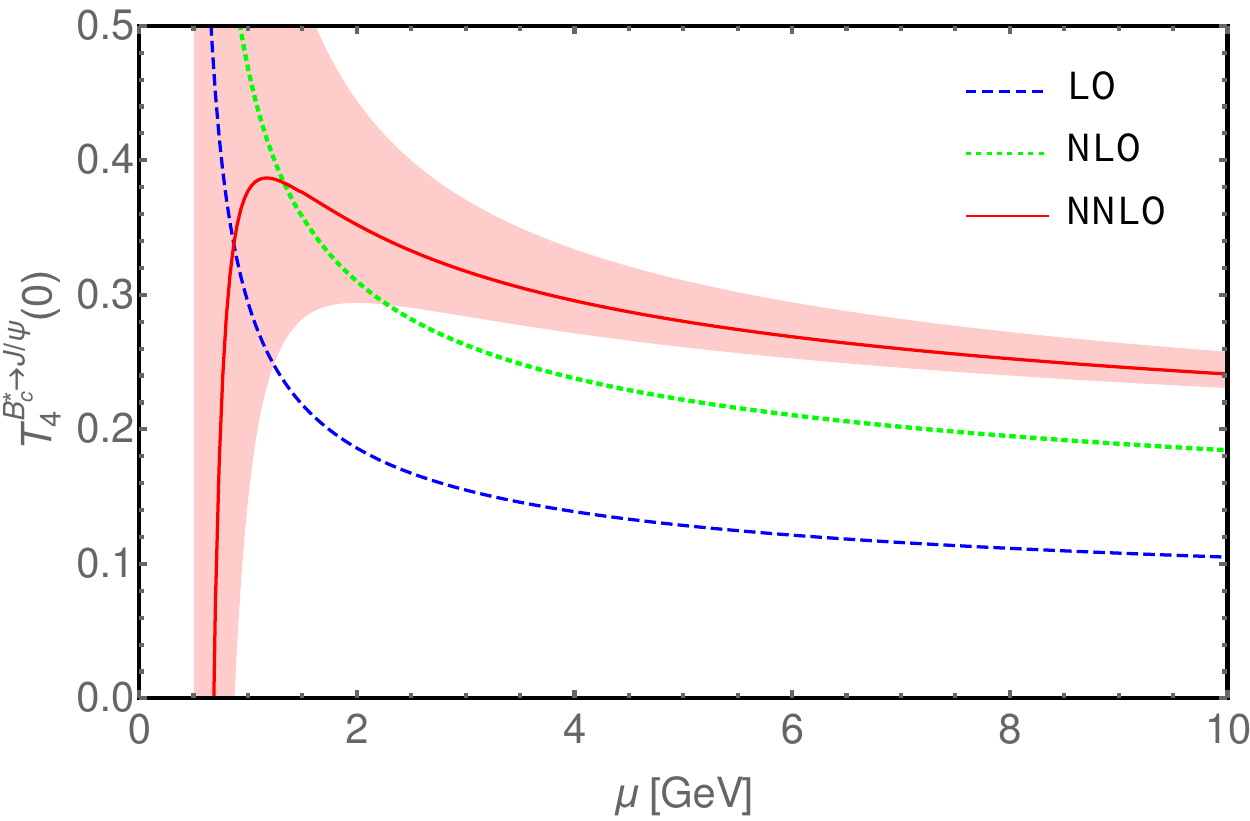}\quad
	\includegraphics[width=0.315\textwidth]{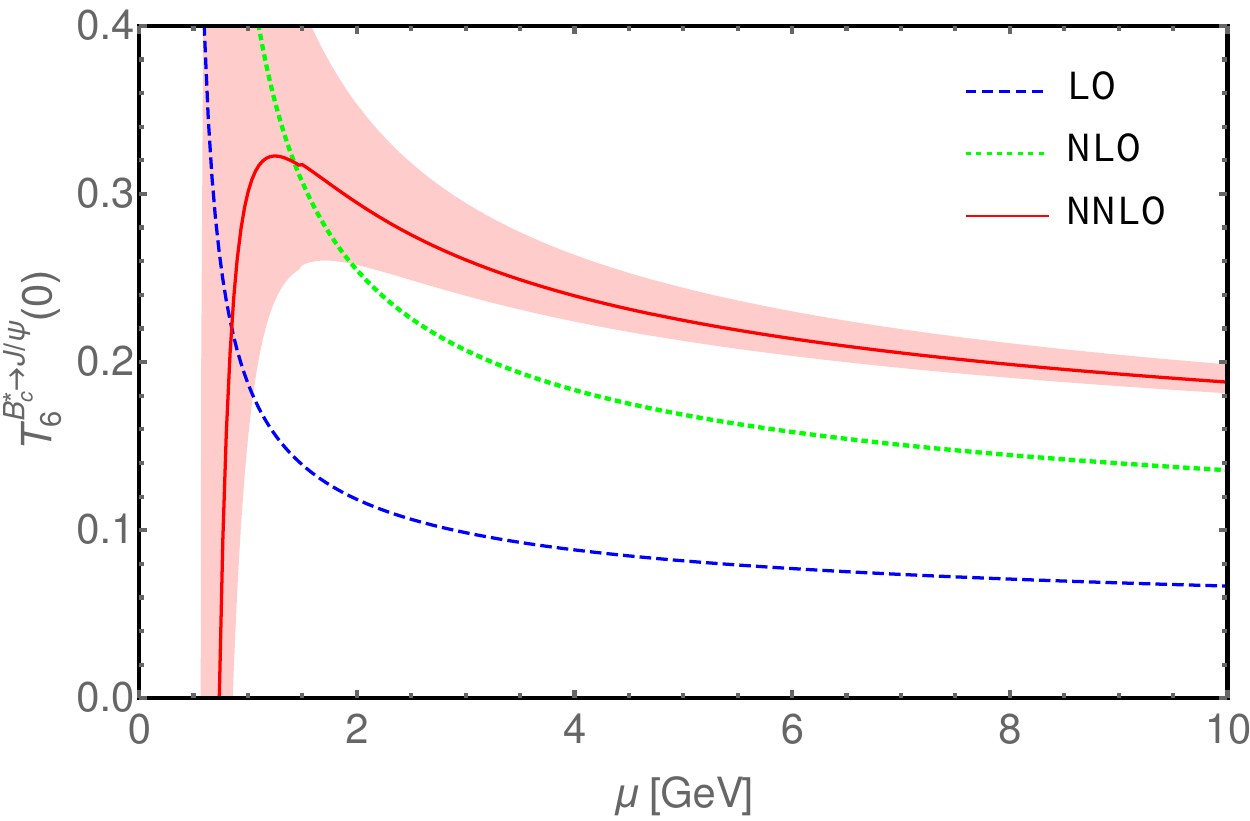}\quad
	\includegraphics[width=0.315\textwidth]{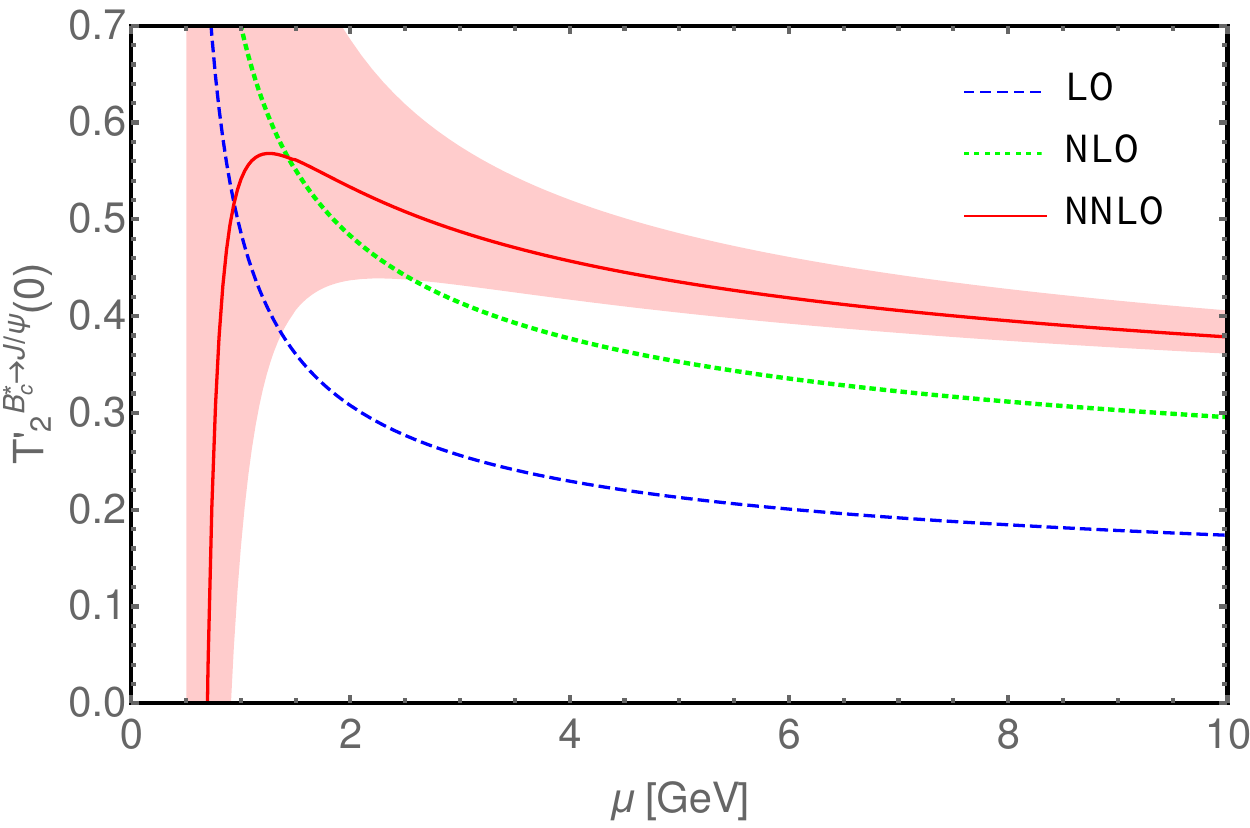}\quad
	\includegraphics[width=0.315\textwidth]{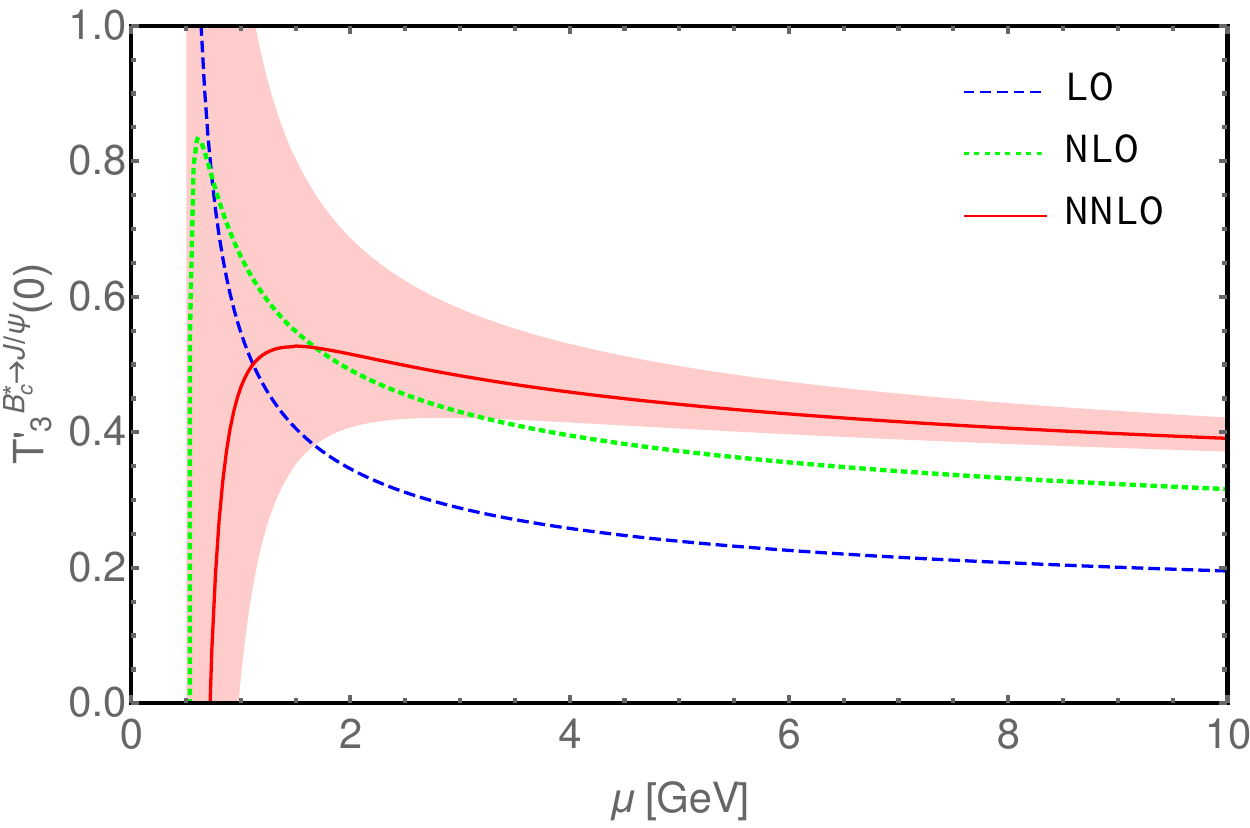}\quad
	\includegraphics[width=0.315\textwidth]{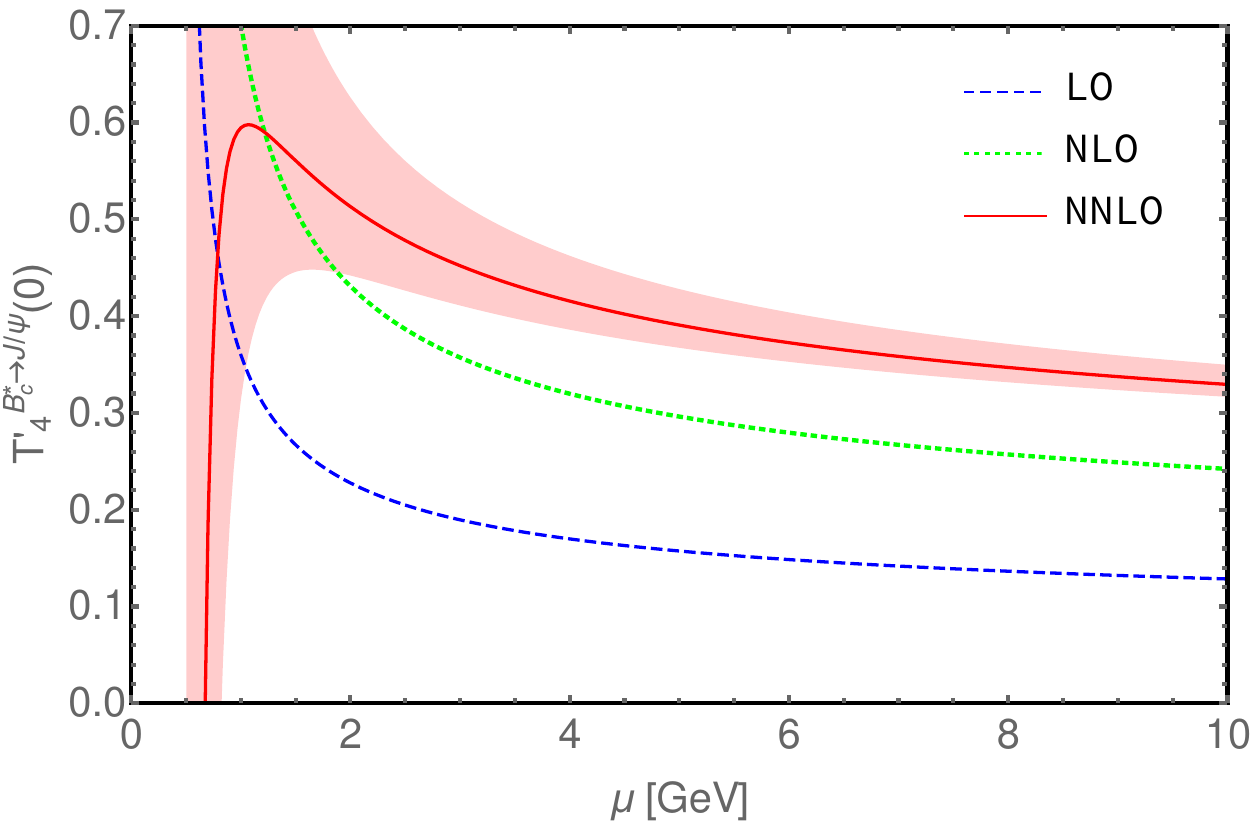}
	\caption{Same as figure~\ref{fig:ff0mubcjpsi}, but for $B_c^*\to J/\psi$, with $V_5^{B_c^*\to J/\psi}(0)=V_6^{B_c^*\to J/\psi}(0)$.
		}
\end{figure}

\begin{figure}[htbp]%
	\centering
	\includegraphics[width=0.44\textwidth]{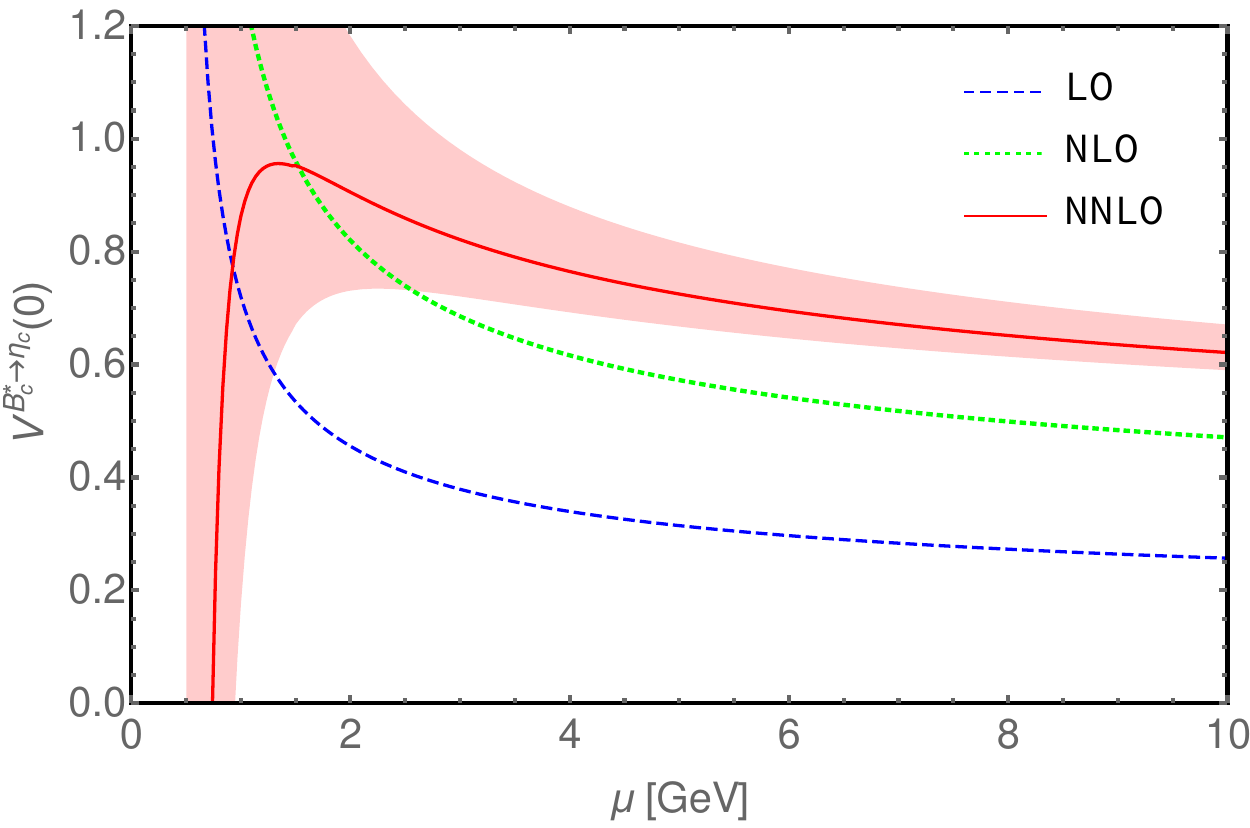}\qquad %
	\includegraphics[width=0.44\textwidth]{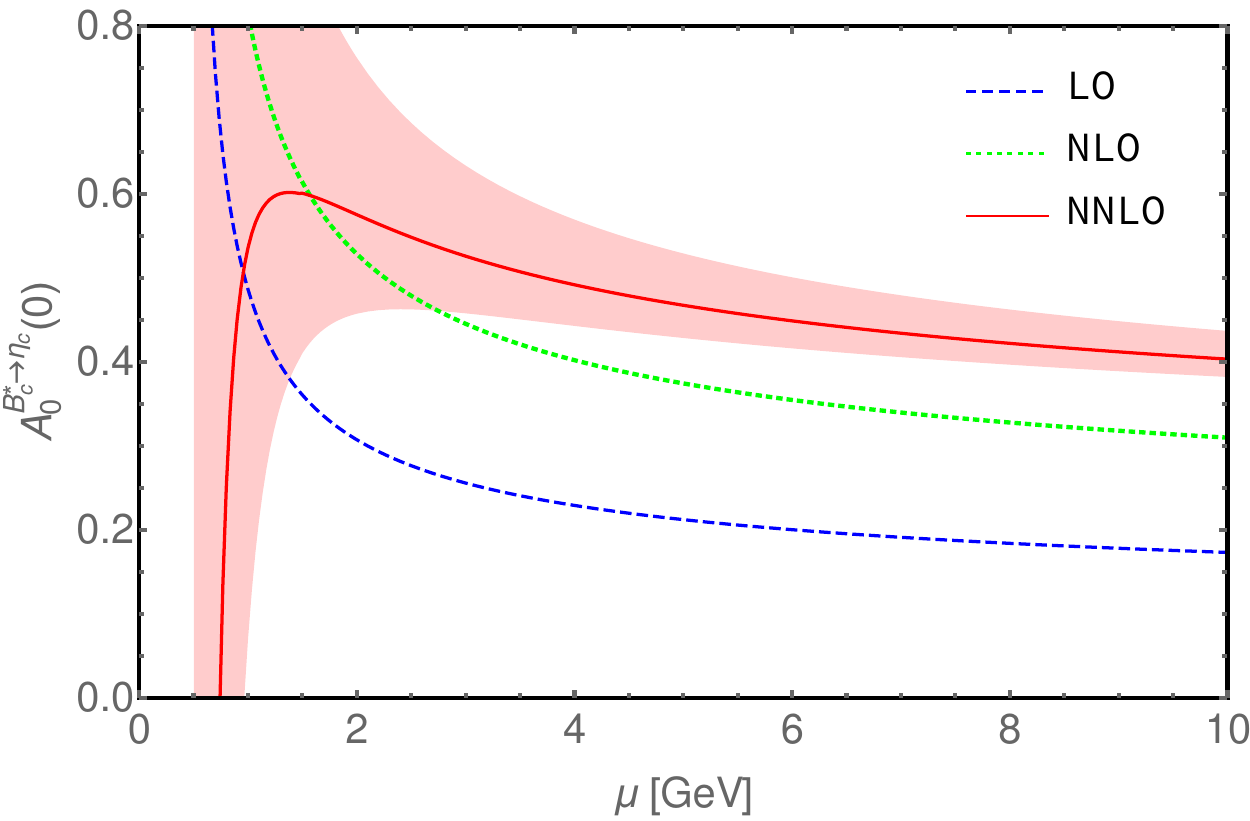}\qquad
	\includegraphics[width=0.44\textwidth]{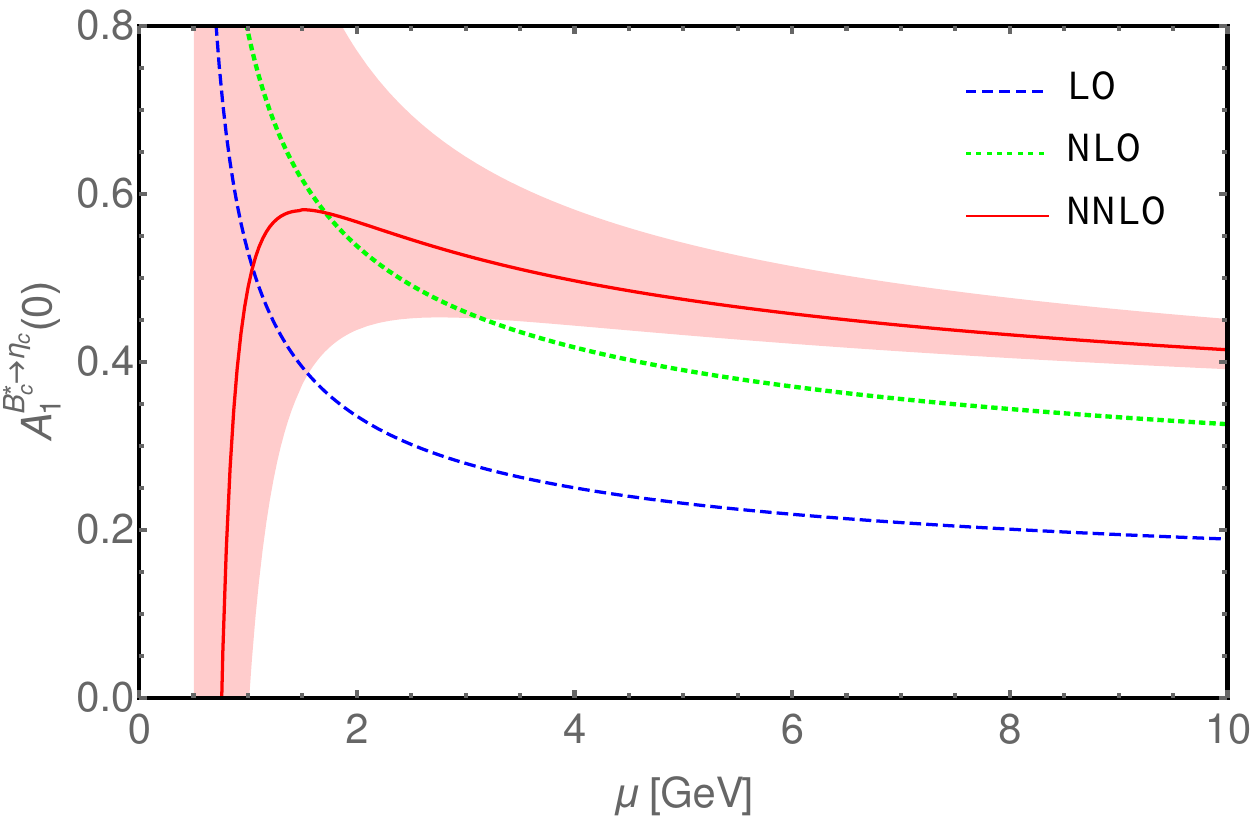}\qquad
	\includegraphics[width=0.44\textwidth]{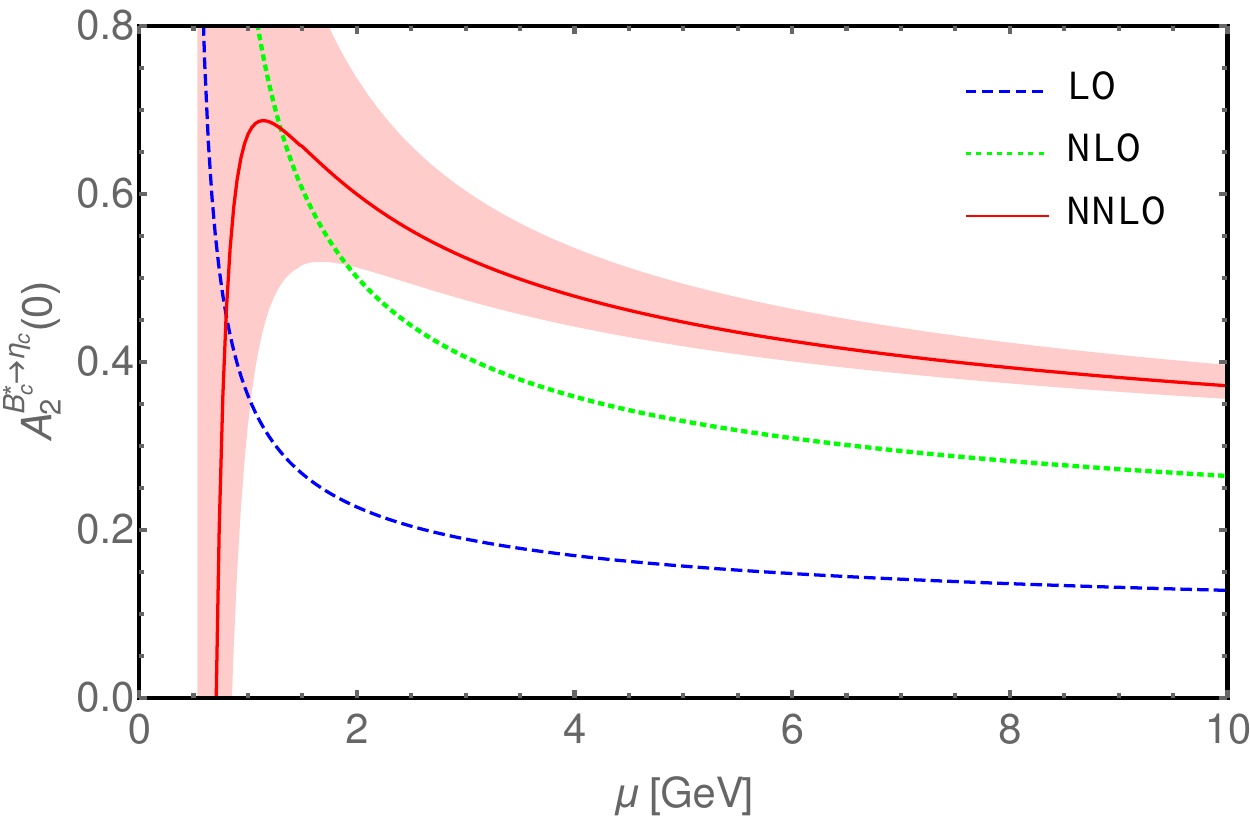}\qquad
	\includegraphics[width=0.44\textwidth]{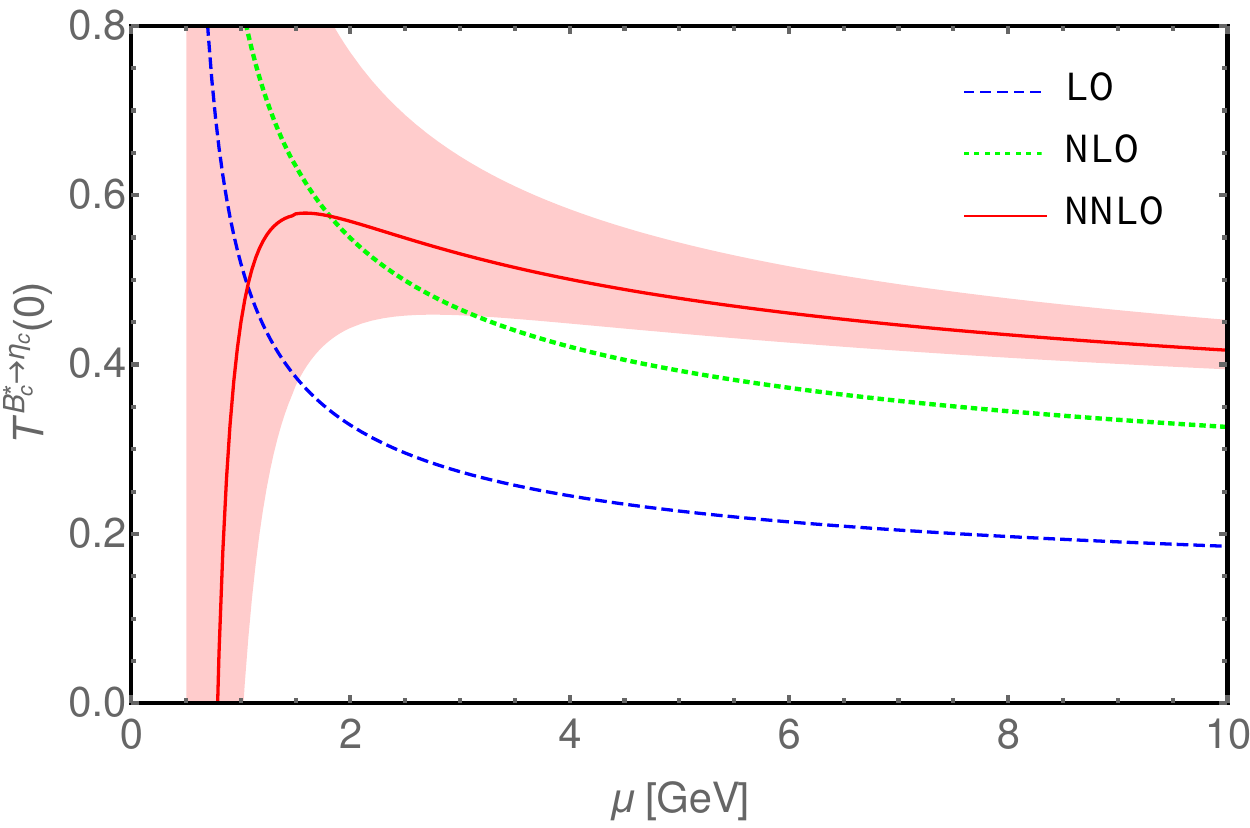}\qquad
	\includegraphics[width=0.44\textwidth]{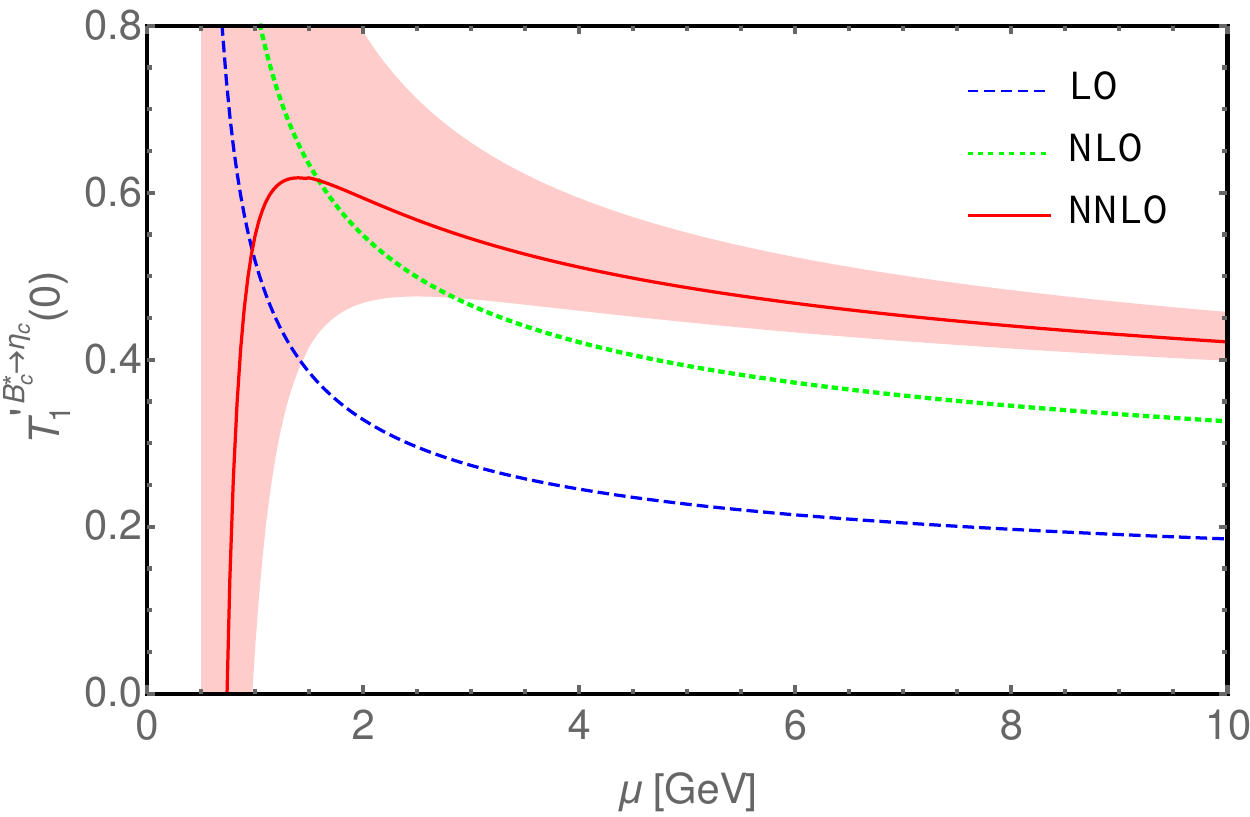}\qquad	
	\includegraphics[width=0.44\textwidth]{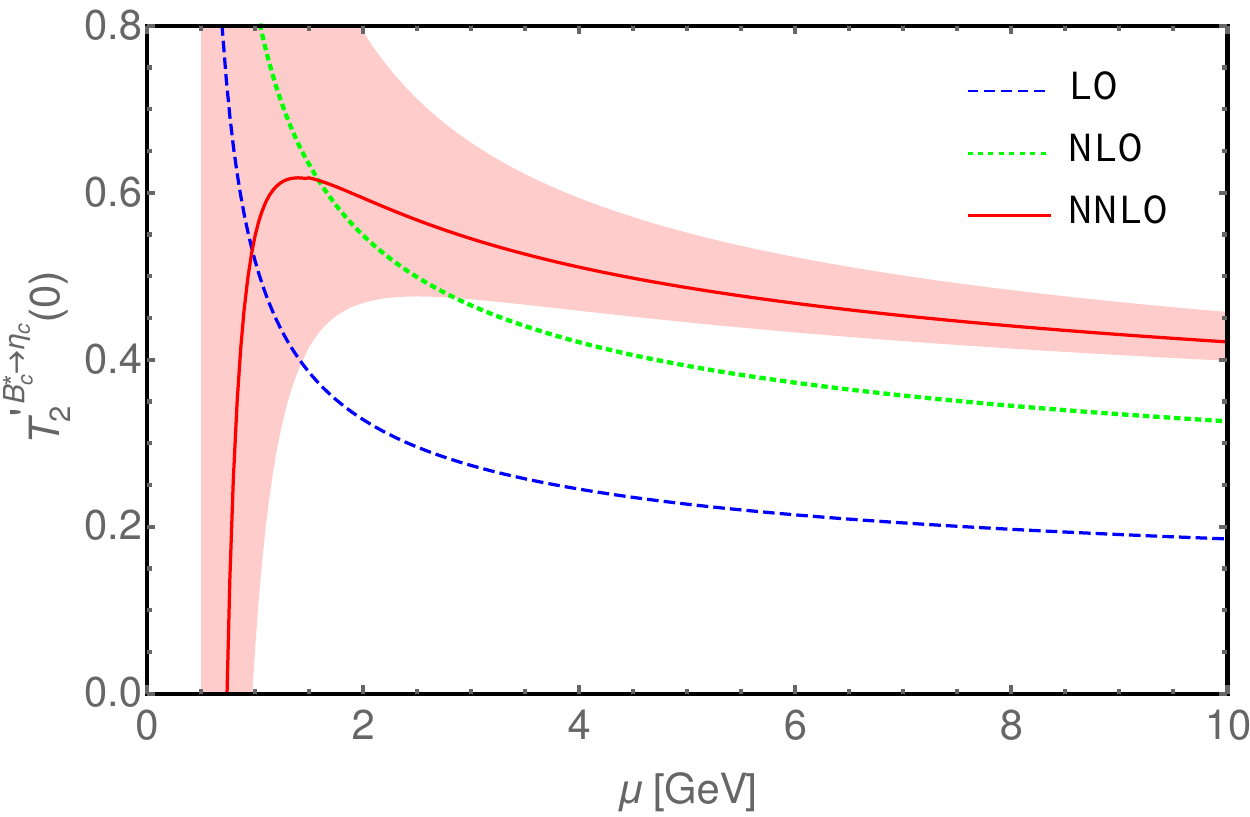}
	\caption{Same as figure~\ref{fig:ff0mubcjpsi}, but for $B_c^*\to \eta_c$, with $T_1^{\prime B_c^*\to \eta_c}(0)=T_2^{\prime B_c^*\to \eta_c}(0)$.
	The minor discrepancy between $T^{B_c^*\to \eta_c}(0)$ and $T_1'^{B_c^*\to \eta_c}(0)$ at the NNLO level is due to the $\gamma_5$-trace prescription (see the discussion in the caption of table~\ref{tab:ff0numbcjpsi}).
  }
\label{fig:ff0mubcvetac}
\end{figure}

With these parameter settings, we first investigate the behavior of the NNLO QCD corrections to the transition form factors at the maximum-recoil point, $q^2=0$.
Figures~\ref{fig:ff0mubcjpsi}--\ref{fig:ff0mubcvetac} show the renormalization-scale dependence of the form factors for $B_c\to J/\psi$, $B_c\to\eta_c$, $B_c^*\to J/\psi$, and $B_c^*\to\eta_c$, respectively, at LO, NLO, and NNLO accuracy, with the heavy-quark masses and wave functions at the origin fixed at their central values given above.
The renormalization scale $\mu$ is varied continuously over the range $\mu\in[0.5,10]~\mathrm{GeV}$, while the shaded error bands attached to the NNLO predictions represent the residual factorization-scale uncertainty obtained by varying $\mu_f\in[0.5,3.0]~\mathrm{GeV}$. 

As shown in figures~\ref{fig:ff0mubcjpsi}--\ref{fig:ff0mubcvetac}, the LO predictions exhibit a pronounced dependence on the renormalization scale $\mu$, reflecting the uncompensated scale dependence introduced by $\alpha_s(\mu)$. 
The inclusion of higher-order QCD corrections significantly enhances the central values and improves the perturbative stability. 
In particular, the NNLO curves develop a broad, relatively flat plateau over the perturbatively reliable range of $\mu$, demonstrating a substantially reduced renormalization-scale dependence relative to lower orders. 
The successive perturbative corrections decrease in size from NLO to NNLO, indicating good perturbative convergence of the $\alpha_s$ expansion series. As $\mu$ increases, the shaded NNLO error bands gradually narrow, reflecting the decreasing $\alpha_s(\mu)$ and the corresponding suppression of the factorization-scale dependence. The relatively narrow bands throughout the physical domain also indicate that the factorization-scale dependence is comparatively weak and well under control at NNLO. At very low scales, the perturbative predictions become highly sensitive to $\mu$ and $\mu_f$, reflecting the deteriorating perturbative convergence and the enhanced nonperturbative effects; we therefore adopt the renormalization-scale range specified in eq.~\eqref{mbmcmumufval} for the uncertainty analysis.

\begin{table}[htbp]
	\centering
	\caption{%
		The central values and individual theoretical uncertainties for the $B_c \to J/\psi$ form factors at $q^2 = 0$, evaluated at LO, NLO, and NNLO accuracy.
The five successive uncertainties associated with each central value propagate from the variations of $m_b = 4.75\pm 0.50~\text{GeV}$, $m_c = 1.50\pm 0.50~\text{GeV}$, $\mu = 4.75^{+5.25}_{-3.25}~\text{GeV}$, $\mu_f = 1.5^{+1.5}_{-1.0}~\text{GeV}$, and $\psi_{B_c}(0) \psi_{J/\psi}(0)$ (with $\psi_{B_c}(0) = 0.325\pm 0.025~\text{GeV}^{3/2}$ and $\psi_{J/\psi}(0) = 0.225\pm 0.025~\text{GeV}^{3/2}$), respectively.
It can be verified that %
 $A_{0,1,2}^{B_c\to J/\psi}(0)$ satisfy the relation in eq.~\eqref{ff0relationbcjpsiav}.		
			At the two-loop level, the small numerical mismatch between $T_1^{B_c \to J/\psi}(0)$ and $T_2^{B_c \to J/\psi}(0)$ is most likely an unphysical artifact originating from the different
	$\gamma_5$-trace prescriptions (i.e., the naive $\gamma_5$ scheme vs. the fixed-reading-point scheme) employed for the even and odd fermion chains in
	$D$ dimensions. A systematic study of the symmetry-restoring finite renormalization constants, such as $Z_{5,p}^{\mathrm{fin}}=1+\alpha_s^2 C_F^2 \frac{2x}{(1+x)^2}$ with $x=m_c/m_b$ or 1, including both the flavor-singlet and flavor-nonsinglet cases, to analytically enforce $T_1^{B_c \to J/\psi}(0)=T_2^{B_c \to J/\psi}(0)$ and $T^{B_c^* \to \eta_c}(0)=T_1'^{B_c^* \to \eta_c}(0)$ at the NNLO level remains to be addressed in future work.}
	\label{tab:ff0numbcjpsi}
	\setlength{\tabcolsep}{1.0mm}
	\renewcommand{\arraystretch}{1.8}%
	\resizebox{\textwidth}{!}
	{
		\begin{tabular}{c|c|c|c}
			\hline
			& LO & NLO & NNLO\\
			\hline
			
			$V^{B_c\to J/\psi}(0)$
			&
			$0.3201_{+0.2011-0.0748+0.2142+0-0.0575}^{-0.1065+0.2089-0.0631-0+0.0629}$
			&
			$0.5565_{+0.3660-0.1252+0.3332+0-0.0999}^{-0.1907+0.3640-0.1019-0+0.1094}$
			&
			$0.6916_{+0.4685-0.1686+0.1926+0.0913-0.1241}^{-0.2424+0.4567-0.1012-0.0576+0.1360}$
			\\
			\hline
			
			$A_0^{B_c\to J/\psi}(0)$
			&
			$0.2163_{+0.1262-0.0343+0.1447+0-0.0388}^{-0.0684+0.1159-0.0426-0+0.0425}$
			&
			$0.3665_{+0.2232-0.0518+0.2082+0-0.0658}^{-0.1192+0.1932-0.0655-0+0.0720}$
			&
			$0.4499_{+0.2809-0.0710+0.1146+0.0617-0.0808}^{-0.1494+0.2372-0.0634-0.0389+0.0884}$
			\\
			\hline
			
			$A_1^{B_c\to J/\psi}(0)$
			&
			$0.2258_{+0.1302-0.0338+0.1510+0-0.0405}^{-0.0708+0.1167-0.0445-0+0.0444}$
			&
			$0.3733_{+0.2262-0.0513+0.2007+0-0.0670}^{-0.1211+0.1930-0.0650-0+0.0734}$
			&
			$0.4515_{+0.2821-0.0735+0.0999+0.0644-0.0810}^{-0.1501+0.2384-0.0607-0.0406+0.0888}$
			\\
			\hline
			
			$A_2^{B_c\to J/\psi}(0)$
			&
			$0.2433_{+0.1420-0.0406+0.1628+0-0.0437}^{-0.0771+0.1290-0.0480-0+0.0478}$
			&
			$0.3858_{+0.2351-0.0560+0.1870+0-0.0693}^{-0.1258+0.1998-0.0642-0+0.0758}$
			&
			$0.4544_{+0.2852-0.0774+0.0728+0.0694-0.0816}^{-0.1516+0.2437-0.0557-0.0438+0.0893}$
			\\
			\hline
			
			$T_1^{B_c\to J/\psi}(0)$
			&
			$0.2309_{+0.1312-0.0296+0.1545+0-0.0414}^{-0.0716+0.1151-0.0455-0+0.0454}$
			&
			$0.3812_{+0.2296-0.0480+0.2043+0-0.0684}^{-0.1230+0.1946-0.0663-0+0.0749}$
			&
			$0.4550_{+0.2844-0.0724+0.0801+0.0658-0.0817}^{-0.1511+0.2407-0.0590-0.0415+0.0894}$
			\\
			\hline
			
			$T_2^{B_c\to J/\psi}(0)$
			&
			$0.2309_{+0.1312-0.0296+0.1545+0-0.0414}^{-0.0716+0.1151-0.0455-0+0.0454}$
			&
			$0.3812_{+0.2296-0.0480+0.2043+0-0.0684}^{-0.1230+0.1946-0.0663-0+0.0749}$
			&
			$0.4613_{+0.2886-0.0744+0.1033+0.0658-0.0828}^{-0.1533+0.2452-0.0620-0.0415+0.0907}$
			\\
			\hline
			
			$T_3^{B_c\to J/\psi}(0)$
			&
			$0.1271_{+0.0563+0.0109+0.0850+0-0.0228}^{-0.0335+0.0220-0.0251-0+0.0250}$
			&
			$0.1505_{+0.0585+0.0396+0.0066+0-0.0270}^{-0.0370-0.0099-0.0154-0+0.0296}$
			&
			$0.1301_{+0.0390+0.0490-0.0970+0.0362-0.0234}^{-0.0282-0.0331+0.0046-0.0229+0.0256}$
			\\
			
			\hline
		\end{tabular}
	}
\end{table}

\begin{table}[htbp]
	\centering
	\caption{Same as table~\ref{tab:ff0numbcjpsi}, but for $B_c\to \eta_c$, with 	$F_p^{B_c\to \eta_c}(0)=F_0^{B_c\to \eta_c}(0)$.}
	\setlength{\tabcolsep}{1.0mm}
	\renewcommand{\arraystretch}{1.8}%
	\resizebox{\textwidth}{!}
	{

\begin{tabular}{c|c|c|c}
	\hline
	& LO & NLO & NNLO\\
	\hline
	
	$F_p^{B_c\to\eta_c}(0)$
	&
	$0.2455_{+0.1362-0.0248+0.1643+0-0.0441}
	^{-0.0748+0.1142-0.0484-0+0.0483}$
	&
	$0.4297_{+0.2484-0.0344+0.2608+0-0.0771}
	^{-0.1346+0.1961-0.0792-0+0.0845}$
	&
	$0.5279_{+0.3128-0.0518+0.1235+0.0835-0.0947}
	^{-0.1688+0.2427-0.0753-0.0527+0.1038}$
	\\
	\hline
	
	$F_0^{B_c\to\eta_c}(0)$
	&
	$0.2455_{+0.1362-0.0248+0.1643+0-0.0441}
	^{-0.0748+0.1142-0.0484-0+0.0483}$
	&
	$0.4297_{+0.2484-0.0344+0.2608+0-0.0771}
	^{-0.1346+0.1961-0.0792-0+0.0845}$
	&
	$0.5279_{+0.3128-0.0518+0.1235+0.0835-0.0947}
	^{-0.1688+0.2427-0.0753-0.0527+0.1038}$
	\\
	\hline
	
	$F_t^{B_c\to\eta_c}(0)$
	&
	$0.3201_{+0.2011-0.0748+0.2142+0-0.0575}
	^{-0.1065+0.2089-0.0631-0+0.0629}$
	&
	$0.6177_{+0.3965-0.1251+0.4426+0-0.1109}
	^{-0.2081+0.3834-0.1237-0+0.1214}$
	&
	$0.8183_{+0.5329-0.1638+0.3469+0.1089-0.1469}
	^{-0.2791+0.4926-0.1407-0.0687+0.1609}$
	\\
	
	\hline
\end{tabular}

	}
\end{table}

\begin{table}[htbp]
	\centering
	\caption{Same as table~\ref{tab:ff0numbcjpsi}, but for $B_c^*\to J/\psi$, with 	
			$V_5^{B_c^*\to J/\psi}(0)=V_6^{B_c^*\to J/\psi}(0)$. It can be verified that $T_{3,4}^{B_c^*\to J/\psi}(0)$ and $T_{2,3,4}^{\prime B_c^*\to J/\psi}(0)$ satisfy the relations in eqs.~\eqref{ff0relationbcvjpsit3} and~\eqref{ff0relationbcvjpsit4}.}
	\setlength{\tabcolsep}{1.0mm}
	\renewcommand{\arraystretch}{1.7}%
	\resizebox{\textwidth}{!}
	{
	\begin{tabular}{c|c|c|c}
		\hline
		& LO & NLO & NNLO\\
		\hline
		
		$V_1^{B_c^*\to J/\psi}(0)$
		&
		$0.2163_{+0.1262-0.0343+0.1447+0-0.0388}^{-0.0684+0.1159-0.0426-0+0.0425}$
		&
		$0.3537_{+0.2177-0.0524+0.1853+0-0.0635}^{-0.1158+0.1919-0.0609-0+0.0695}$
		&
		$0.4296_{+0.2748-0.0761+0.1046+0.0530-0.0771}^{-0.1448+0.2412-0.0581-0.0334+0.0845}$
		\\
		\hline
		
		$V_2^{B_c^*\to J/\psi}(0)$
		&
		$0.1125_{+0.0513+0.0062+0.0752+0-0.0202}^{-0.0303+0.0229-0.0222-0+0.0221}$
		&
		$0.1871_{+0.0914+0.0114+0.1020+0-0.0336}^{-0.0529+0.0404-0.0328-0+0.0368}$
		&
		$0.2315_{+0.1187+0.0048+0.0683+0.0275-0.0415}^{-0.0680+0.0537-0.0330-0.0174+0.0455}$
		\\
		\hline
		
		$V_3^{B_c^*\to J/\psi}(0)$
		&
		$0.4326_{+0.2525-0.0686+0.2894+0-0.0776}^{-0.1368+0.2318-0.0853-0+0.0850}$
		&
		$0.7220_{+0.4442-0.1075+0.3968+0-0.1296}^{-0.2364+0.3924-0.1270-0+0.1419}$
		&
		$0.8960_{+0.5731-0.1584+0.2699+0.1059-0.1608}^{-0.3020+0.5037-0.1288-0.0668+0.1761}$
		\\
		\hline
		
		$V_4^{B_c^*\to J/\psi}(0)$
		&
		$0.2076_{+0.1498-0.0810+0.1389+0-0.0373}^{-0.0762+0.1861-0.0409-0+0.0408}$
		&
		$0.3491_{+0.2631-0.1330+0.1950+0-0.0627}^{-0.1315+0.3144-0.0619-0+0.0686}$
		&
		$0.4358_{+0.3392-0.1734+0.1377+0.0508-0.0782}^{-0.1677+0.4001-0.0637-0.0321+0.0857}$
		\\
		\hline
		
		$V_5^{B_c^*\to J/\psi}(0)$
		&
		$0$
		&
		$0.0849_{+0.0567-0.0264+0.1515+0-0.0152}^{-0.0295+0.0634-0.0302-0+0.0167}$
		&
		$0.1760_{+0.1184-0.0446+0.2160+0-0.0316}^{-0.0619+0.1059-0.0549-0+0.0346}$
		\\
		\hline
		
		$V_6^{B_c^*\to J/\psi}(0)$
		&
		$0$
		&
		$0.0849_{+0.0567-0.0264+0.1515+0-0.0152}^{-0.0295+0.0634-0.0302-0+0.0167}$
		&
		$0.1760_{+0.1184-0.0446+0.2160+0-0.0316}^{-0.0619+0.1059-0.0549-0+0.0346}$
		\\
		\hline
		
		$A_1^{B_c^*\to J/\psi}(0)$
		&
		$0.2163_{+0.1262-0.0343+0.1447+0-0.0388}^{-0.0684+0.1159-0.0426-0+0.0425}$
		&
		$0.3438_{+0.2113-0.0495+0.1676+0-0.0617}^{-0.1125+0.1853-0.0574-0+0.0676}$
		&
		$0.4100_{+0.2615-0.0709+0.0829+0.0530-0.0736}^{-0.1379+0.2289-0.0521-0.0334+0.0806}$
		\\
		\hline

	$A_2^{B_c^*\to J/\psi}(0)$
	&
	$0.1125_{+0.0513+0.0062+0.0752+0-0.0202}^{-0.0303+0.0229-0.0222-0+0.0221}$
	&
	$0.1913_{+0.0936+0.0111+0.1095+0-0.0343}^{-0.0541+0.0420-0.0343-0+0.0376}$
	&
	$0.2408_{+0.1240+0.0043+0.0810+0.0275-0.0432}^{-0.0709+0.0572-0.0360-0.0174+0.0473}$
	\\
	\hline
	
	$A_3^{B_c^*\to J/\psi}(0)$
	&
	$0$
	&
	$0.0613_{+0.0417-0.0202+0.1095+0-0.0110}^{-0.0216+0.0477-0.0218-0+0.0121}$
	&
	$0.1263_{+0.0869-0.0350+0.1526+0-0.0227}^{-0.0451+0.0811-0.0392-0+0.0248}$
	\\
	\hline
	
	$A_4^{B_c^*\to J/\psi}(0)$
	&
	$0$
	&
	$0.0712_{+0.0466-0.0206+0.1271+0-0.0128}^{-0.0244+0.0507-0.0253-0+0.0140}$
	&
	$0.1438_{+0.0962-0.0347+0.1672+0-0.0258}^{-0.0503+0.0851-0.0442-0+0.0283}$
	\\
	\hline

	$T_2^{B_c^*\to J/\psi}(0)$
	&
	$0.1125_{+0.0513+0.0062+0.0752+0-0.0202}^{-0.0303+0.0229-0.0222-0+0.0221}$
	&
	$0.1908_{+0.0927+0.0125+0.1087+0-0.0342}^{-0.0537+0.0403-0.0341-0+0.0375}$
	&
	$0.2401_{+0.1222+0.0073+0.0808+0.0275-0.0431}^{-0.0702+0.0544-0.0358-0.0174+0.0472}$
	\\
	\hline
	
	$T_3^{B_c^*\to J/\psi}(0)$
	&
	$0.2725_{+0.1519-0.0312+0.1823+0-0.0489}^{-0.0835+0.1273-0.0537-0+0.0536}$
	&
	$0.4639_{+0.2698-0.0423+0.2662+0-0.0833}^{-0.1462+0.2128-0.0832-0+0.0912}$
	&
	$0.5794_{+0.3476-0.0622+0.1791+0.0667-0.1040}^{-0.1872+0.2673-0.0851-0.0421+0.1139}$
	\\
	\hline
	
	$T_4^{B_c^*\to J/\psi}(0)$
	&
	$0.1308_{+0.0906-0.0469+0.0875+0-0.0235}^{-0.0468+0.1061-0.0258-0+0.0257}$
	&
	$0.2255_{+0.1615-0.0764+0.1327+0-0.0405}^{-0.0823+0.1790-0.0409-0+0.0443}$
	&
	$0.2837_{+0.2082-0.0982+0.0925+0.0320-0.0509}^{-0.1054+0.2241-0.0426-0.0202+0.0558}$
	\\
	\hline

	$T_6^{B_c^*\to J/\psi}(0)$
	&
	$0.0832_{+0.0414-0.0032+0.0557+0-0.0149}^{-0.0239+0.0245-0.0164-0+0.0164}$
	&
	$0.1719_{+0.1013-0.0251+0.1352+0-0.0308}^{-0.0552+0.0819-0.0362-0+0.0338}$
	&
	$0.2281_{+0.1393-0.0411+0.0894+0.0204-0.0409}^{-0.0758+0.0985-0.0400-0.0129+0.0448}$
	\\
	\hline

$T_2^{\prime\,B_c^*\to J/\psi}(0)$
&
$0.2163_{+0.1262-0.0343+0.1447+0-0.0388}^{-0.0684+0.1159-0.0426-0+0.0425}$
&
$0.3581_{+0.2200-0.0528+0.1931+0-0.0643}^{-0.1171+0.1934-0.0625-0+0.0704}$
&
$0.4402_{+0.2805-0.0765+0.1213+0.0530-0.0790}^{-0.1481+0.2442-0.0617-0.0334+0.0865}$
\\
\hline	
	
$T_3^{\prime\,B_c^*\to J/\psi}(0)$
&
$0.2433_{+0.1420-0.0406+0.1628+0-0.0437}^{-0.0771+0.1290-0.0480-0+0.0478}$
&
$0.3774_{+0.2306-0.0544+0.1720+0-0.0677}^{-0.1232+0.1970-0.0613-0+0.0742}$
&
$0.4454_{+0.2826-0.0771+0.0820+0.0596-0.0799}^{-0.1495+0.2453-0.0543-0.0376+0.0875}$
\\
\hline

$T_4^{\prime\,B_c^*\to J/\psi}(0)$
&
$0.1600_{+0.1006-0.0374+0.1071+0-0.0287}^{-0.0532+0.1045-0.0316-0+0.0315}$
&
$0.3013_{+0.1940-0.0609+0.2078+0-0.0541}^{-0.1017+0.1887-0.0591-0+0.0592}$
&
$0.3962_{+0.2603-0.0790+0.1645+0.0392-0.0711}^{-0.1359+0.2382-0.0668-0.0247+0.0779}$
\\
\hline

\end{tabular}

	}
\end{table}

\begin{table}[htbp]
	\centering
	\caption{Same as table~\ref{tab:ff0numbcjpsi}, but for $B_c^*\to\eta_c$, with $T_1'^{B_c^*\to \eta_c}(0)=T_2'^{B_c^*\to \eta_c}(0)$.	
It can be verified that %
 $A_{0,1,2}^{B_c^*\to \eta_c}(0)$ satisfy the relation in eq.~\eqref{ff0relationbcvetacav}.		
  The minor discrepancy between $T^{B_c^*\to \eta_c}(0)$ and $T_1'^{B_c^*\to \eta_c}(0)$ at the NNLO level is due to the $\gamma_5$-trace prescription (see the discussion in the caption of table~\ref{tab:ff0numbcjpsi}).}
\label{tab:ff0numbcvetac}
	\setlength{\tabcolsep}{1.0mm}
	\renewcommand{\arraystretch}{1.8}%
	\resizebox{\textwidth}{!}
	{
	\begin{tabular}{c|c|c|c}
		\hline
		& LO & NLO & NNLO\\
		\hline
		
		$V^{B_c^*\to\eta_c}(0)$
		&
		$0.3201_{+0.2011-0.0748+0.2142+0-0.0575}^{-0.1065+0.2089-0.0631-0+0.0629}$
		&
		$0.5819_{+0.3745-0.1161+0.3787+0-0.1044}^{-0.1962+0.3625-0.1109-0+0.1144}$
		&
		$0.7338_{+0.4817-0.1463+0.2186+0.0960-0.1317}^{-0.2512+0.4491-0.1126-0.0606+0.1442}$
		\\
		\hline
		
		$A_0^{B_c^*\to\eta_c}(0)$
		&
		$0.2163_{+0.1262-0.0343+0.1447+0-0.0388}^{-0.0684+0.1159-0.0426-0+0.0425}$
		&
		$0.3806_{+0.2273-0.0444+0.2334+0-0.0683}^{-0.1219+0.1912-0.0705-0+0.0748}$
		&
		$0.4730_{+0.2870-0.0542+0.1277+0.0649-0.0849}^{-0.1536+0.2310-0.0695-0.0409+0.0930}$
		\\
		\hline
		
		$A_1^{B_c^*\to\eta_c}(0)$
		&
		$0.2360_{+0.1322-0.0253+0.1579+0-0.0424}^{-0.0723+0.1135-0.0465-0+0.0464}$
		&
		$0.3963_{+0.2314-0.0325+0.2206+0-0.0711}^{-0.1248+0.1864-0.0701-0+0.0779}$
		&
		$0.4796_{+0.2879-0.0485+0.1016+0.0708-0.0861}^{-0.1544+0.2285-0.0649-0.0447+0.0943}$
		\\
		\hline
		
		$A_2^{B_c^*\to\eta_c}(0)$
		&
		$0.1600_{+0.1006-0.0374+0.1071+0-0.0287}^{-0.0532+0.1045-0.0316-0+0.0315}$
		&
		$0.3361_{+0.2090-0.0568+0.2699+0-0.0603}^{-0.1106+0.1935-0.0715-0+0.0661}$
		&
		$0.4543_{+0.2821-0.0609+0.2020+0.0480-0.0815}^{-0.1498+0.2339-0.0825-0.0303+0.0893}$
		\\
		\hline
		
		$T^{B_c^*\to\eta_c}(0)$
		&
		$0.2309_{+0.1312-0.0296+0.1545+0-0.0414}^{-0.0716+0.1151-0.0455-0+0.0454}$
		&
		$0.3989_{+0.2348-0.0390+0.2359+0-0.0716}^{-0.1265+0.1923-0.0726-0+0.0784}$
		&
		$0.4833_{+0.2922-0.0528+0.0946+0.0693-0.0867}^{-0.1565+0.2337-0.0664-0.0437+0.0950}$
		\\
		\hline

		$T_1^{\prime\,B_c^*\to\eta_c}(0)$
		&
		$0.2309_{+0.1312-0.0296+0.1545+0-0.0414}^{-0.0716+0.1151-0.0455-0+0.0454}$
		&
		$0.3989_{+0.2348-0.0390+0.2359+0-0.0716}^{-0.1265+0.1923-0.0726-0+0.0784}$
		&
		$0.4919_{+0.2971-0.0541+0.1260+0.0693-0.0883}^{-0.1591+0.2381-0.0705-0.0437+0.0967}$
		\\
		\hline
		
		$T_2^{\prime\,B_c^*\to\eta_c}(0)$
		&
		$0.2309_{+0.1312-0.0296+0.1545+0-0.0414}^{-0.0716+0.1151-0.0455-0+0.0454}$
		&
		$0.3989_{+0.2348-0.0390+0.2359+0-0.0716}^{-0.1265+0.1923-0.0726-0+0.0784}$
		&
		$0.4919_{+0.2971-0.0541+0.1260+0.0693-0.0883}^{-0.1591+0.2381-0.0705-0.0437+0.0967}$
		\\
		
		\hline
	\end{tabular}

	}
\end{table}

The numerical values of the form factors at $q^2=0$ are summarized in tables~\ref{tab:ff0numbcjpsi}--\ref{tab:ff0numbcvetac}. The five individual uncertainties attached to each central value correspond, from left to right, to the variations of $m_b$, $m_c$, $\mu$, $\mu_f$, and $\psi_{B_c^{(*)}}(0) \psi_{J/\psi(\eta_c)}(0)$, respectively. 
The central values exhibit a clear hierarchy of perturbative corrections: 
the NLO contributions enhance the LO results by $\mathcal{O}(70\%)$, while the NNLO corrections are substantially smaller, amounting to $\mathcal{O}(20\%)$ of the NLO predictions. The decreasing size of successive corrections indicates improved perturbative convergence, although the sizable NNLO contribution also shows that NNLO effects remain important for precision predictions.

The uncertainty analysis shows that the uncertainties arising from the heavy-quark masses are the most prominent, whereas the uncertainty associated with the renormalization scale is substantially suppressed at NNLO compared to NLO, and the factorization-scale uncertainty remains the smallest. 
The relatively large mass-induced uncertainties are attributed to our conservative choice of sizable variation ranges ($m_b = 4.75 \pm 0.50~\text{GeV}, m_c = 1.50 \pm 0.50~\text{GeV}$), as well as the universal denominator factor $[(m_b - m_c)^2 - q^2]$ governing the form-factor expressions (see, e.g., eqs.~(4)--(7) of ref.~\cite{Tao:2022yur}, eq.~(12) of ref.~\cite{Chang:2025rna}, and eq.~(10) of ref.~\cite{Tao:2025kqf}). As an intrinsic kinematic pole dictated by the $b \to c$ decay geometry, this propagator denominator inevitably appears in the hard-scattering amplitudes across arbitrary perturbative orders. 
At $q^2=0$, this structure reduces to inverse powers of $(m_b - m_c)$, 
inducing non-negligible dependence on quark-mass fluctuations. 
Unlike the %
timelike region ($q^2 > 0$) where the denominator shrinks and amplifies mass-induced instabilities near the threshold, moving into the spacelike domain ($q^2 < 0$) enlarges the denominator to $(m_b - m_c)^2 + |q^2|$, which monotonically suppresses the inverse-power quark-mass sensitivity and improves numerical stability.  
This feature constitutes an important supporting factor for restricting the input matching points predominantly to the spacelike domain and its immediate vicinity ($q^2 \lesssim 0\,\text{GeV}^2$) for the subsequent $z$-series fit.

Overall, the reduction of the renormalization-scale dependence, the mild factorization-scale dependence, and the decreasing size of successive perturbative corrections indicate that the NNLO predictions are more stable and reliable than the lower-order results. These results at $q^2=0$ therefore provide a stable starting point for the subsequent $z$-series analysis of the full $q^2$ dependence.

\subsection{$F(q^2)$}

\begin{figure}[htbp]%
	\centering
	\includegraphics[width=0.44\textwidth]{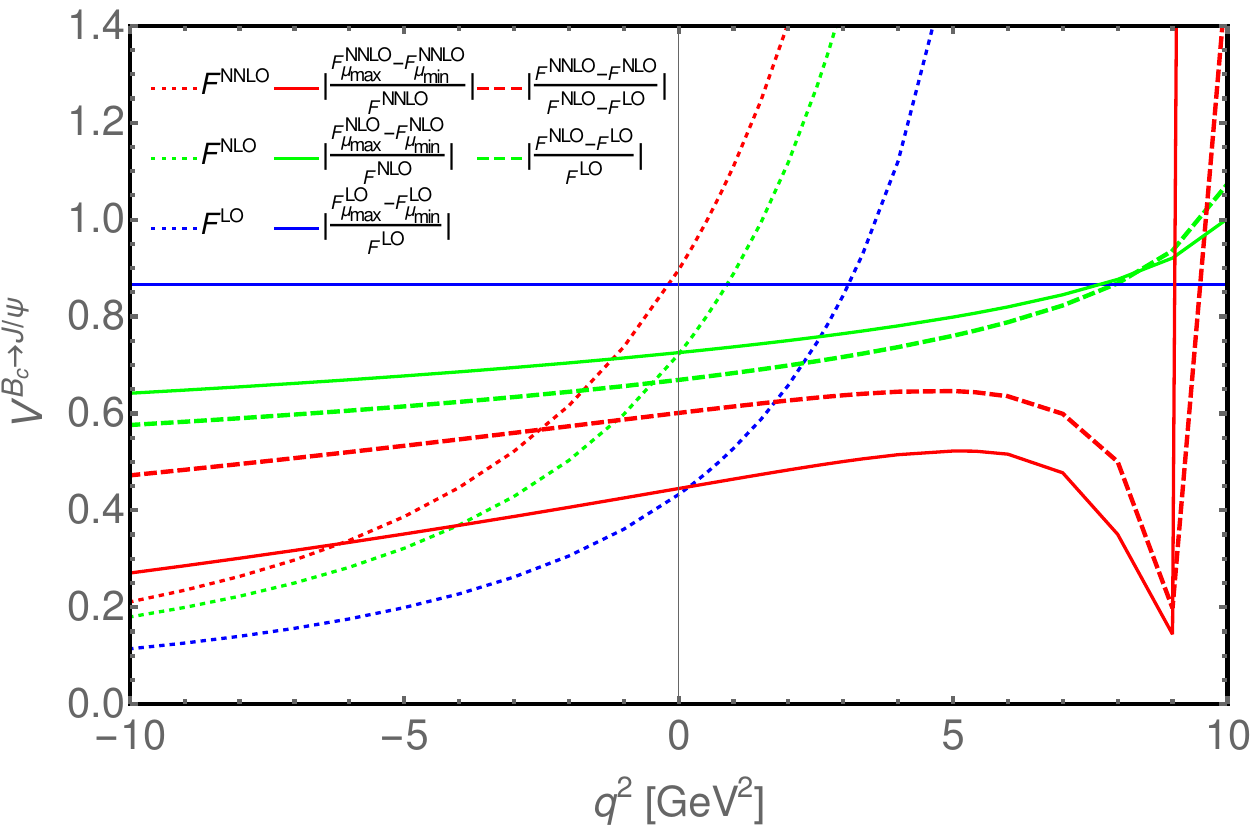}\qquad %
	\includegraphics[width=0.44\textwidth]{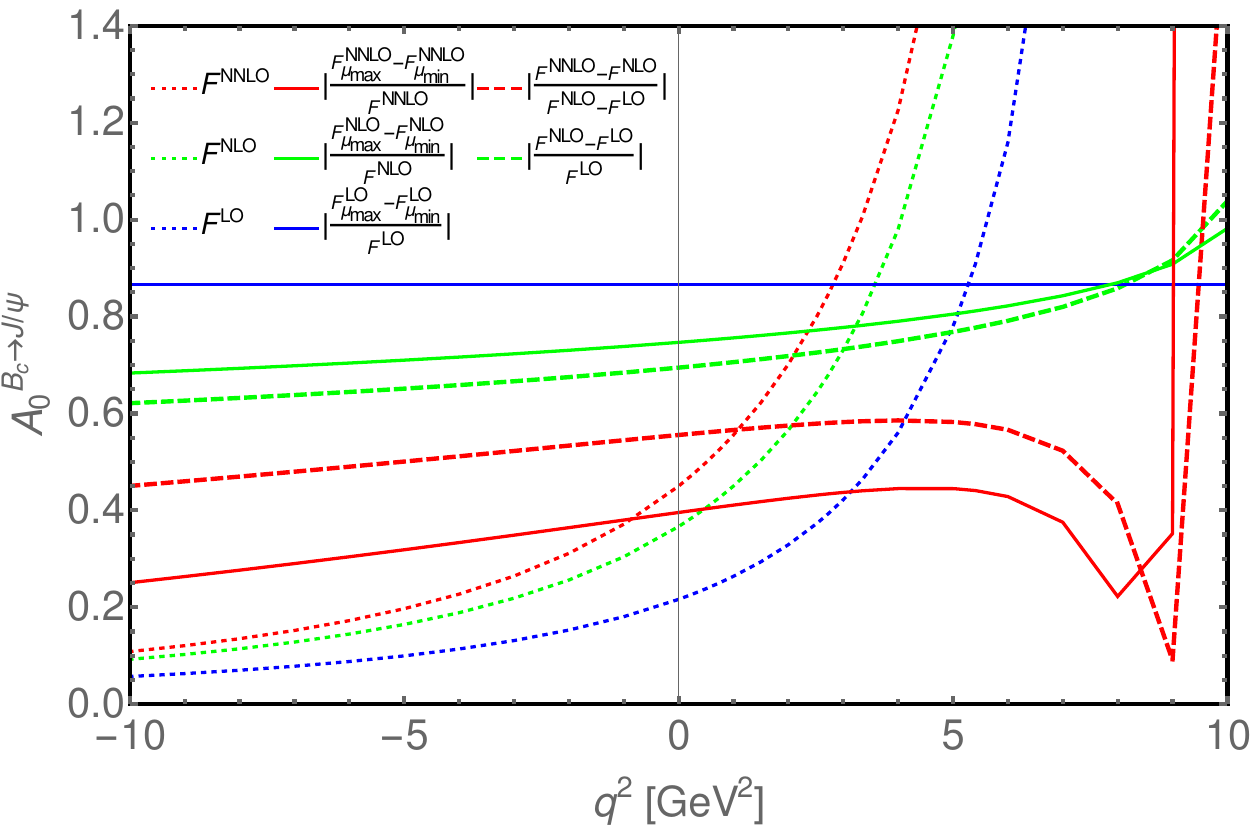}\qquad
	\includegraphics[width=0.44\textwidth]{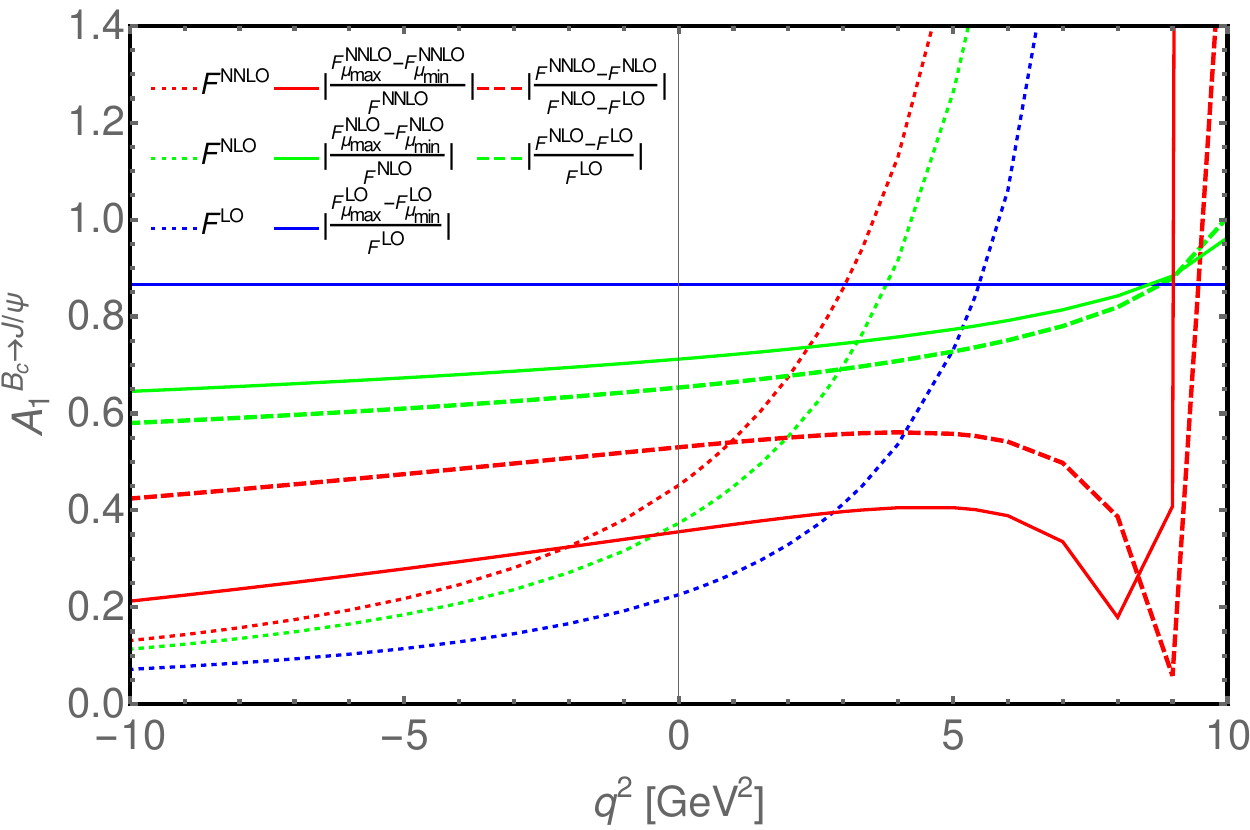}\qquad
	\includegraphics[width=0.44\textwidth]{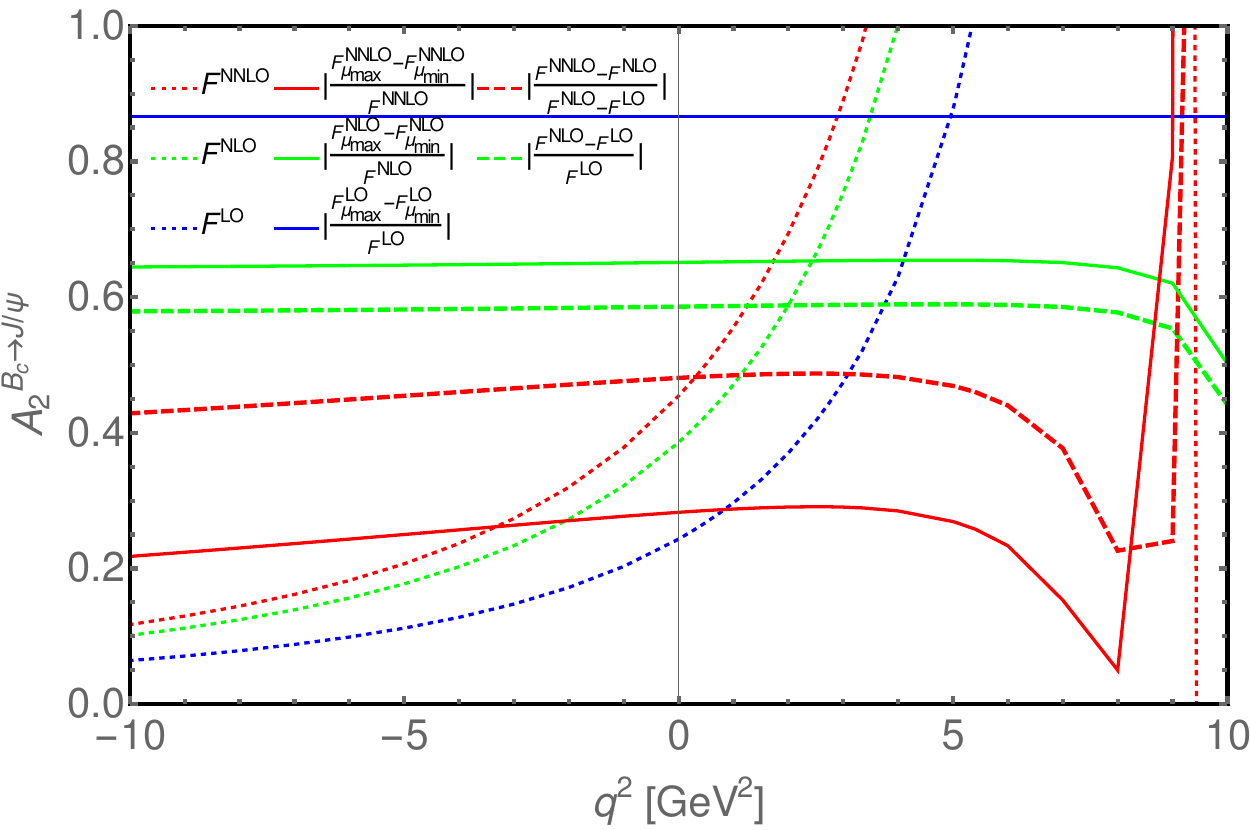}\qquad
	\includegraphics[width=0.44\textwidth]{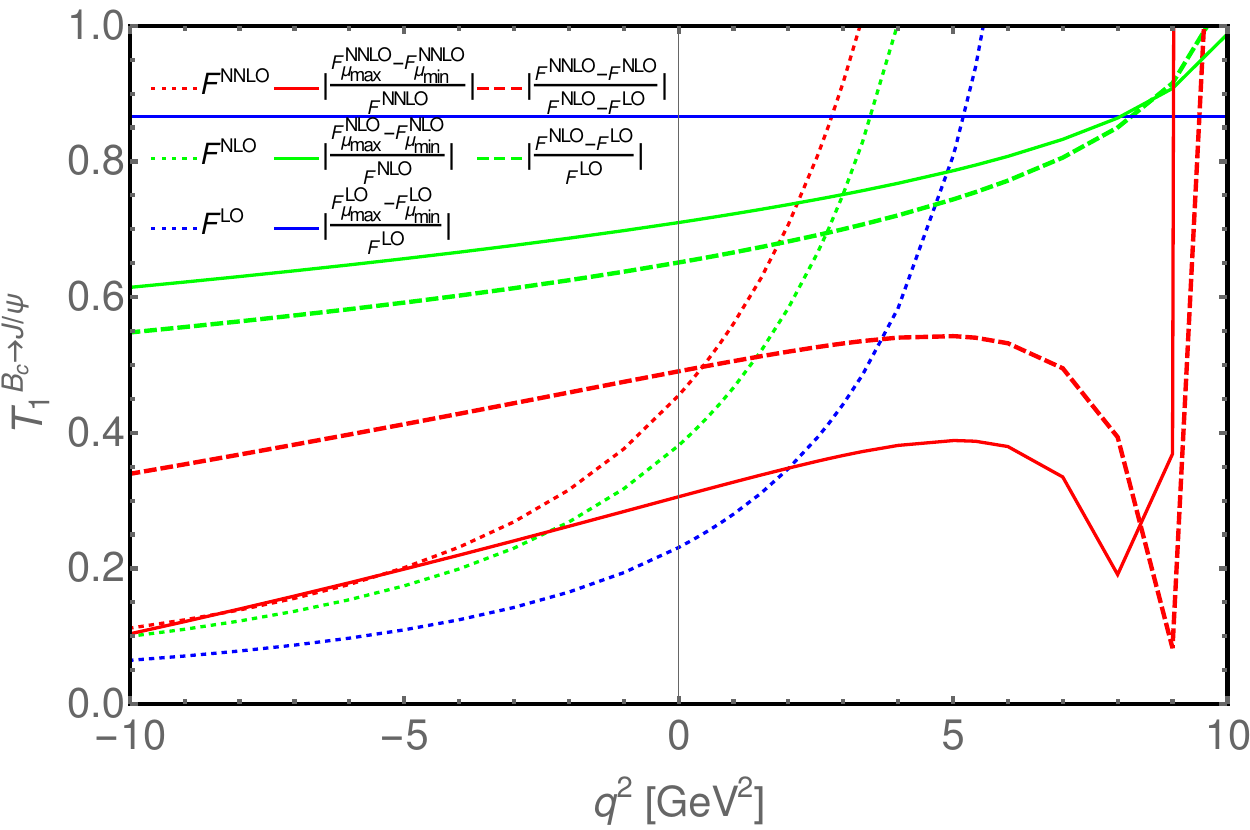}\qquad
	\includegraphics[width=0.44\textwidth]{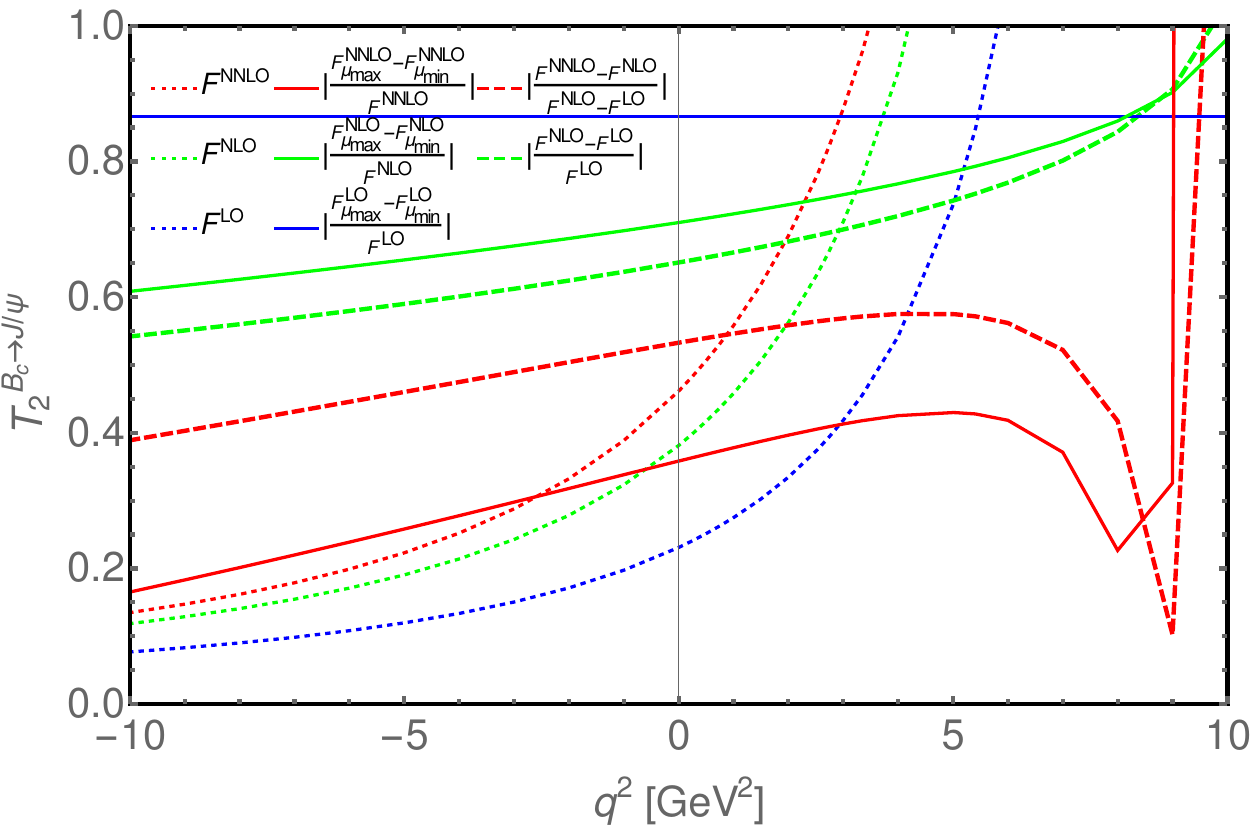}\qquad	
	\includegraphics[width=0.44\textwidth]{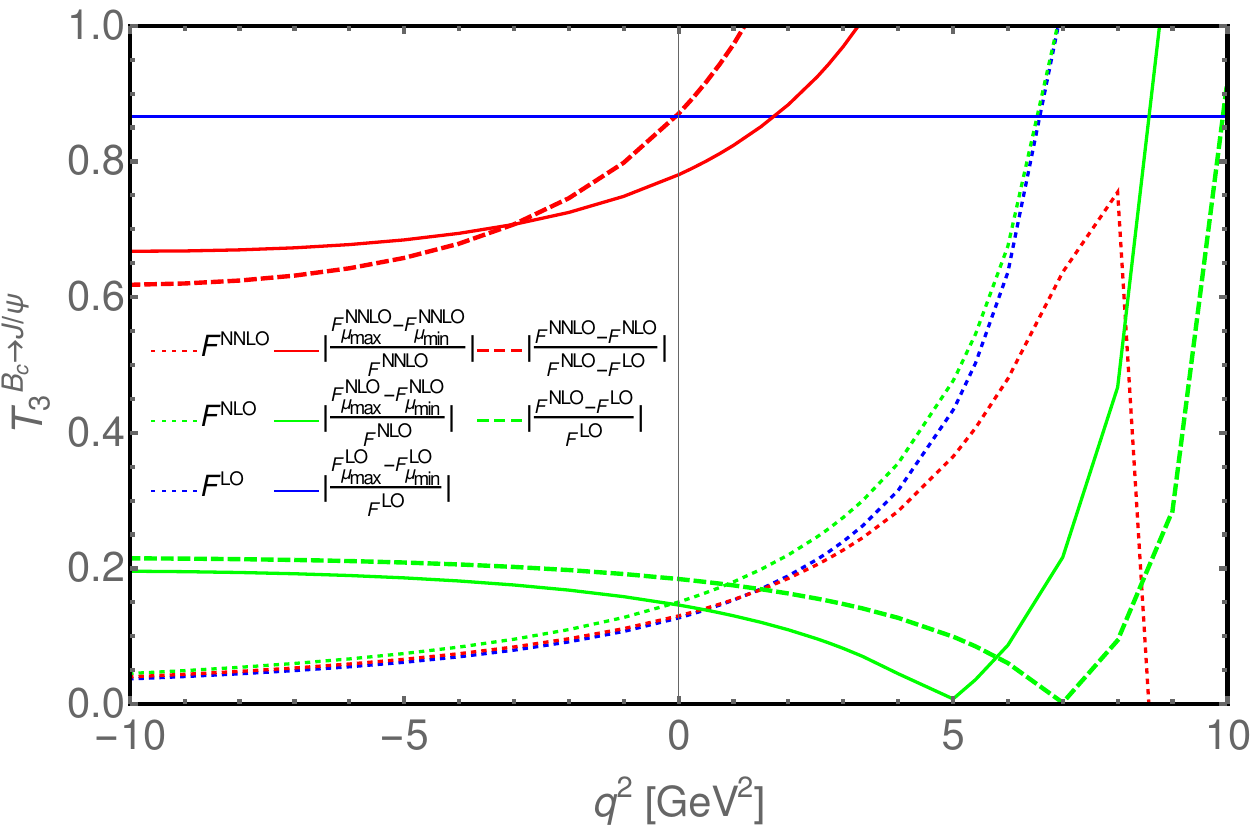}
	\caption{NRQCD results for the $B_c \to J/\psi$ form factors at LO, NLO, and NNLO accuracy, together with their perturbative convergence and renormalization-scale dependence as functions of $q^2$. All input parameters ($m_b, m_c, \mu, \mu_f, \psi_{B_c}(0)$, and $\psi_{J/\psi}(0)$) are fixed to their central values, except when evaluating the renormalization-scale dependence, where $\mu = 4.75_{-3.25}^{+5.25}\,\text{GeV}$ is used.}
	\label{fig:ffq2nrbcjpsi}
\end{figure}

\begin{figure}[htbp]%
	\centering
	\includegraphics[width=0.44\textwidth]{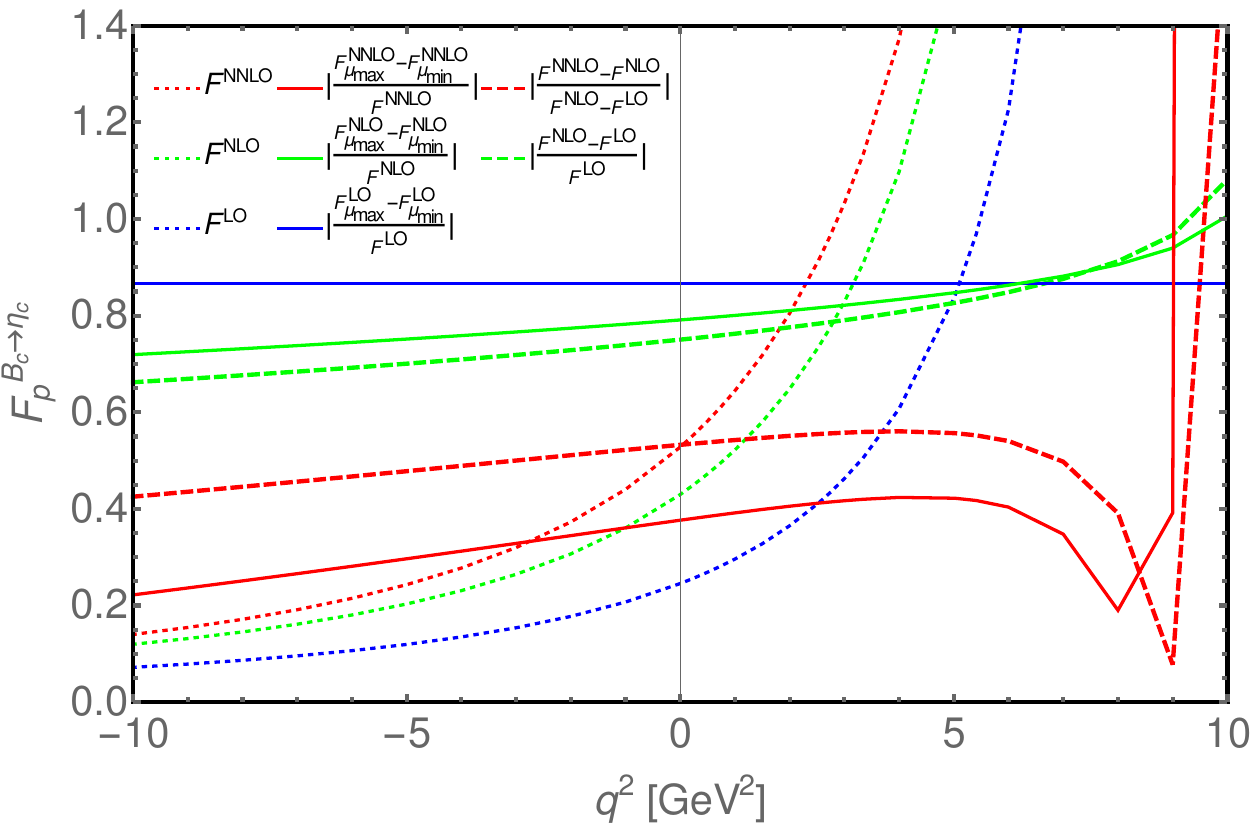}\qquad %
	\includegraphics[width=0.44\textwidth]{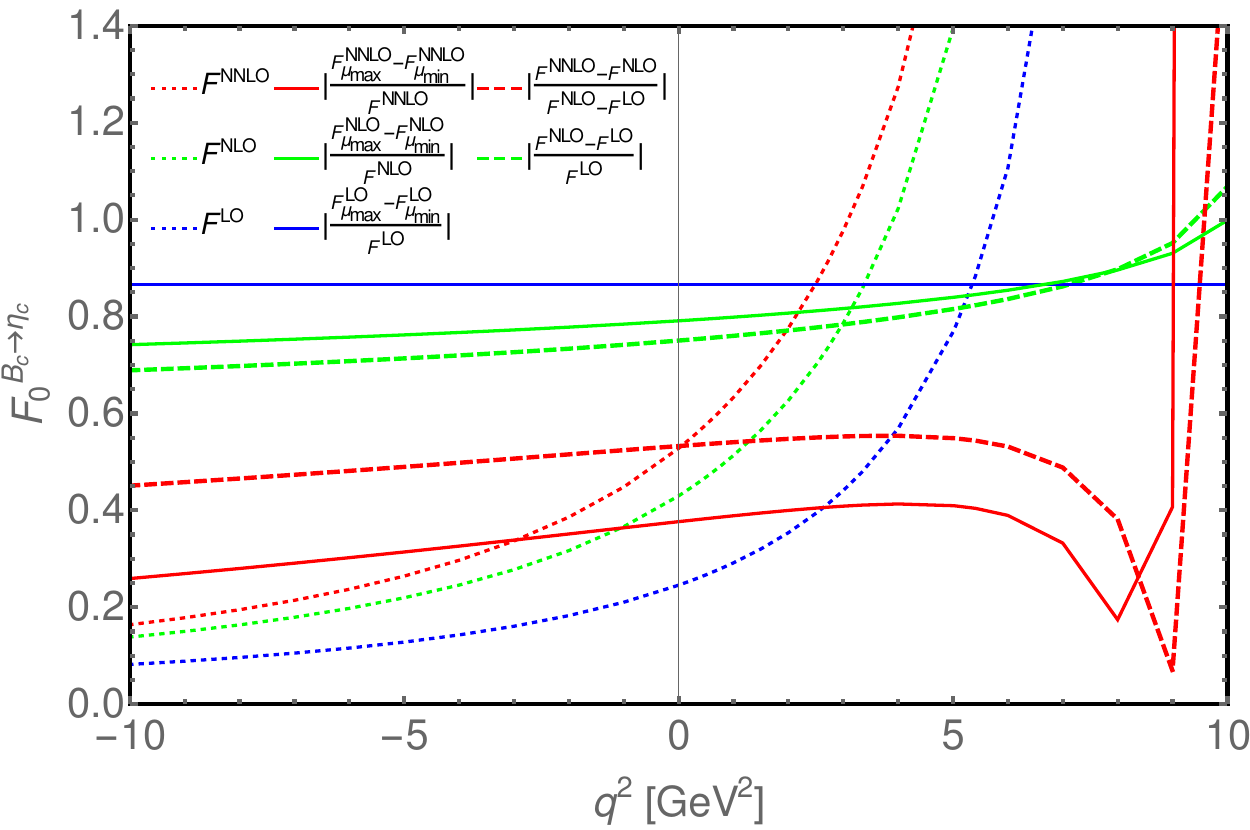}\qquad
	\includegraphics[width=0.44\textwidth]{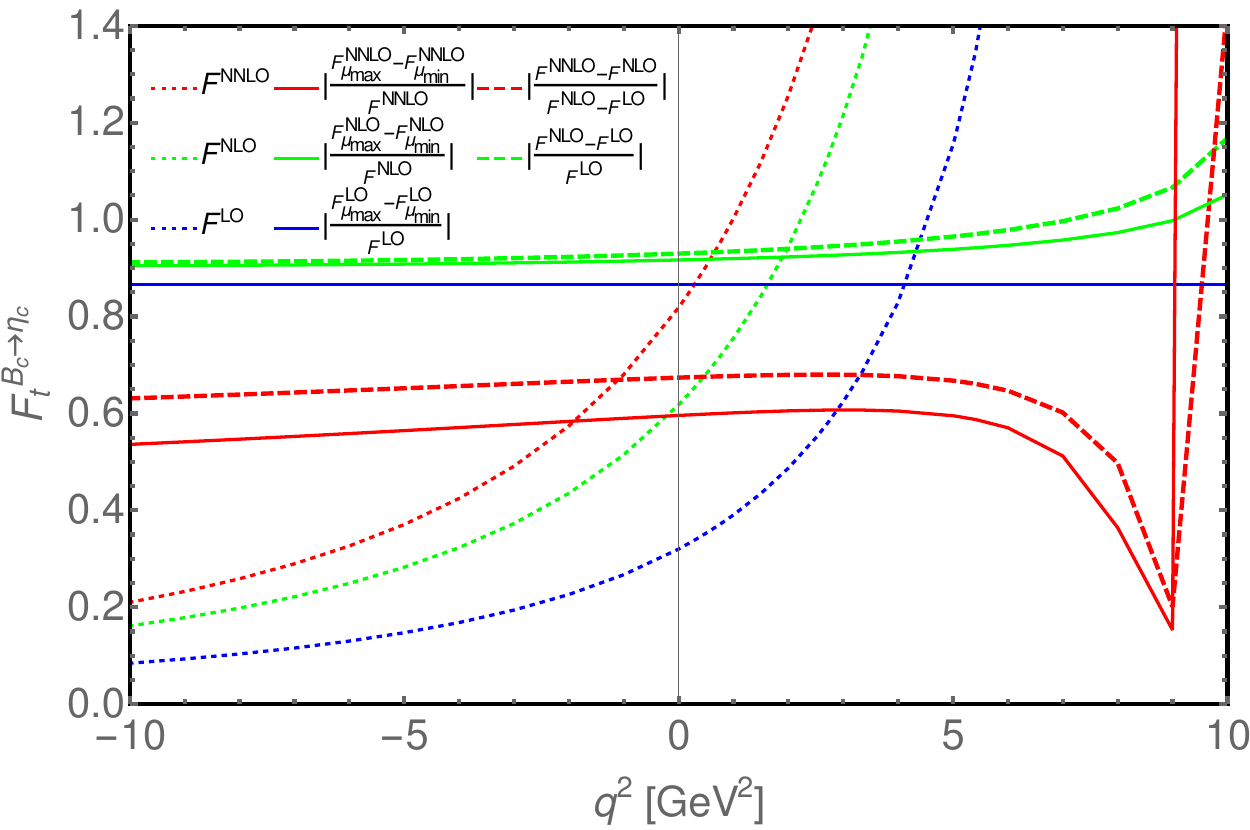}
	\caption{Same as figure~\ref{fig:ffq2nrbcjpsi}, but for $B_c \to \eta_c$.}
	\label{fig:ffq2nrbcetac}
\end{figure}

\begin{figure}[htbp]%
	\centering
	\includegraphics[width=0.315\textwidth]{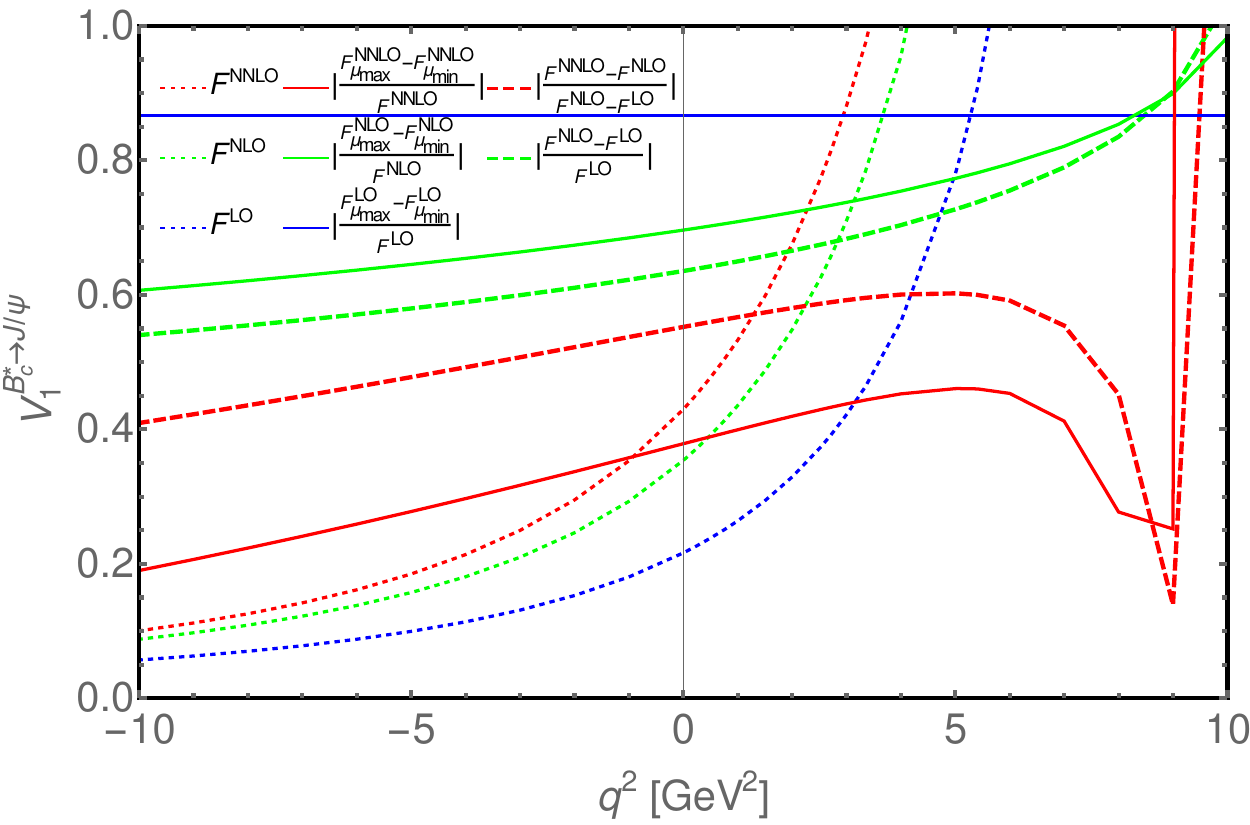}\quad %
	\includegraphics[width=0.315\textwidth]{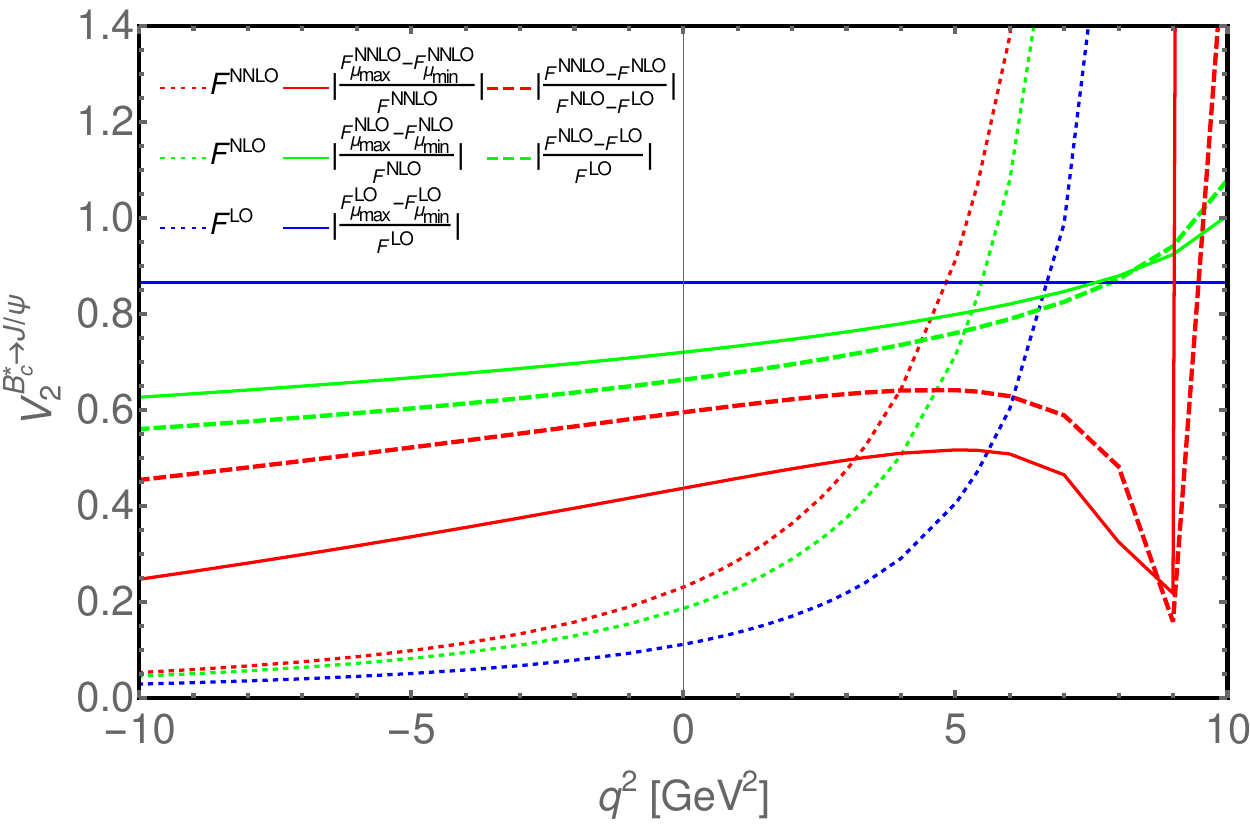}\quad
	\includegraphics[width=0.315\textwidth]{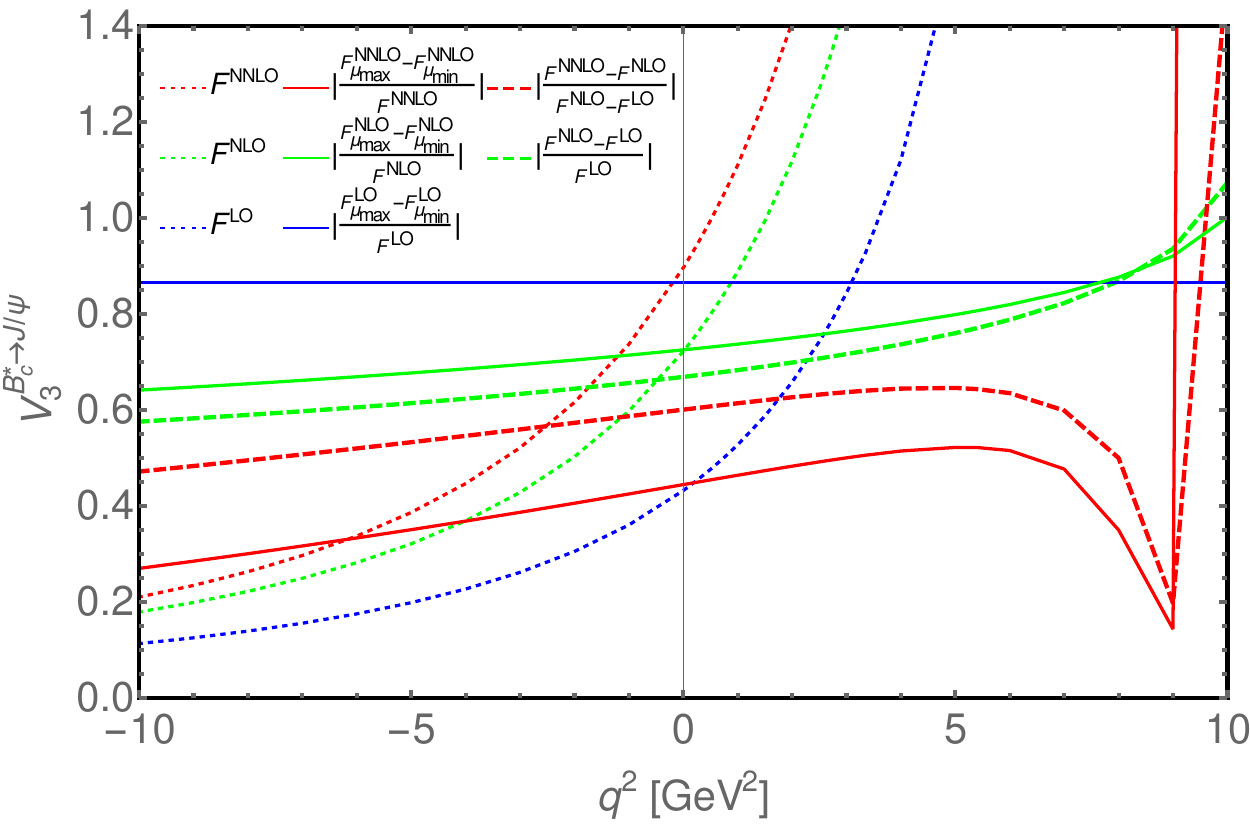}\quad
	\includegraphics[width=0.315\textwidth]{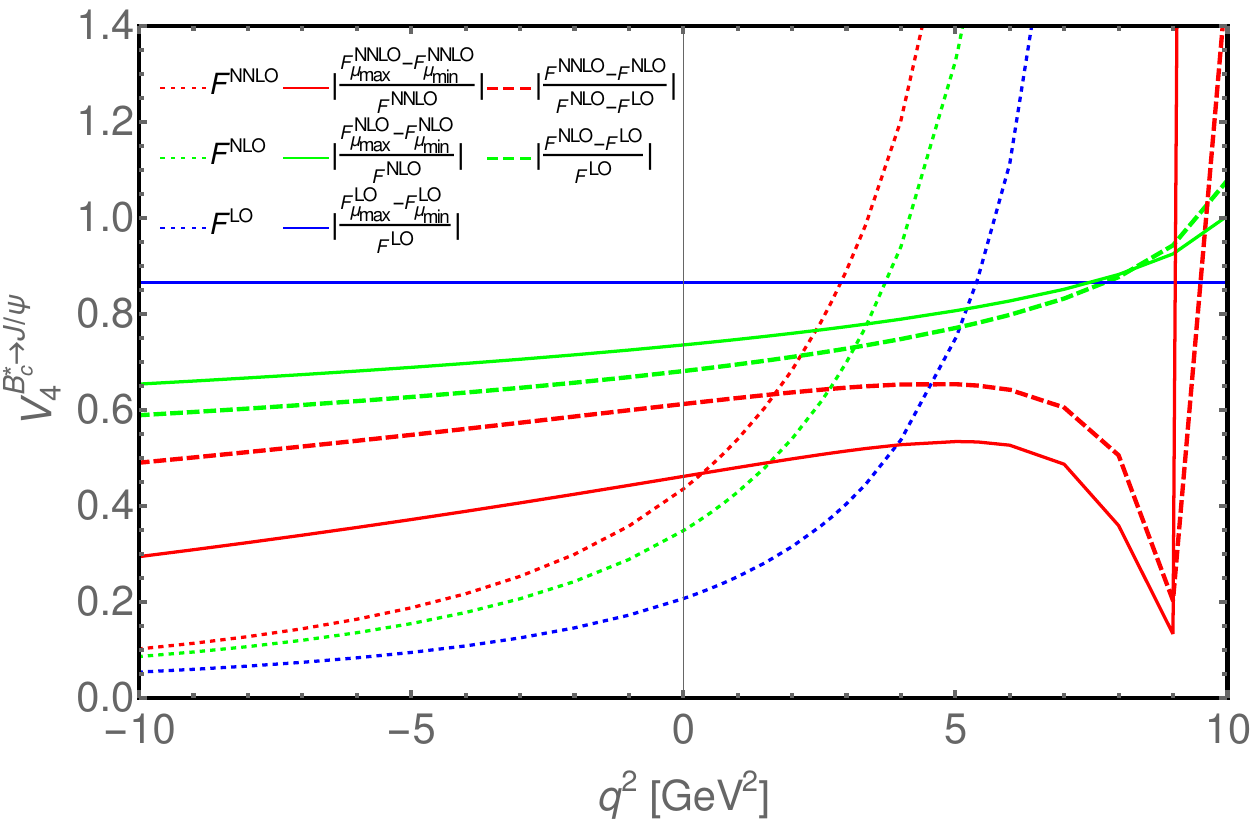}\quad
	\includegraphics[width=0.315\textwidth]{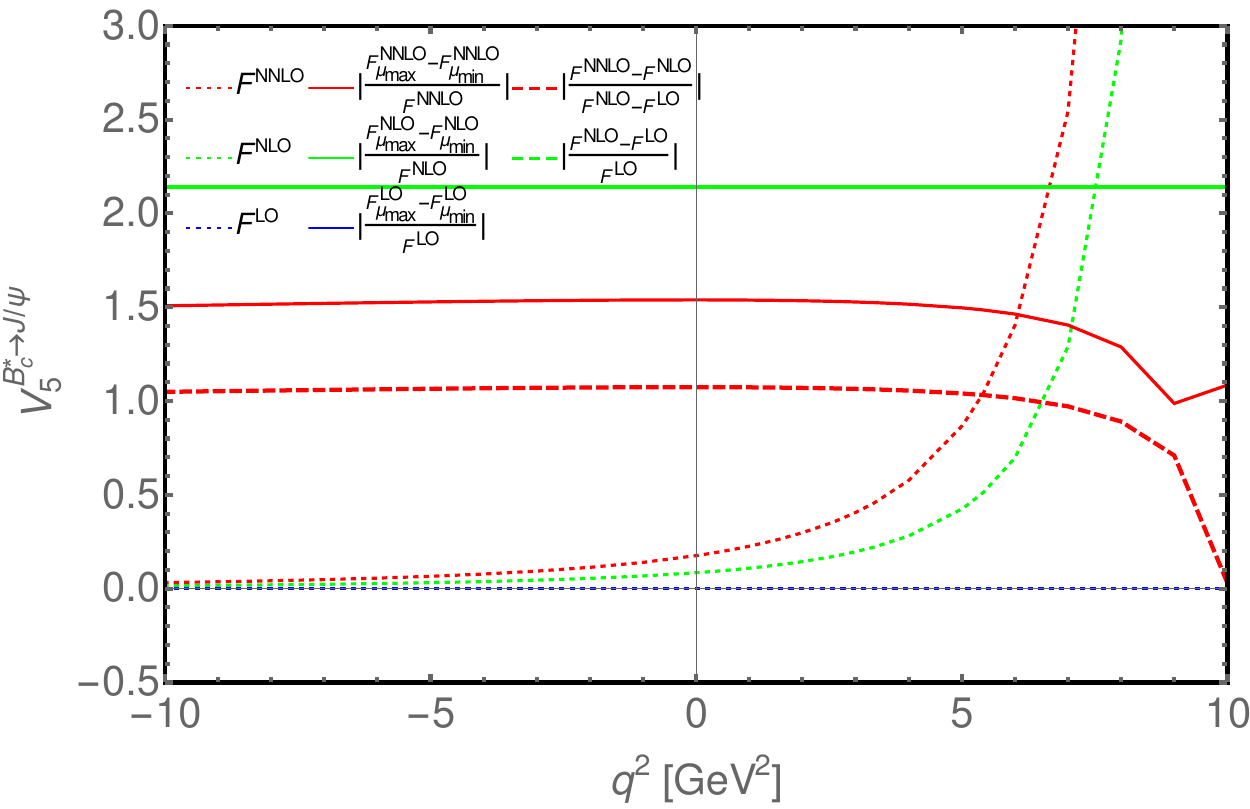}\quad
	\includegraphics[width=0.315\textwidth]{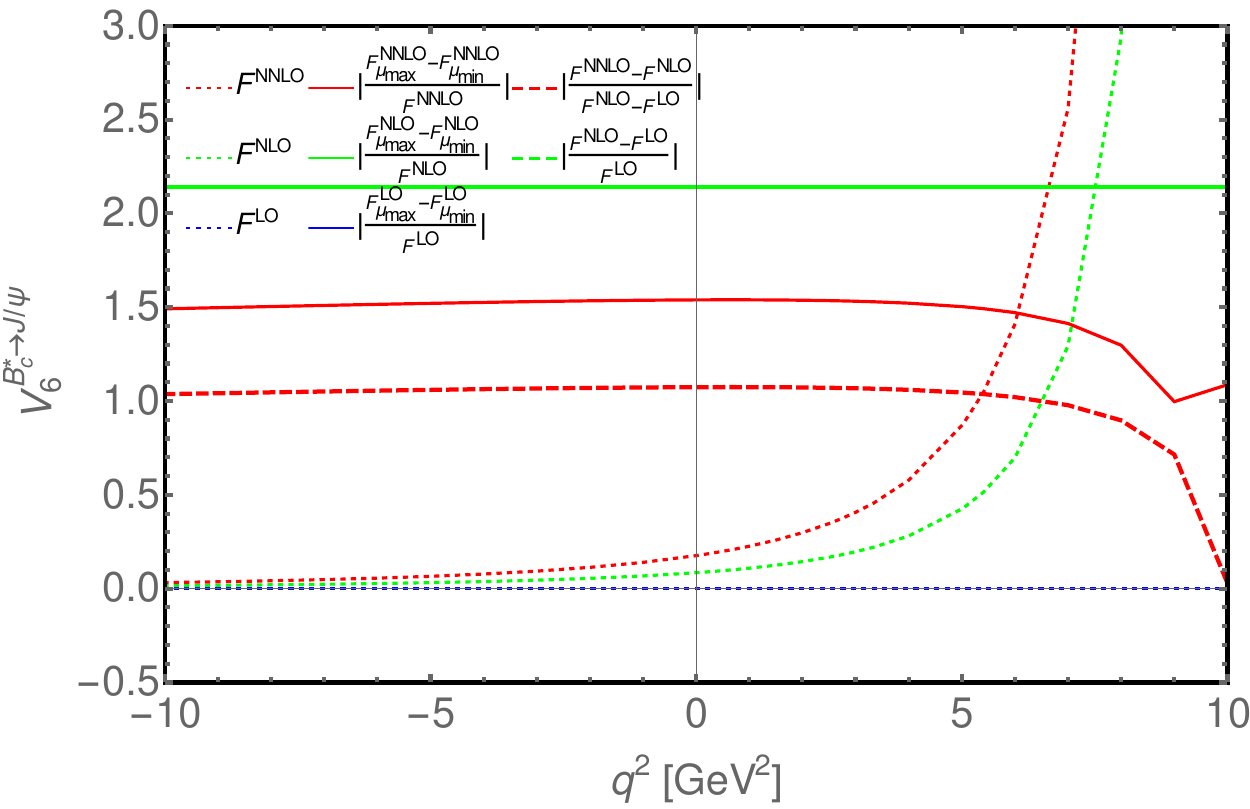}\quad
	\includegraphics[width=0.315\textwidth]{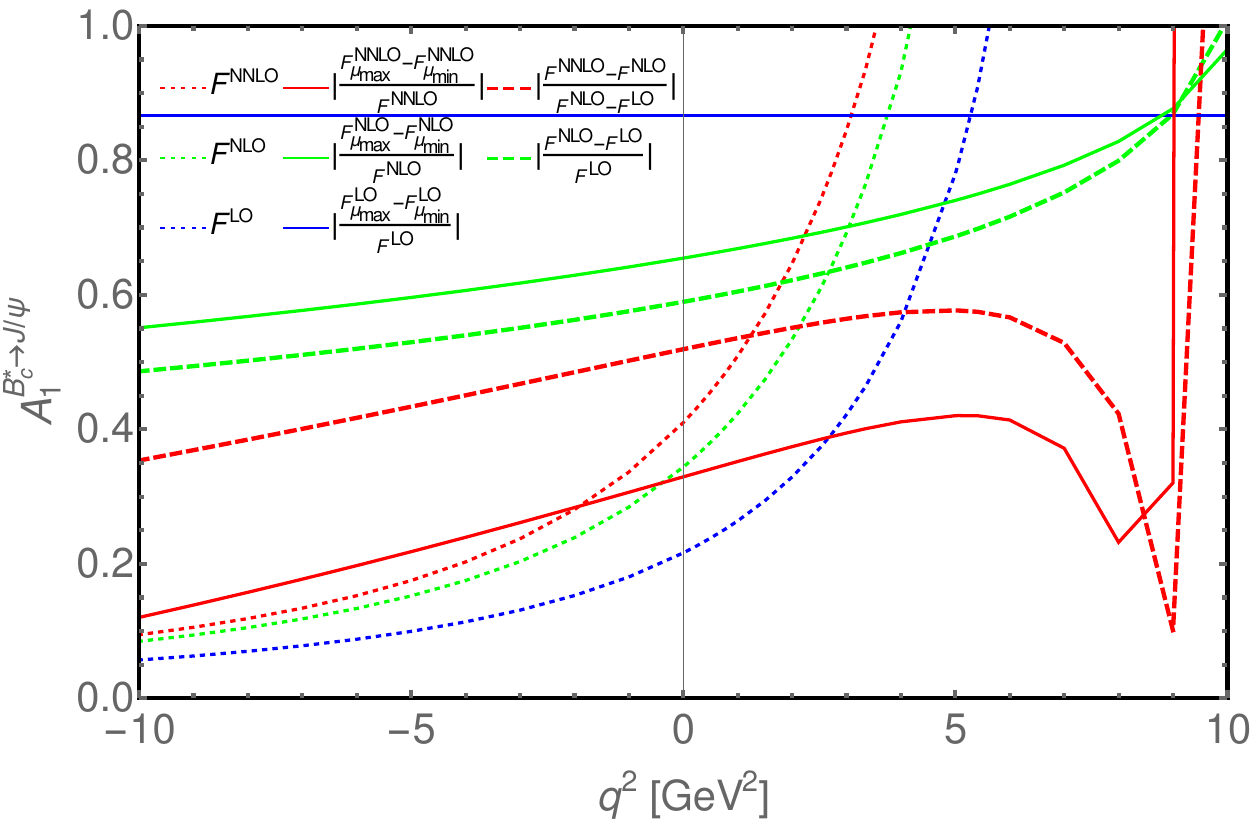}\quad	
	\includegraphics[width=0.315\textwidth]{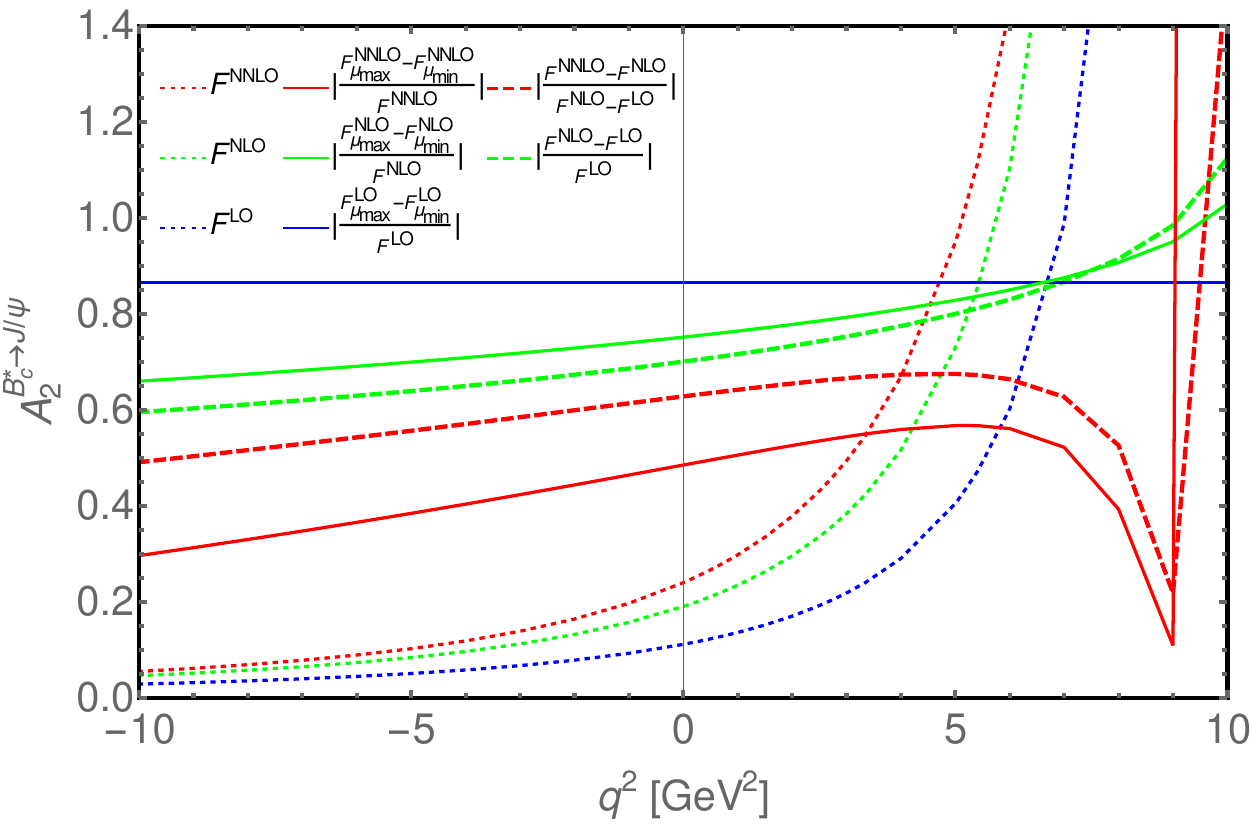}\quad
	\includegraphics[width=0.315\textwidth]{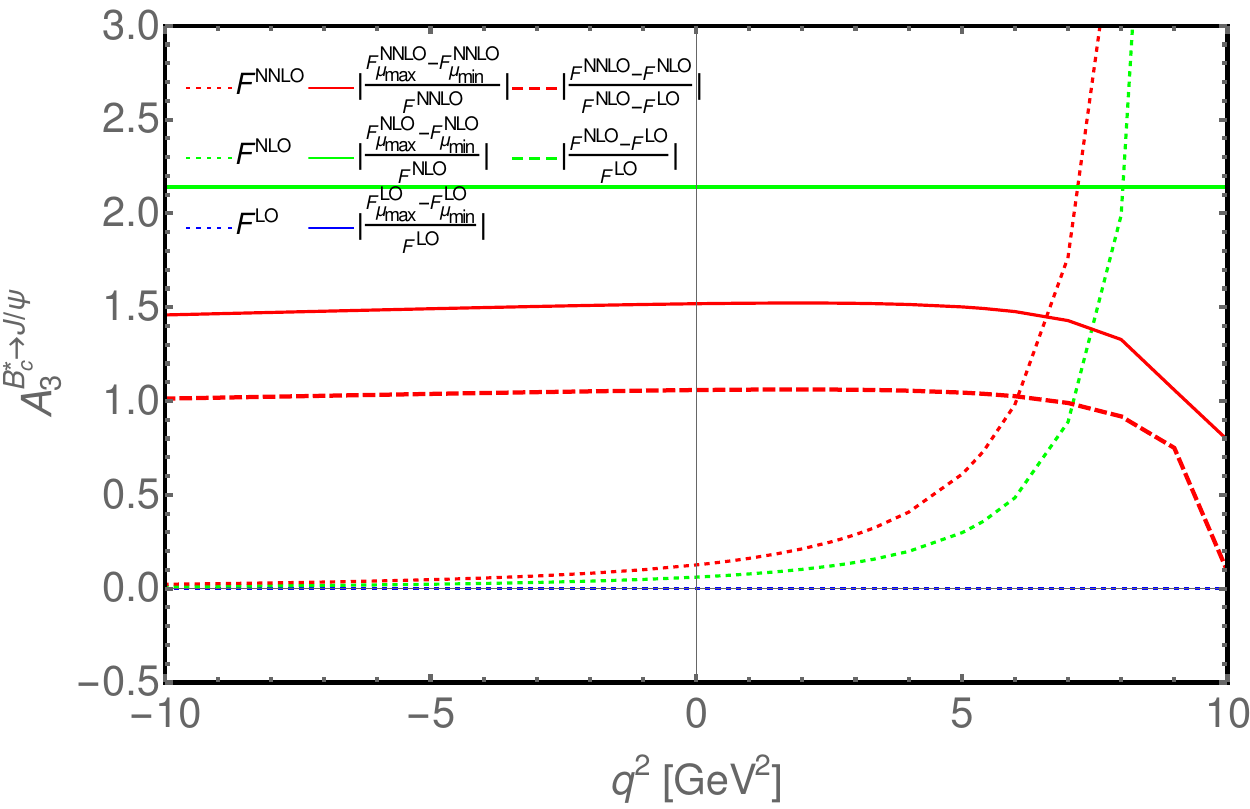}\quad
	\includegraphics[width=0.315\textwidth]{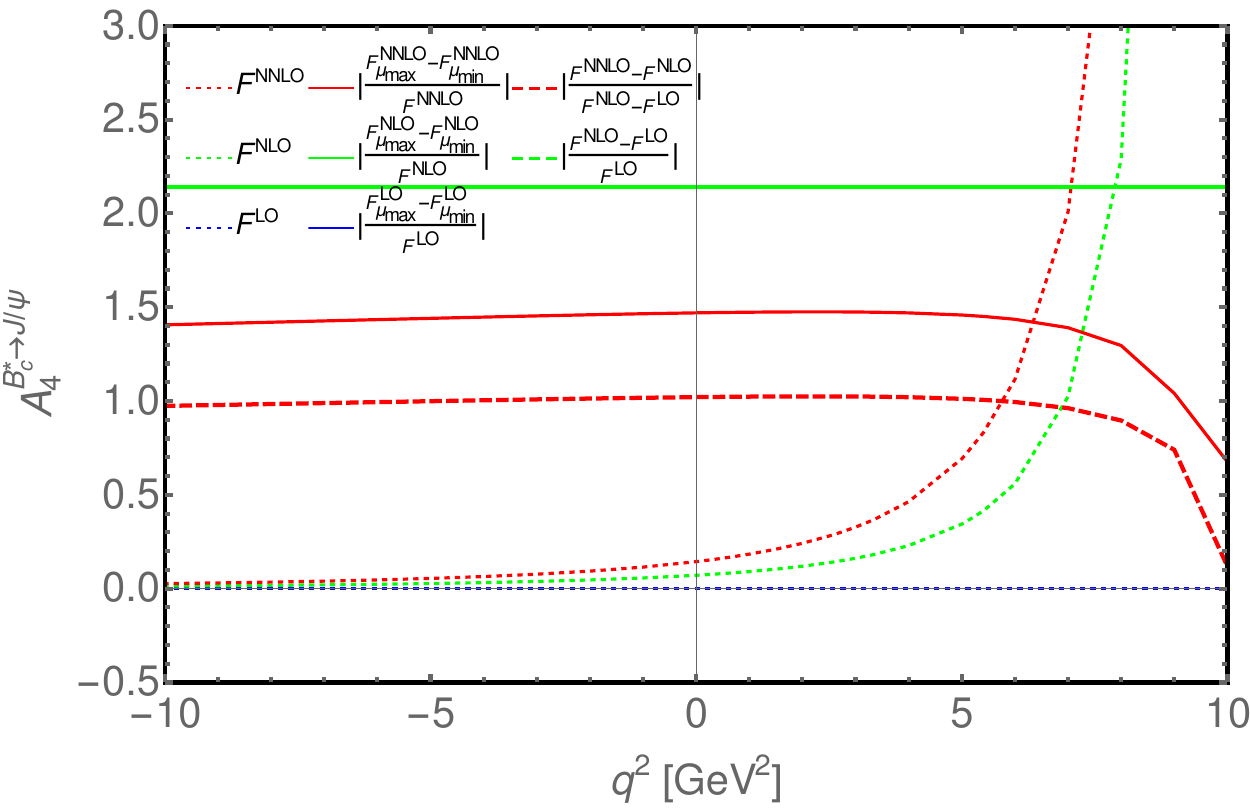}\quad		
	\includegraphics[width=0.315\textwidth]{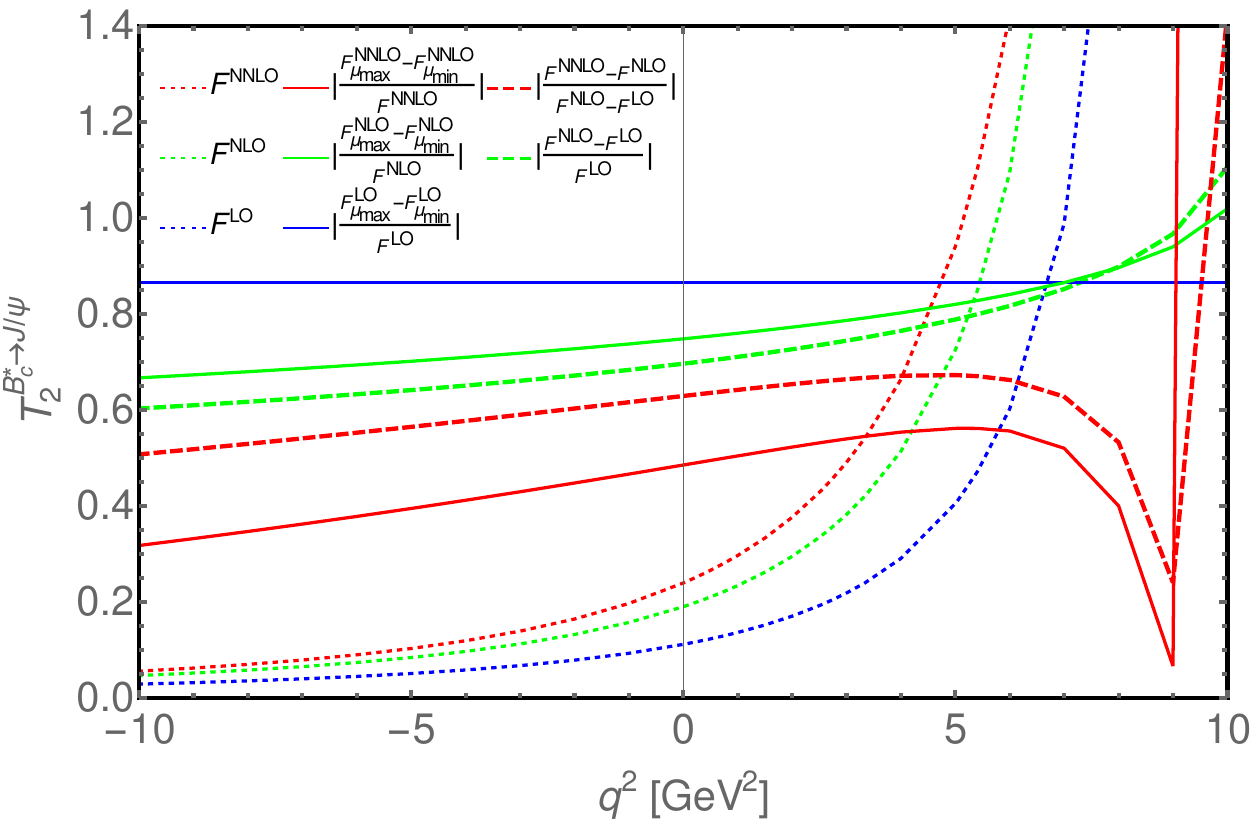}\quad
	\includegraphics[width=0.315\textwidth]{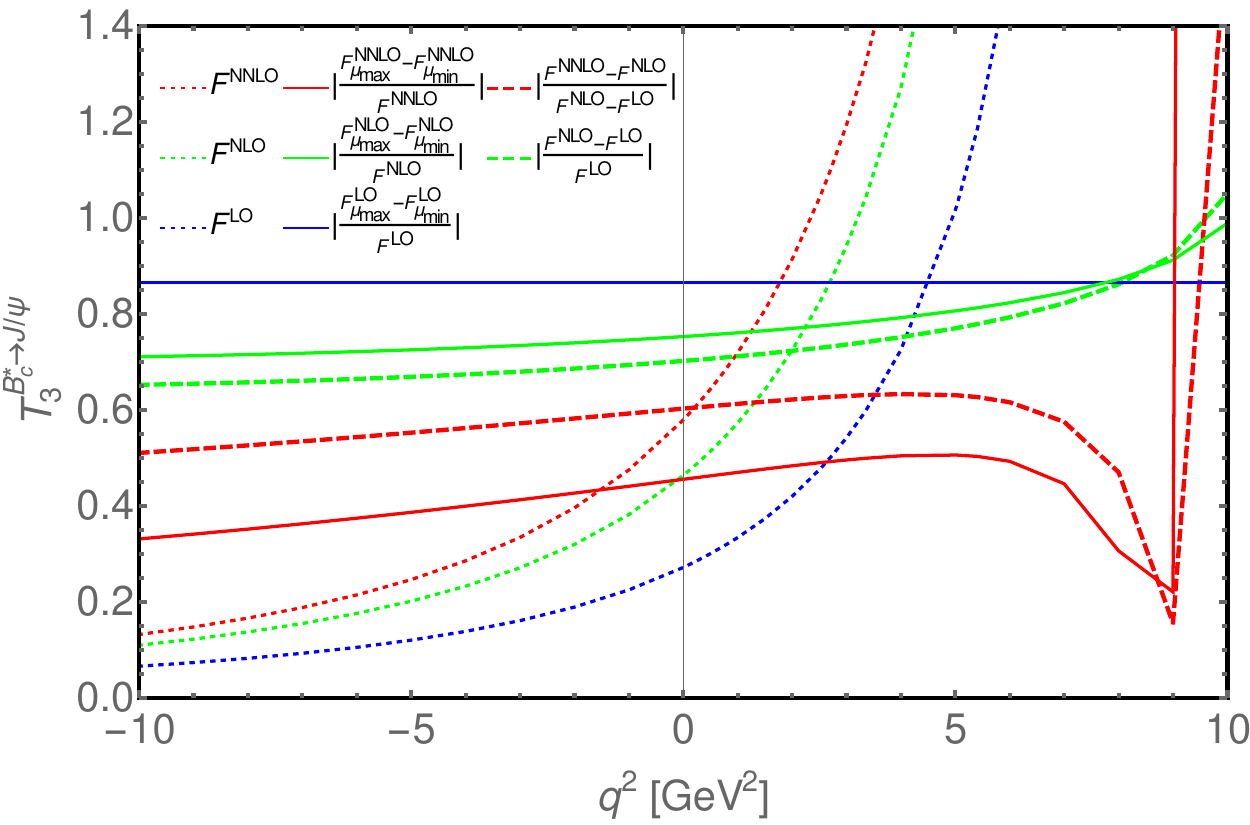}\quad
	\includegraphics[width=0.315\textwidth]{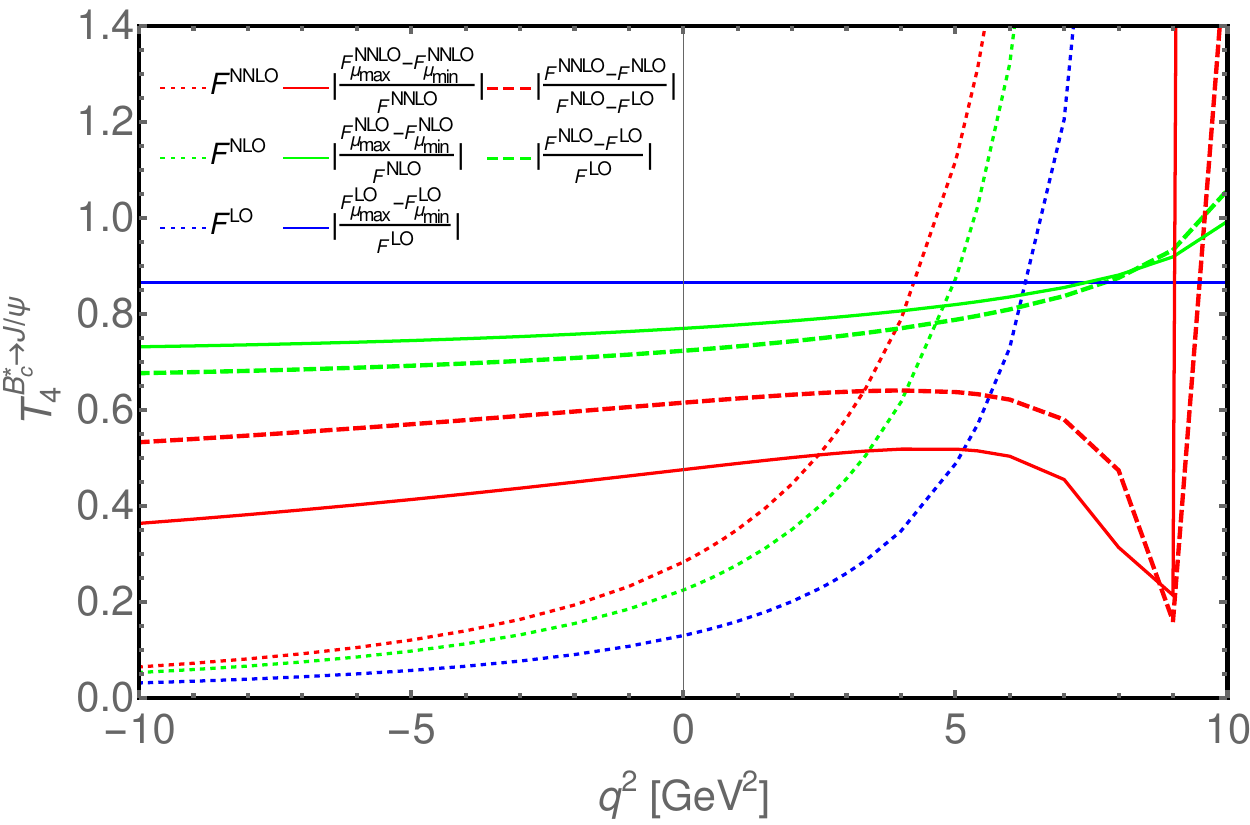}\quad
	\includegraphics[width=0.315\textwidth]{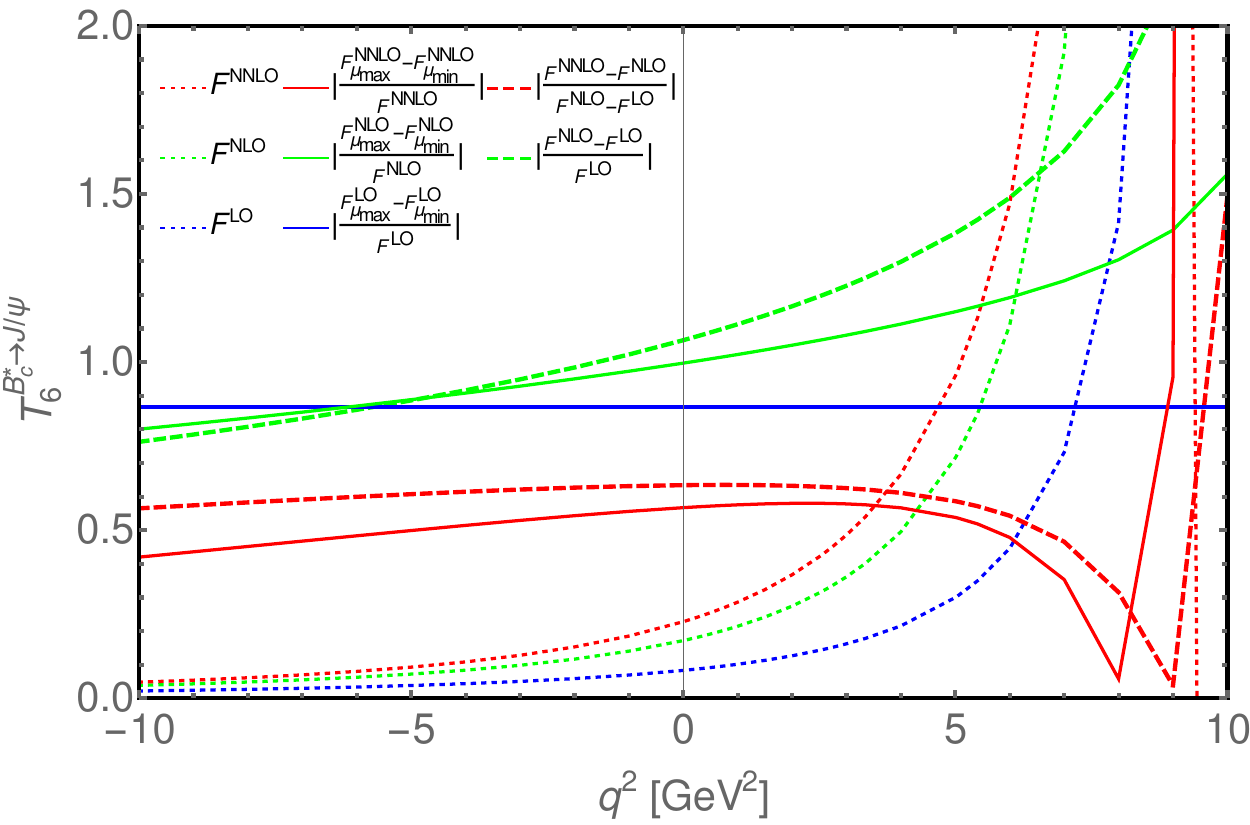}\quad
	\includegraphics[width=0.315\textwidth]{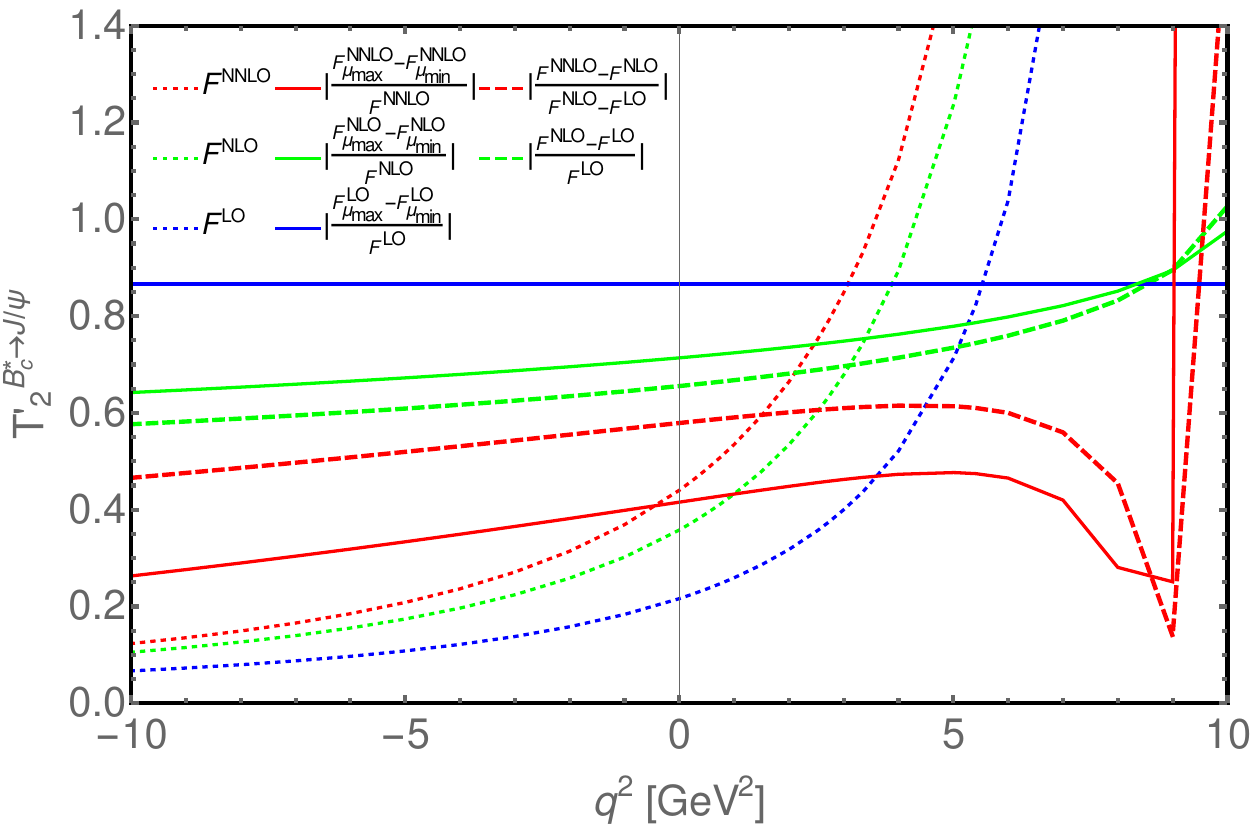}\quad
	\includegraphics[width=0.315\textwidth]{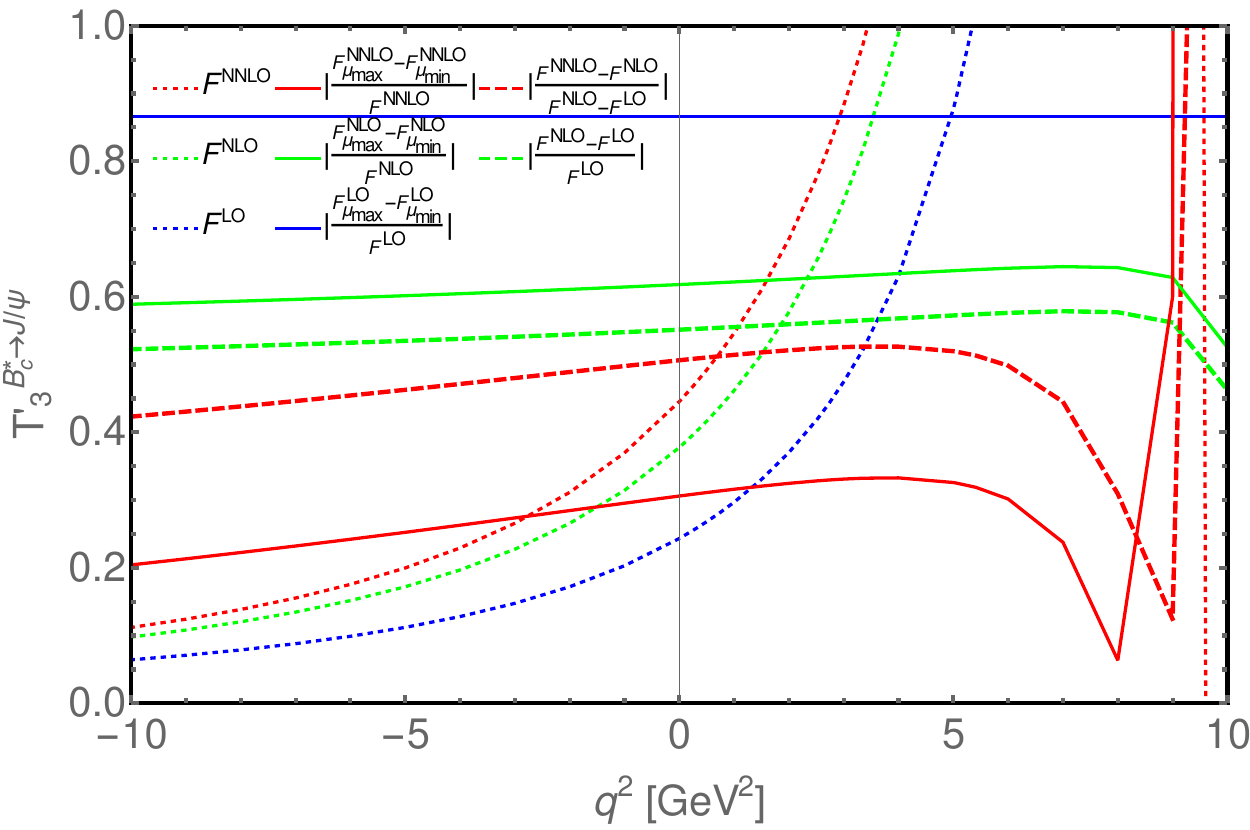}\quad
	\includegraphics[width=0.315\textwidth]{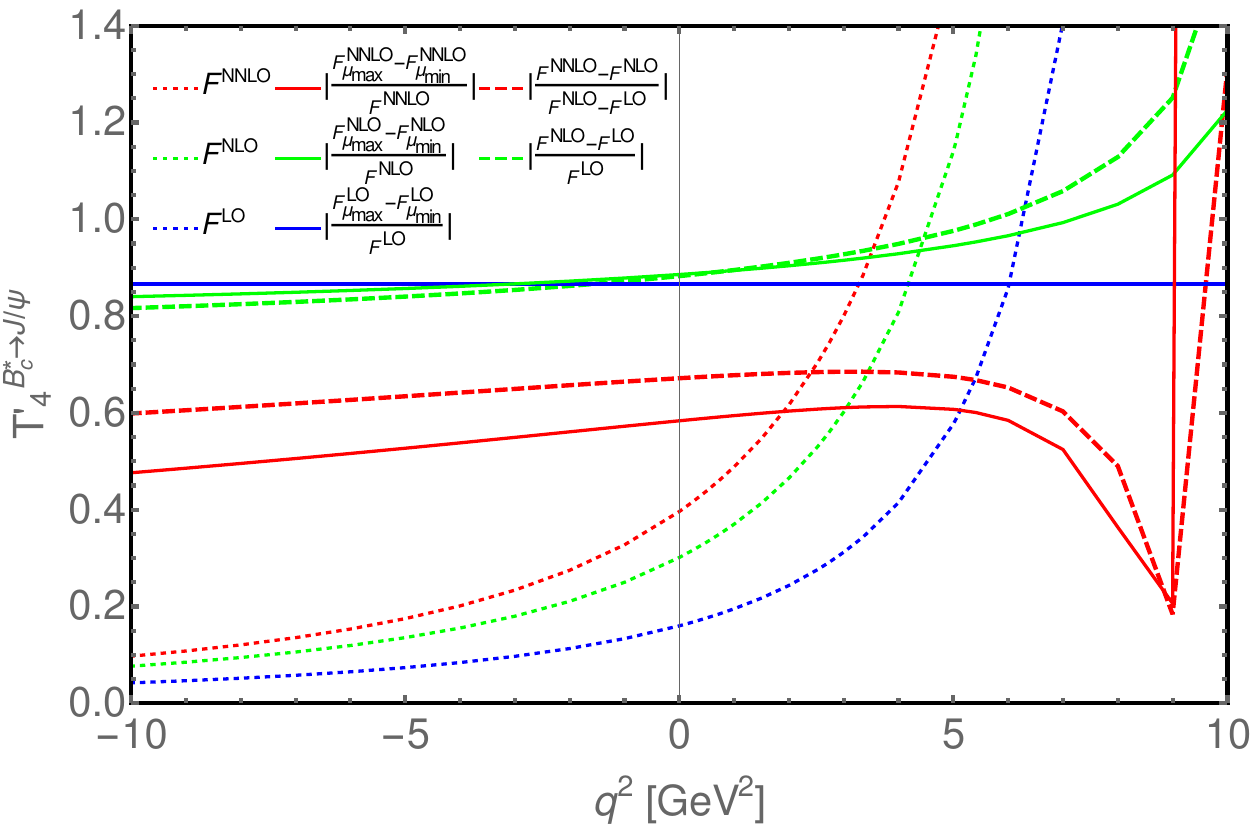}
	\caption{Same as figure~\ref{fig:ffq2nrbcjpsi}, but for $B_c^* \to J/\psi$.}
	\label{fig:ffq2nrbcvjpsi}
\end{figure}

\begin{figure}[htbp]%
	\centering
	\includegraphics[width=0.44\textwidth]{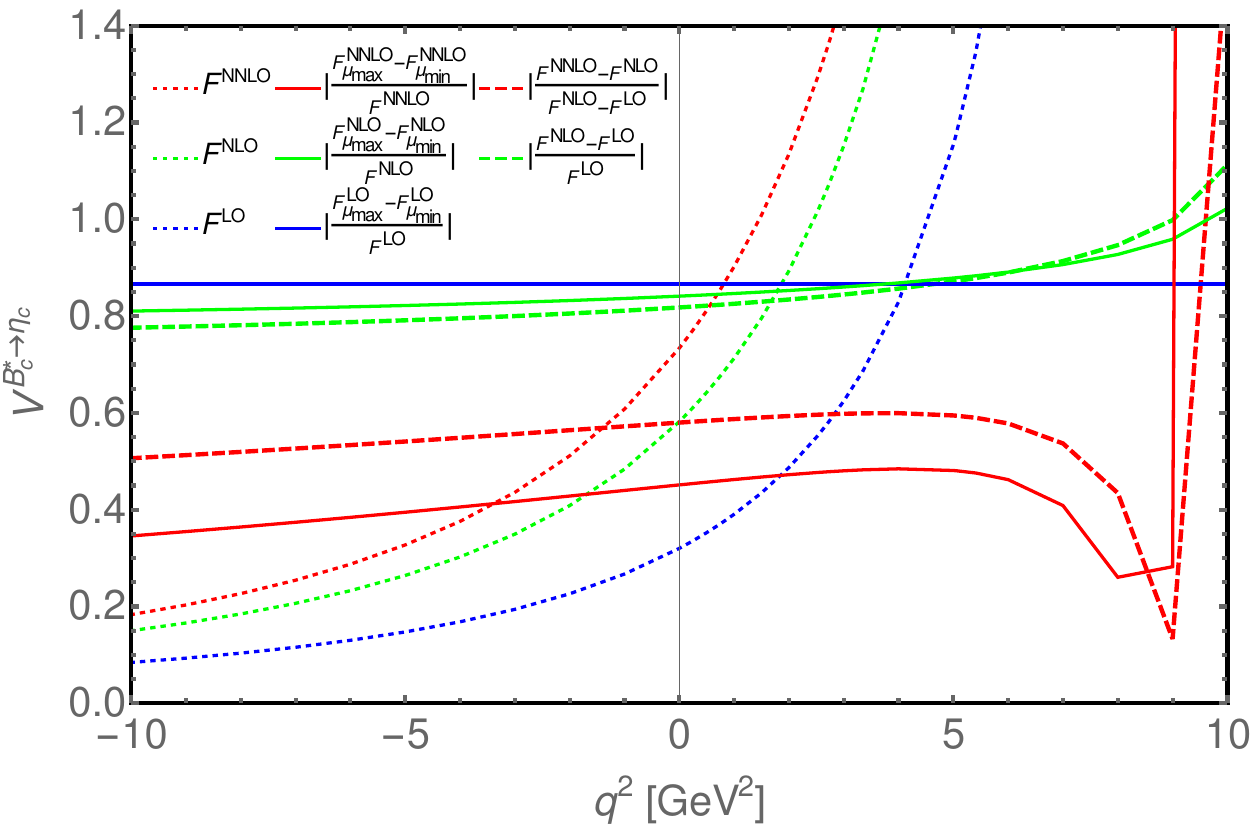}\qquad %
	\includegraphics[width=0.44\textwidth]{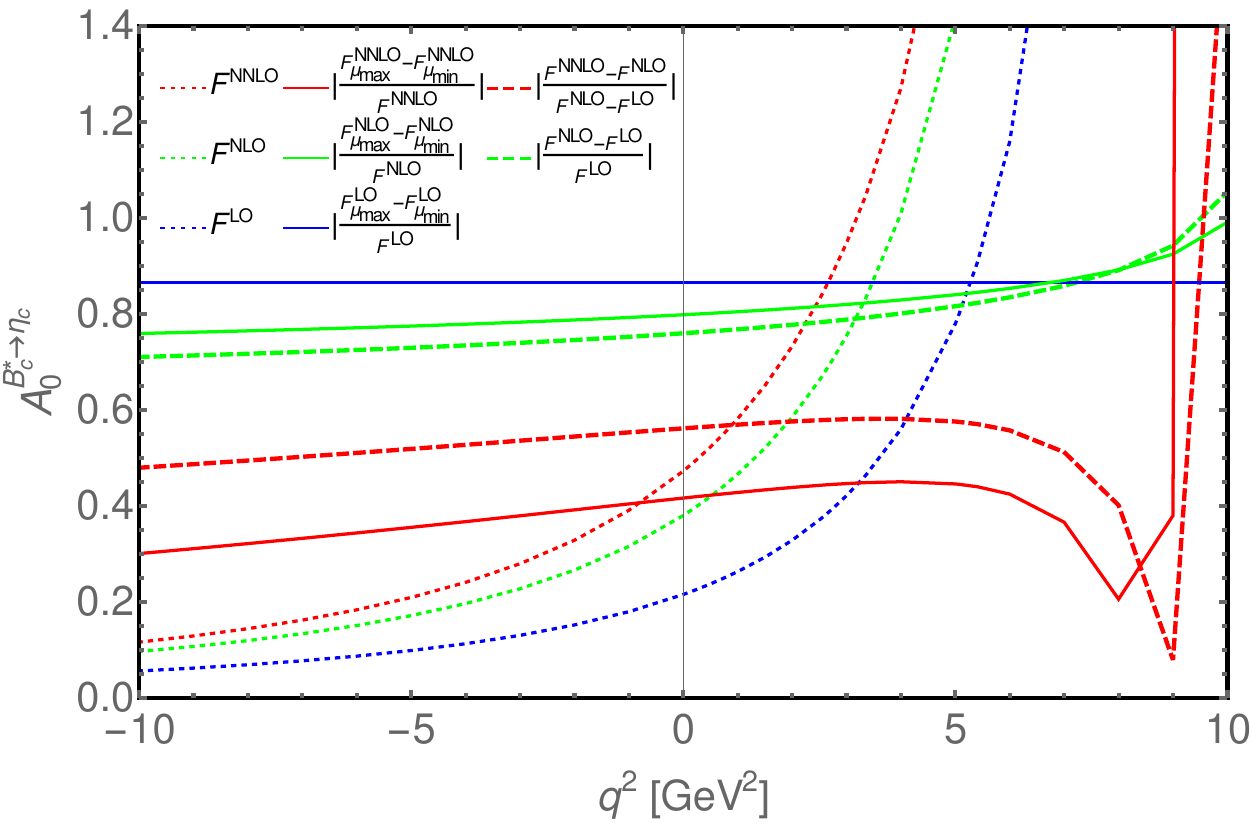}\qquad
	\includegraphics[width=0.44\textwidth]{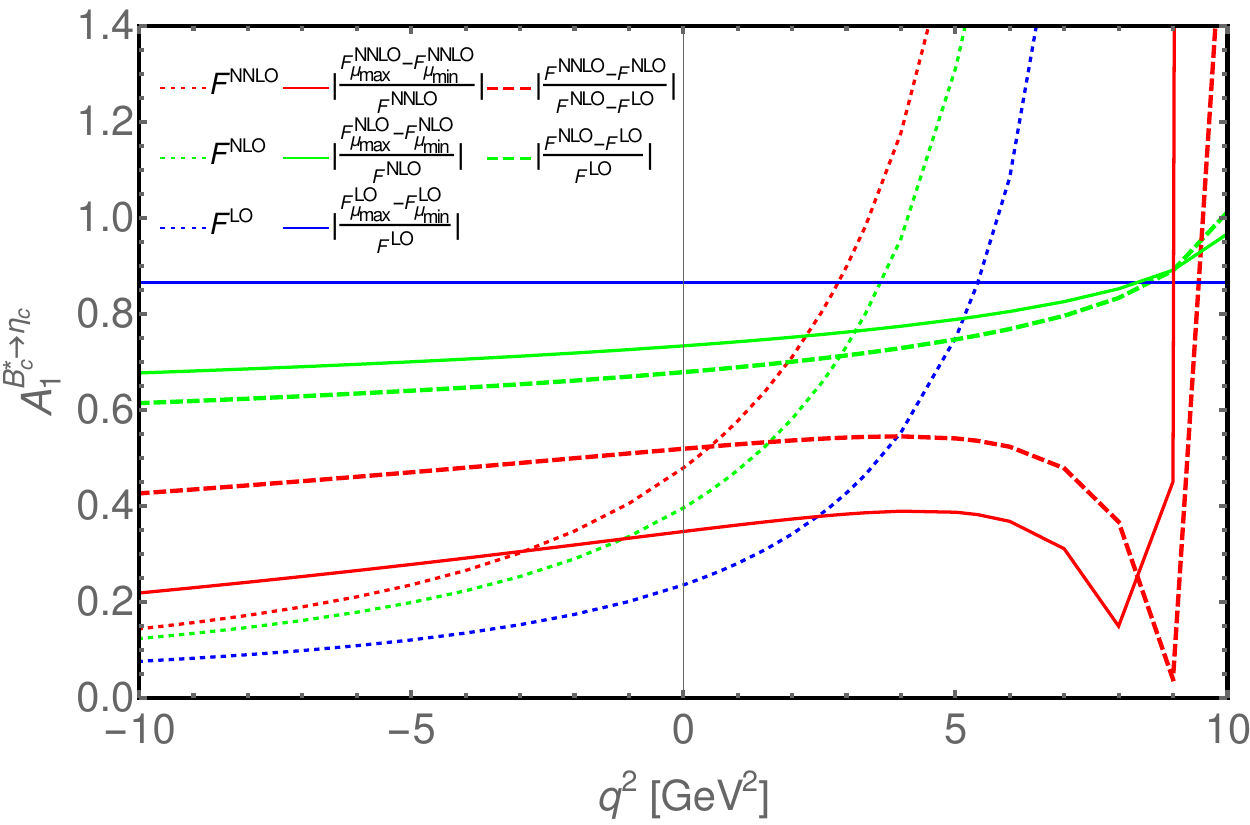}\qquad
	\includegraphics[width=0.44\textwidth]{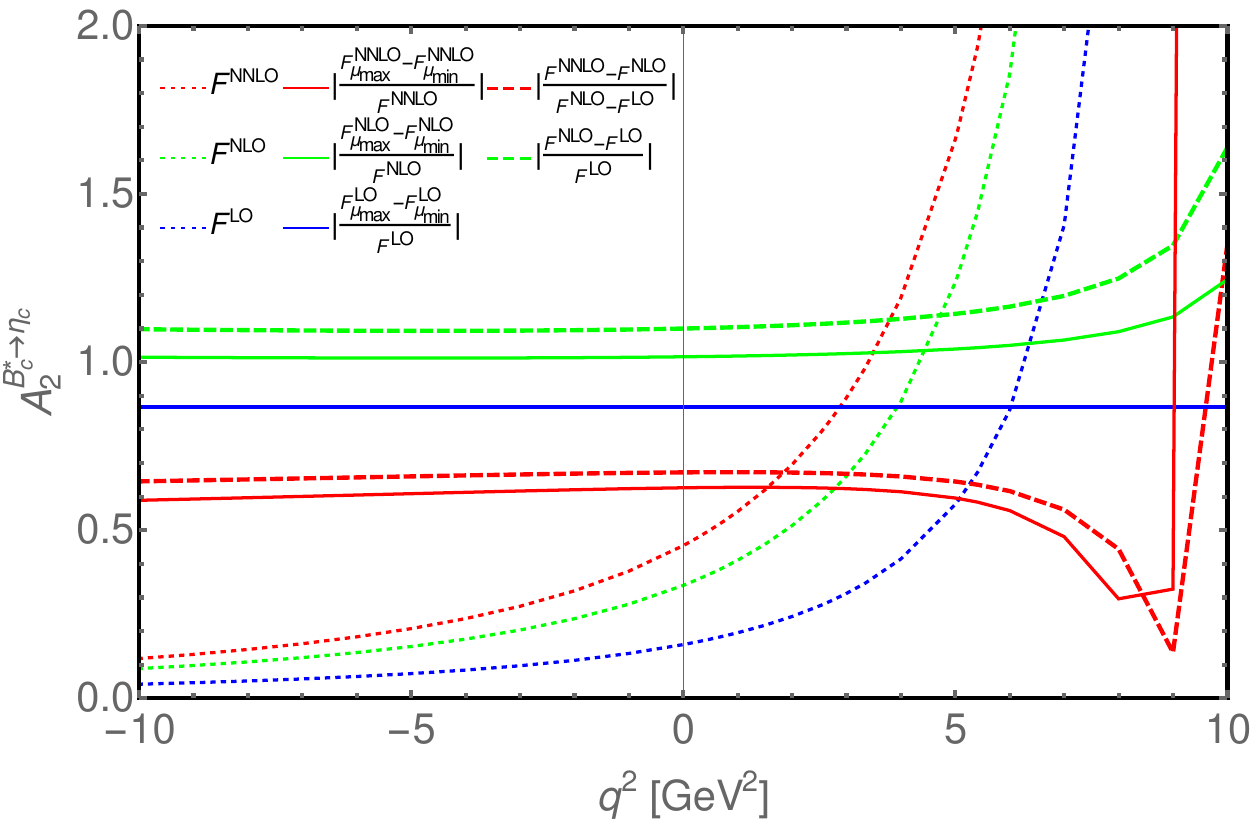}\qquad
	\includegraphics[width=0.44\textwidth]{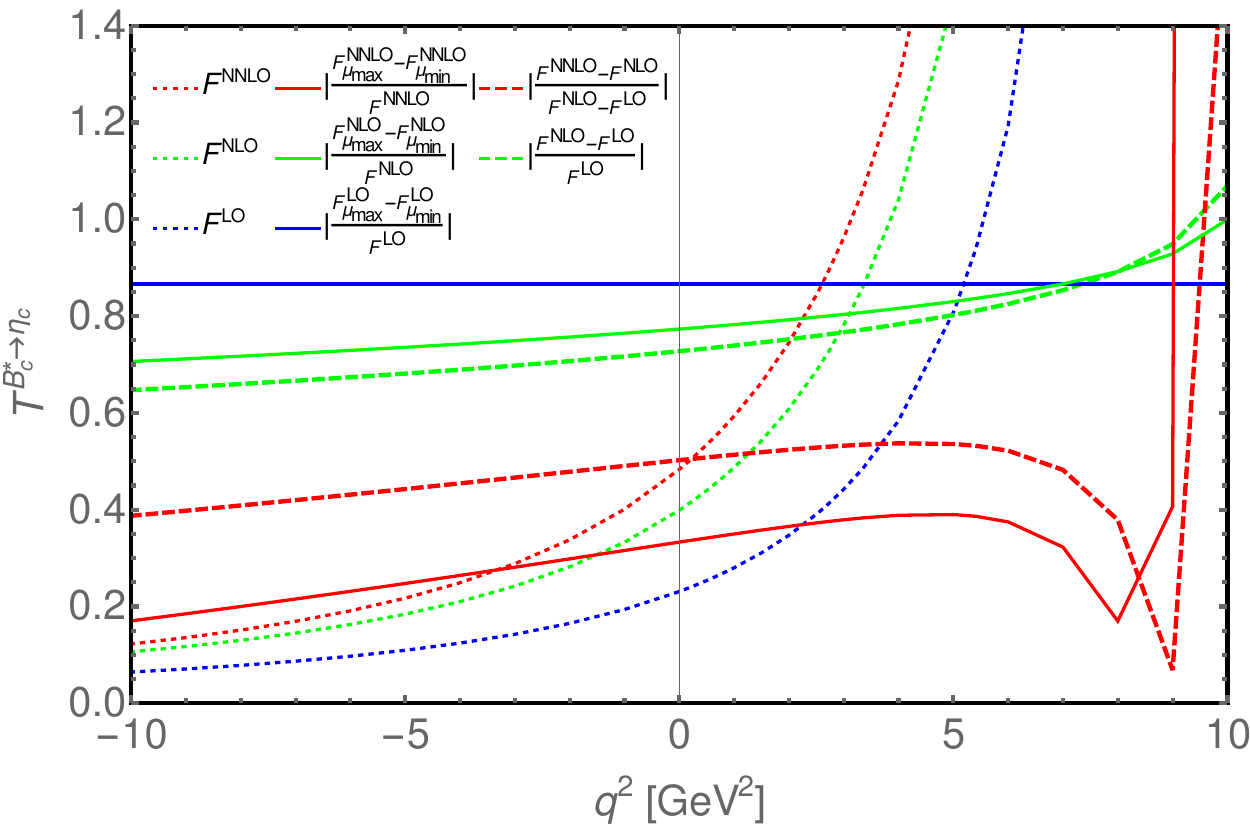}\qquad
	\includegraphics[width=0.44\textwidth]{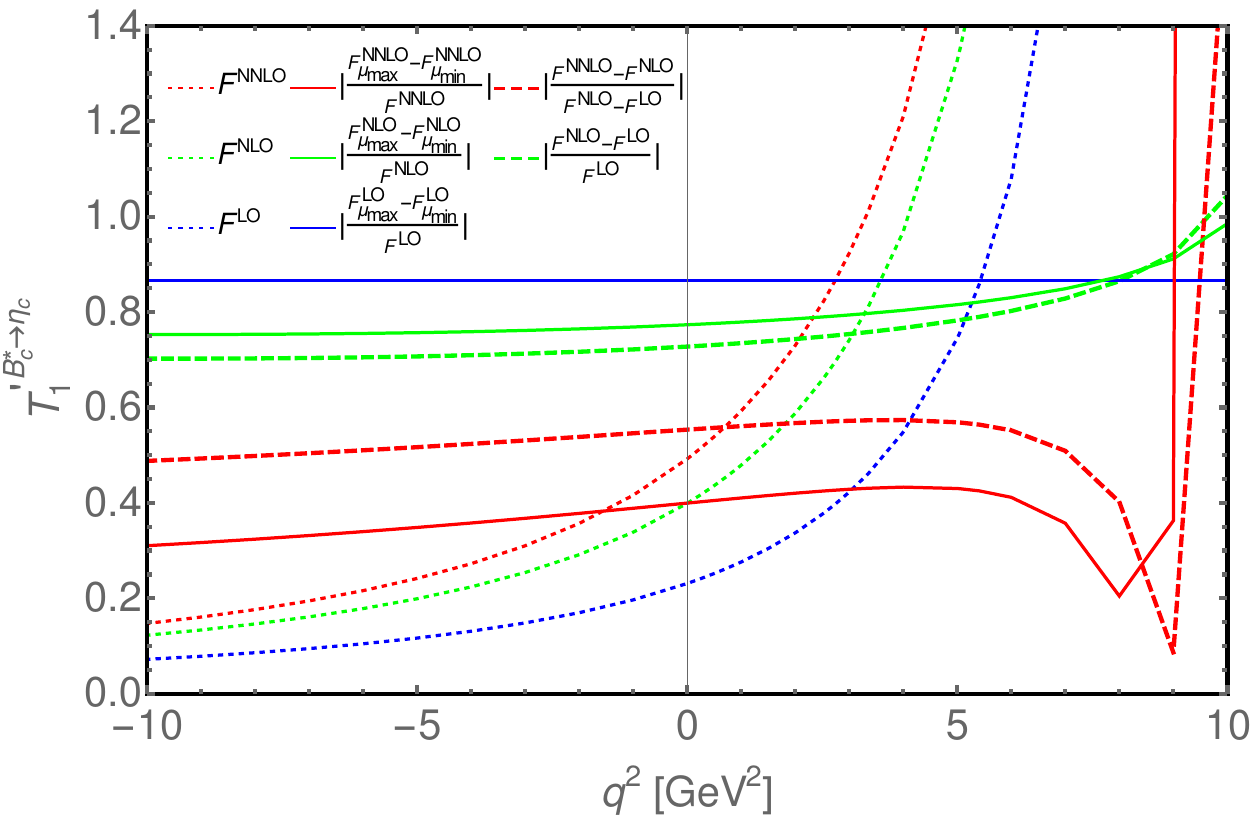}\qquad	
	\includegraphics[width=0.44\textwidth]{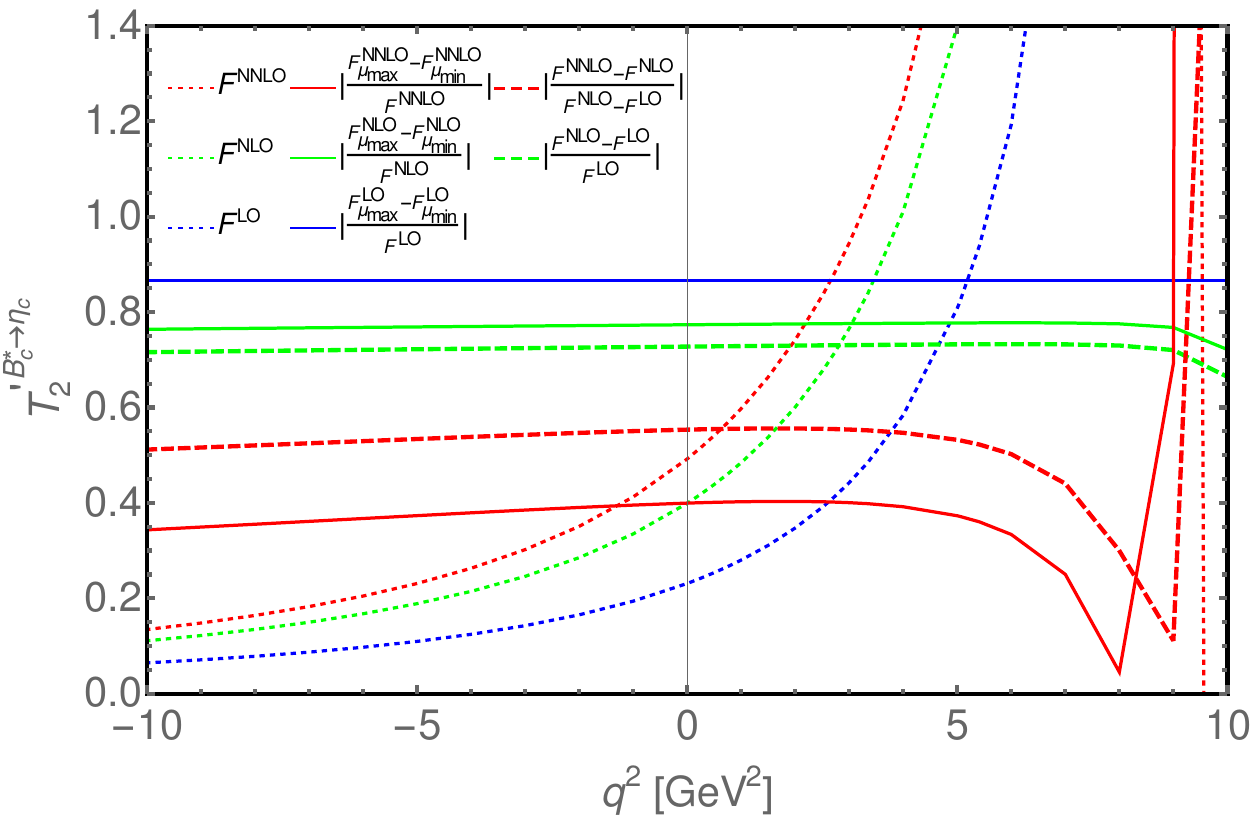}
	\caption{Same as figure~\ref{fig:ffq2nrbcjpsi}, but for $B_c^* \to \eta_c$.}
	\label{fig:ffq2nrbcvetac}
\end{figure}

We next present the NRQCD results for the $B_c^{(*)} \to J/\psi(\eta_c)$ form factors as functions of $q^2$ over the full %
range $q^2 \in [-10, 10]\,\text{GeV}^2$, covering both the spacelike ($q^2 < 0$) and timelike ($q^2 > 0$) regions as well as the maximum-recoil point ($q^2 = 0$).
Using the form-factor data obtained at the 31 rational-valued $q^2$ points specified in eq.~\eqref{datasetq2all}, we connect these discrete points to construct continuous curves. 
Figures~\ref{fig:ffq2nrbcjpsi}--\ref{fig:ffq2nrbcvetac} display the $q^2$ dependence of the form factors at LO, NLO, and NNLO accuracy, together with two complementary diagnostics for each form factor: 
(i) the relative renormalization-scale uncertainty at each perturbative order, arising from varying $\mu = 4.75_{-3.25}^{+5.25}\,\text{GeV}$, and 
(ii) the relative sizes of successive perturbative corrections.

A closer inspection of figures~\ref{fig:ffq2nrbcjpsi}--\ref{fig:ffq2nrbcvetac} reveals a universal contrast between the spacelike ($q^2<0$) and timelike ($q^2>0$) kinematic domains. In the spacelike region ($q^2<0$), the absolute form-factor values (dotted curves) exhibit smooth and well-behaved $q^2$ dependence at all three perturbative orders. More importantly, the successive QCD corrections remain under reasonable control, with the NNLO contributions generally smaller than the NLO ones. Crucially, the NNLO-to-NLO correction ratio (red dashed curve) consistently resides well below the NLO-to-LO correction ratio (green dashed curve) and remains substantially less than unity throughout $q^2\le0$, confirming the orderly asymptotic convergence of the $\alpha_s$ expansion series. As indicated by the solid curves, the relative renormalization-scale uncertainty drops progressively from LO to NLO, and is further suppressed at NNLO. %
Taken together, these features indicate good perturbative stability and convergence of the NRQCD predictions in the spacelike region.

In contrast to the spacelike domain, the behavior of the form factors in the timelike region ($q^2 > 0$) deteriorates noticeably. As $q^2$ increases, the continuous shrinking of the kinematic denominator factor $(m_b - m_c)^2 - q^2$ drives an increasingly steep rise in the absolute form-factor magnitudes across all three perturbative orders (dotted curves). At larger $q^2$ values, typically $q^2\gtrsim 5~\mathrm{GeV}^2$, and particularly toward the kinematic threshold, the relative renormalization-scale uncertainties at NNLO (red solid curves) and the NNLO-to-NLO correction ratios (red dashed curves) may approach unphysical zeros or singularities and develop pronounced anomalous variations, including steep downturns or even divergent behavior, indicating an increasingly unstable and poorly convergent fixed-order expansion. Moreover, for $q^2\gtrsim3~\mathrm{GeV}^2$, the NNLO (red) and NLO (green) solid/dashed curves, which are approximately parallel at lower $q^2$, progressively separate, further signaling the deterioration of perturbative stability. This loss of perturbative reliability is further illustrated by $T_3^{B_c\to J/\psi}$ in figure~\ref{fig:ffq2nrbcjpsi}. Specifically, the red dashed curve exceeds unity for $q^2\gtrsim1~\mathrm{GeV}^2$, indicating that the NNLO correction becomes comparable to or larger than the NLO result and that the perturbative convergence is no longer under satisfactory control. Moreover, for $q^2\gtrsim2~\mathrm{GeV}^2$, the red solid curve rises above the blue solid curve, indicating that the renormalization-scale dependence at NNLO becomes larger than that at LO. 
Overall, the observed steep growth of the form-factor magnitudes, deterioration of perturbative convergence, and enhanced renormalization-scale dependence at $q^2>0$ may be associated with threshold-related effects and the shrinking kinematic denominator $(m_b-m_c)^2-q^2$ within the NRQCD framework, as well as the decreasing recoil kinetic energy and the increasing importance of nonperturbative dynamics as $q^2$ increases from the spacelike region ($q^2<0$) toward the zero-recoil endpoint ($q^2=q^2_{\max}$)~\cite{Tao:2025kqf}.

The unambiguous physical contrast demonstrated in figures~\ref{fig:ffq2nrbcjpsi}--\ref{fig:ffq2nrbcvetac} provides the theoretical foundation for our fitting strategy in the next section. 
Attempting to directly use NRQCD calculations at large timelike $q^2 > 0$ would contaminate global fits with severe truncation and threshold instabilities. 
Conversely, the spacelike region $q^2 < 0$ offers an exceptionally clean, stable, and perturbatively robust environment where the NNLO corrections are under good theoretical control. 
This provides strong numerical support, complementing the analytic arguments based on the shrinking denominator $(m_b-m_c)^2-q^2$ in section~\ref{sec:ff0nr}, as well as on the decreasing recoil kinetic energy and increasingly nonperturbative dynamics as $q^2$ increases toward the zero-recoil endpoint~\cite{Tao:2025kqf}.
We therefore choose the fit input points predominantly in the spacelike region~\cite{Shen:2008zzb,Chang:2019obq,Bordone:2022drp}, supplemented by points in the immediate vicinity of $q^2=0$, where the perturbative expansion remains under good control. This choice retains sensitivity to the onset of the physical timelike region while avoiding the increasingly unstable regime at higher timelike $q^2$~\cite{Tao:2025kqf,Chang:2025rna,Qiao:2011yz}, thereby providing a theoretically well-controlled set of NRQCD input points to anchor the subsequent model-independent $z$-series parameterization in section~\ref{sec:ffq2nrz}.
The analytic continuation provided by the $z$-expansion will then stably and reliably map these high-precision spacelike predictions into the full physical decay region $0 \le q^2 \le q_{\text{max}}^2$, 
thereby avoiding direct fixed-order NRQCD calculations in the higher-$q^2$ region, where NRQCD may progressively lose reliability.

\section{NRQCD+$z$-series results for form factors~\label{sec:ffq2nrz}}

To obtain a reliable description of the form factors over the full physical region $0 \le q^2 \le q_{\max}^2=(M-M')^2$, we employ the model-independent BCL-type $z$-series parameterization~\cite{Bourrely:2008za,Khodjamirian:2011ub,Chang:2019obq}. Since $z(q^2)$ decreases monotonically with $q^2$, the most negative value of $q^2$ determines $|z(q^2)|_{\max}$ and is restricted here to $-q_{\max}^2$ to maintain the accuracy of the truncated $z$-series. Given that $|z(q^2)| \lesssim 0.05 \ll 1$ across both the spacelike input domain and the physical decay region, $-q_{\max}^2 \lesssim q^2 \le q_{\max}^2$, we adopt the truncated BCL form with $K=2$ for each form factor,
\begin{align}\label{eq:zfitformule}
F_i(q^2)=\frac{1}{1-q^2/m_{{\rm pole},i}^2}\left[b_0^{(i)}+b_1^{(i)}\left(z(q^2)+\frac{1}{2}z^2(q^2)\right)\right],
\end{align}
with
\begin{align}
z(q^2)&=
\frac{\sqrt{t_+-q^2}-\sqrt{t_+-t_0}}
{\sqrt{t_+-q^2}+\sqrt{t_+-t_0}},\\
t_0&=t_+\left(1-\sqrt{1-\frac{t_-}{t_+}}\right),\\
t_\pm&=(M\pm M')^2,
\end{align}
where $M$ and $M'$ denote the masses of the initial and final mesons, respectively, for the specific transition associated with $F_i(q^2)$.
The resonance pole mass $m_{\mathrm{pole},i}$ is chosen as the mass of the lowest-lying $c\bar{b}$ bound state carrying the same $J^P$ quantum numbers as the relevant current channel associated with the form factor $F_i(q^2)$~\cite{Boyd:1997kz,Hu:2019qcn,Leljak:2019eyw,Harrison:2025yan}. The explicit pole masses $m_{\mathrm{pole},i}$ assigned to each form factor are specified in table~\ref{tab:ff_mpole}.
For each form factor, the BCL parameters $(b_0^{(i)}, b_1^{(i)})$ are determined by a least-squares fit of eq.~\eqref{eq:zfitformule} to our NRQCD predictions in the spacelike region and its immediate vicinity, $-q_{\max}^2 \lesssim q^2 \lesssim 0$, where the fixed-order $\alpha_s$ expansion is well behaved and free from threshold instabilities. The resulting parameterization is then used to describe the form factor over the entire physical region, $0 \le q^2 \le q_{\max}^2$.
\begin{table}[htbp]
	\centering
	\caption{Pole masses used in the $z$-series parameterization, eq.~\eqref{eq:zfitformule}, for each form factor.}
	\label{tab:ff_mpole}
	\setlength{\tabcolsep}{15pt}
	\renewcommand{\arraystretch}{1.2}
	\begin{tabular}{cl}
		\toprule
		$m_{\text{pole}}$ &  Form Factors \\
		\midrule
		\begin{tabular}[t]{@{}c@{}}
			$m_{B_c}$ 
		\end{tabular} & 
		{\renewcommand{\arraystretch}{1.2}\begin{tabular}[t]{@{}l@{}}
			$A_0^{B_c \to J/\psi}$ \\
			$A_1^{B_c^* \to J/\psi}$ \\
			$A_0^{B_c^* \to \eta_c}$
		\end{tabular}} \\
		\midrule
		\begin{tabular}[t]{@{}c@{}}
			$m_{B_c^*}$
		\end{tabular} & 
		{\renewcommand{\arraystretch}{1.3}\begin{tabular}[t]{@{}l@{}}
			$V^{B_c \to J/\psi}, \, T_1^{B_c \to J/\psi}$ \\
			$F_p^{B_c \to \eta_c}, \, F_t^{B_c \to \eta_c}$ \\
			$V_{1,3,4,5}^{B_c^* \to J/\psi}, \, T_{2,3,4,6}^{B_c^* \to J/\psi}$ \\
			$V^{B_c^* \to \eta_c}, \, T^{B_c^* \to \eta_c}$
		\end{tabular}} \\
		\midrule
		\begin{tabular}[t]{@{}c@{}}
			$m_{B_{c0}^*(0^+)}$ 
		\end{tabular} & 
		{\renewcommand{\arraystretch}{1.2}\begin{tabular}[t]{@{}l@{}}
			$F_0^{B_c \to \eta_c}$ \\
			$V_{2,6}^{B_c^* \to J/\psi}$
		\end{tabular}} \\
		\midrule
		\begin{tabular}[t]{@{}c@{}}
			$m_{B_{c1}(1^+)}$ 
		\end{tabular} & 
		{\renewcommand{\arraystretch}{1.5}\begin{tabular}[t]{@{}l@{}}
			$A_{1,2}^{B_c \to J/\psi}, \, T_{2,3}^{B_c \to J/\psi}$ \\
			$A_{2,3,4}^{B_c^* \to J/\psi}, \, T_{2,3,4}^{\prime\, B_c^* \to J/\psi}$ \\
			$A_{1,2}^{B_c^* \to \eta_c}, \, T_{1,2}^{\prime\, B_c^* \to \eta_c}$
		\end{tabular}} \\
		\bottomrule
	\end{tabular}
\end{table}

\begin{figure}[htbp]%
	\centering
	\includegraphics[width=0.38\textwidth]{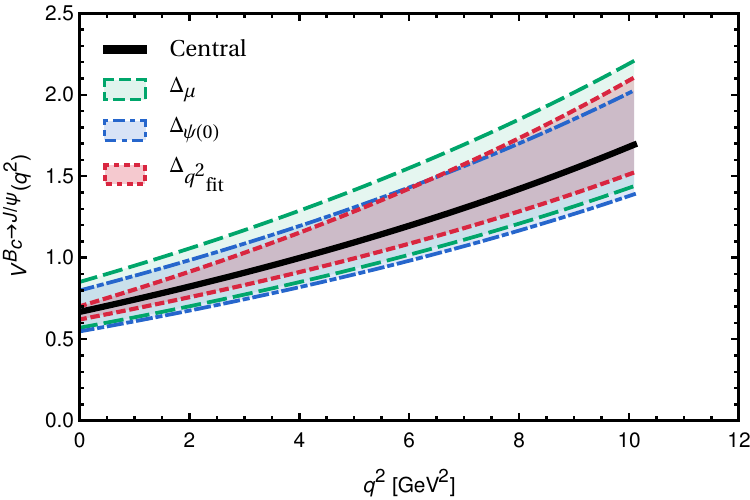}\qquad %
	\includegraphics[width=0.38\textwidth]{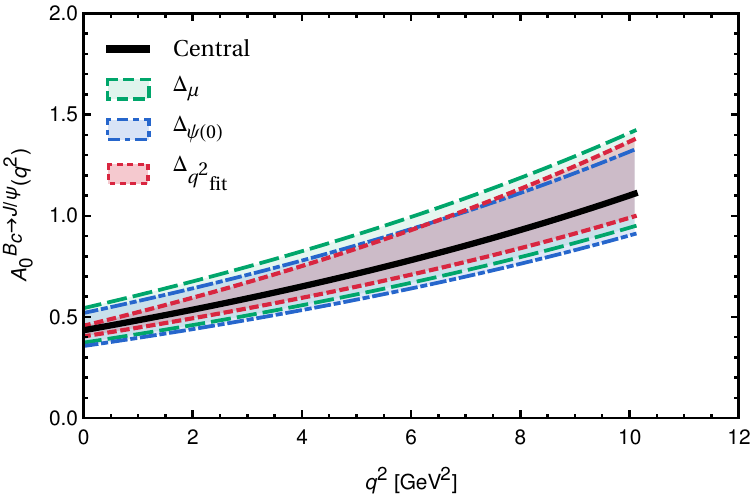}\qquad
	\includegraphics[width=0.38\textwidth]{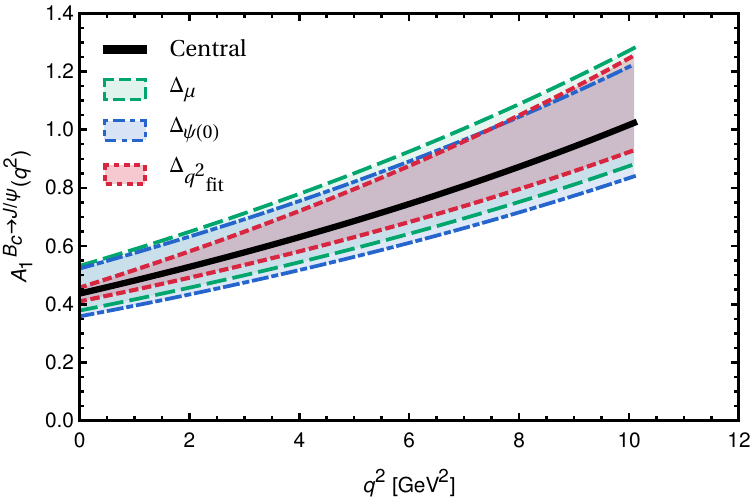}\qquad
	\includegraphics[width=0.38\textwidth]{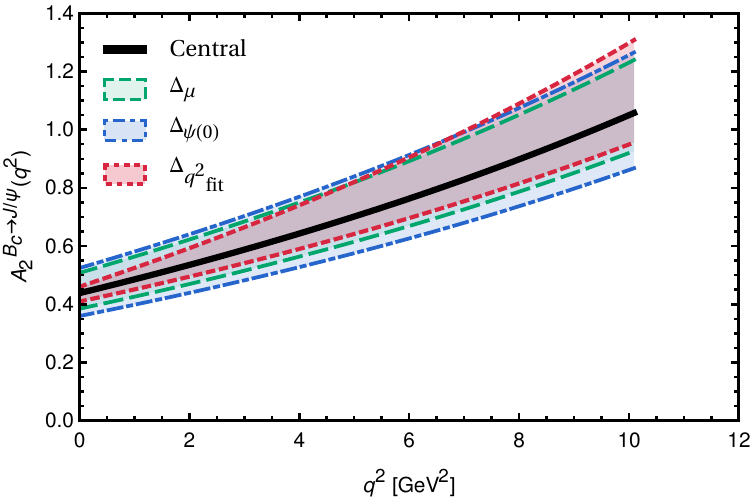}\qquad
	\includegraphics[width=0.38\textwidth]{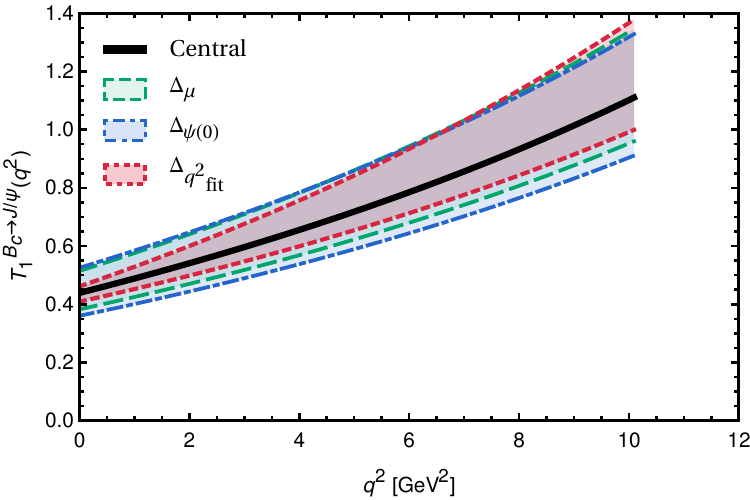}\qquad
	\includegraphics[width=0.38\textwidth]{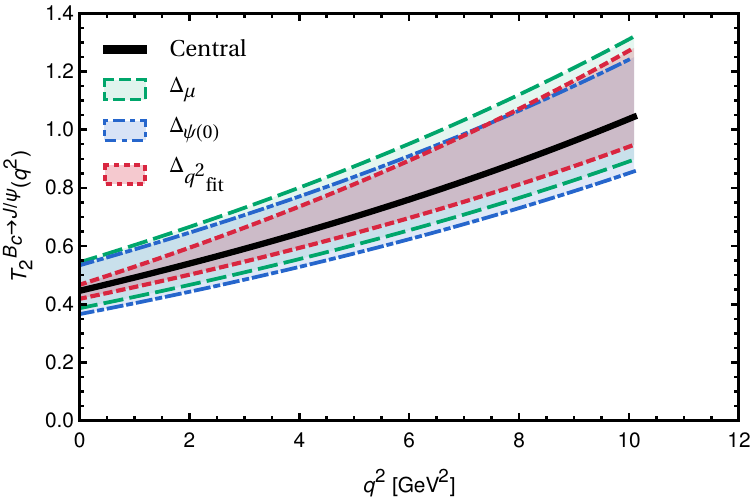}\qquad	
	\includegraphics[width=0.38\textwidth]{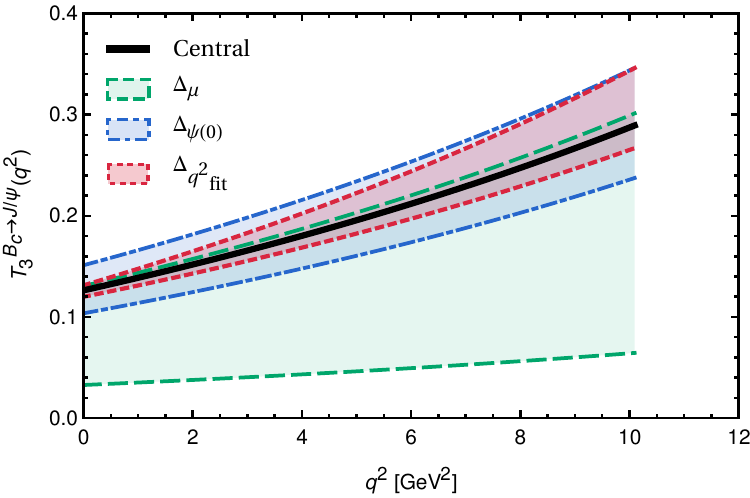}
	\caption{NNLO NRQCD+$z$-series predictions for the $B_c\to J/\psi$ form factors as functions of $q^2$ over the full physical region $0 \le q^2 \le \left(M-M'\right)^2$. For each form factor, the central value and the upper and lower edges of the three uncertainty bands are obtained from seven independent $z$-series fits, corresponding to the central and the six upper-/lower-edge NRQCD input configurations.
		Specifically, the thick solid black central curve is obtained from the $z$-series fit using the NNLO NRQCD input points at $q^2_{\rm fit}=\{-10,-9,-8,-7,-6,-5,-4,-3,-2,-1,0,1\}$~GeV$^2$, with $m_b=4.75$~GeV, $m_c=1.5$~GeV, $\mu=4.75$~GeV, $\mu_f=1.5$~GeV, and $\psi_{B_c}(0)\psi_{J/\psi}(0)=0.325\times0.225~{\rm GeV}^3$.
		The upper (lower) dashed edge of the green band is obtained by instead setting $\mu=1.5$ ($10$)~GeV. The upper (lower) dash-dotted edge of the blue band is obtained by instead setting $\psi_{B_c}(0)\psi_{J/\psi}(0)=0.350\times0.250$ ($0.300\times0.200$)~GeV$^3$. The upper dotted edge of the red band is obtained by instead using the fitting-point set $q^2_{\rm fit}=\{-5,-4,-3,-2,-1,0,1\}$~GeV$^2$, while the lower dotted edge of the red band is obtained by instead using $q^2_{\rm fit}=\{-10,-9,-8,-7,-6,-5,-4,-3,-2,-1,0\}$~GeV$^2$.
		}
	\label{fig:ffq2nrzbcjpsi}
\end{figure}

\begin{figure}[htbp]%
	\centering
	\includegraphics[width=0.44\textwidth]{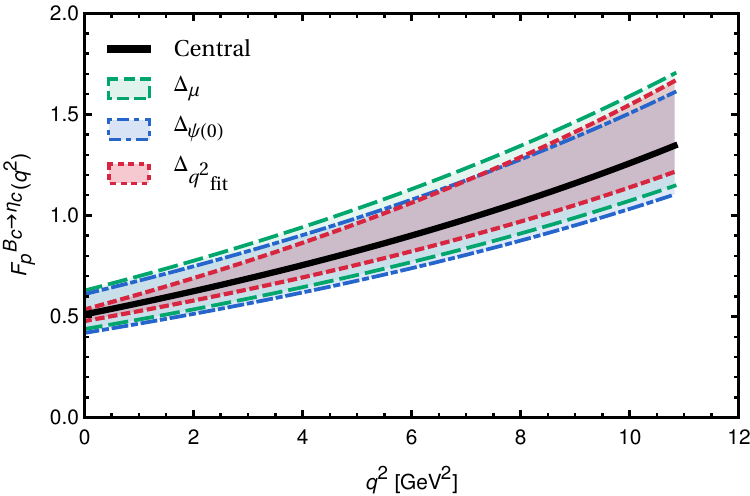}\qquad %
	\includegraphics[width=0.44\textwidth]{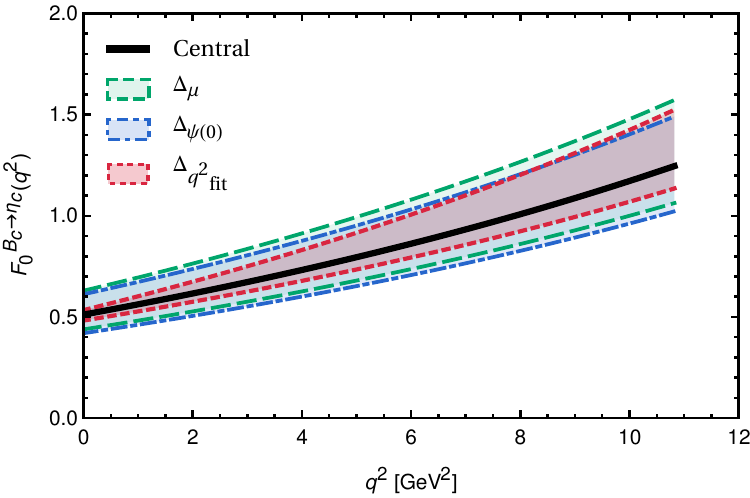}\qquad
	\includegraphics[width=0.44\textwidth]{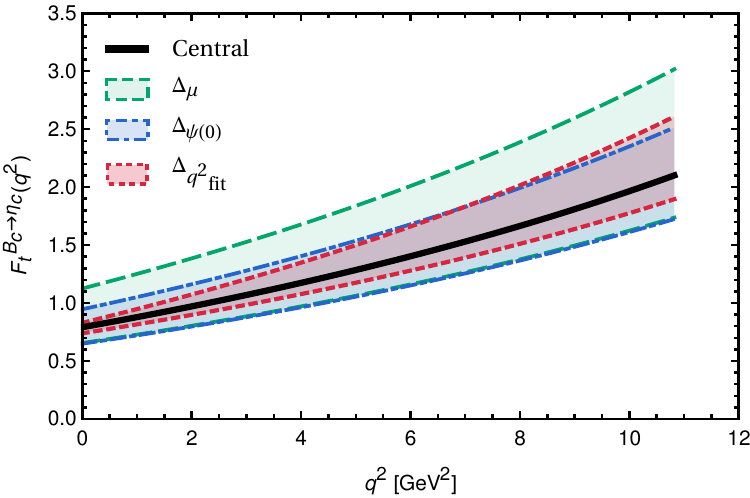}
	\caption{Same as figure~\ref{fig:ffq2nrzbcjpsi}, but for $B_c\to \eta_c$.}
	\label{fig:ffq2nrzbcetac}
\end{figure}

\begin{figure}[htbp]%
	\centering
	\includegraphics[width=0.315\textwidth]{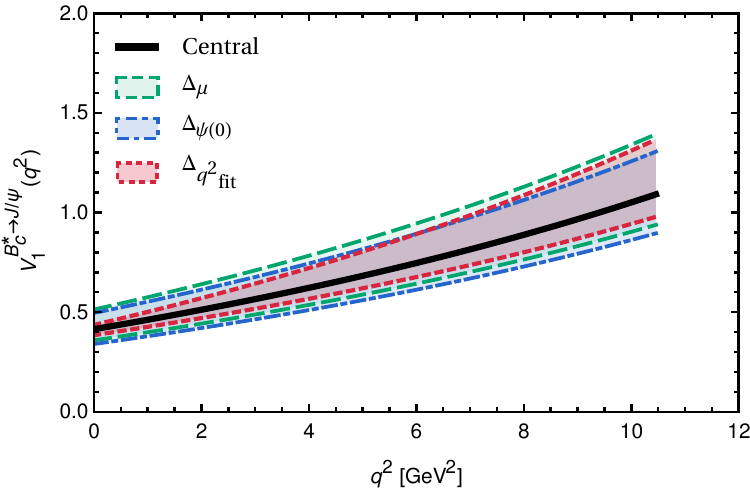}\quad %
	\includegraphics[width=0.315\textwidth]{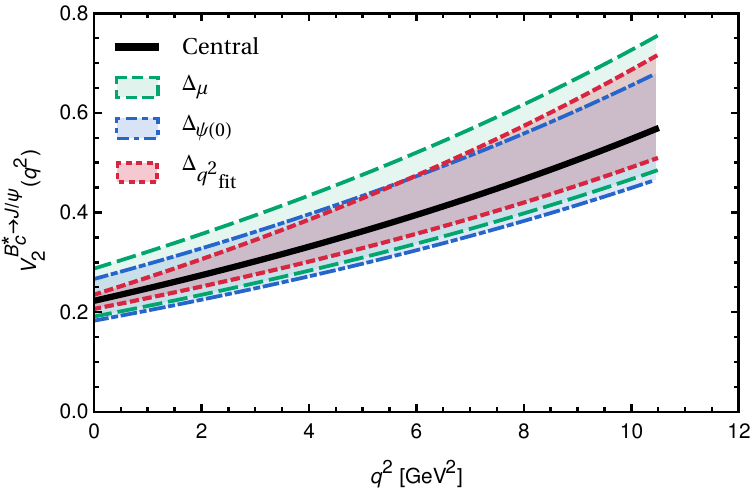}\quad
	\includegraphics[width=0.315\textwidth]{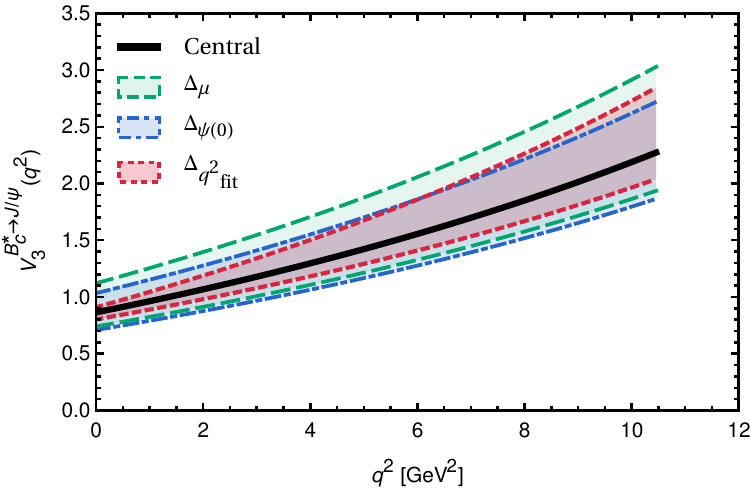}\quad
	\includegraphics[width=0.315\textwidth]{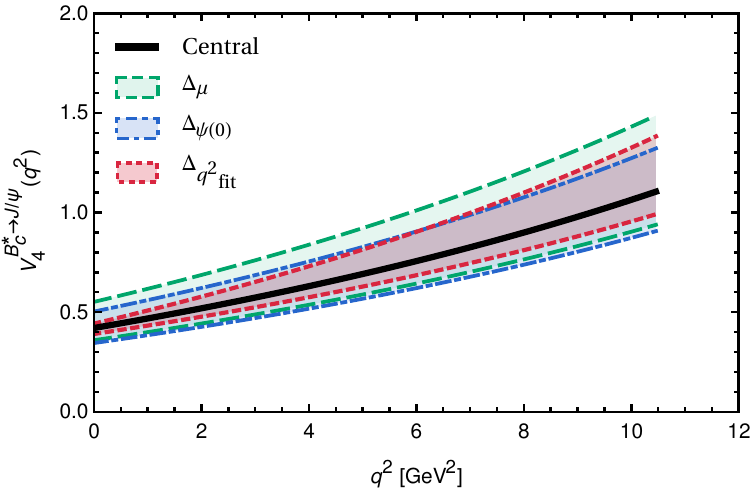}\quad
	\includegraphics[width=0.315\textwidth]{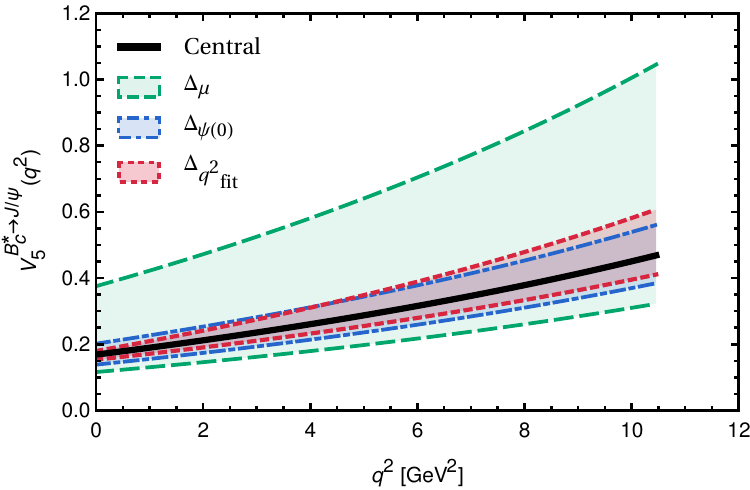}\quad
	\includegraphics[width=0.315\textwidth]{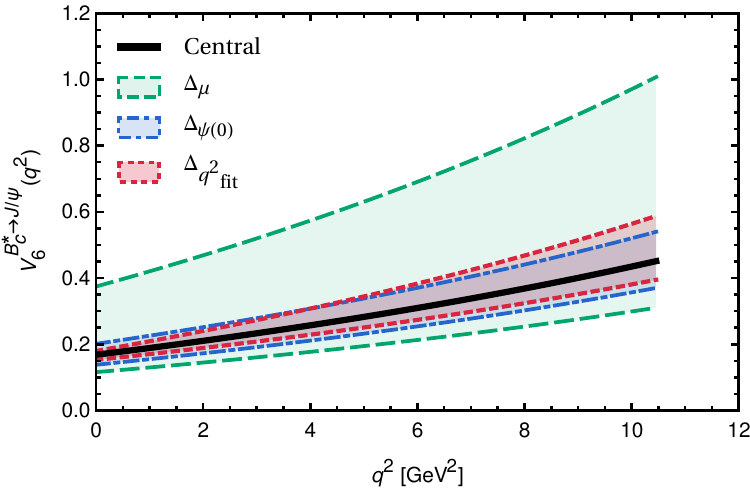}\quad
	\includegraphics[width=0.315\textwidth]{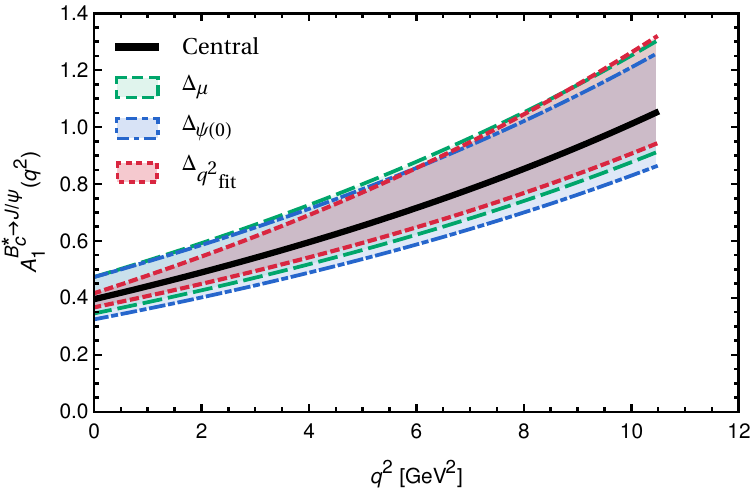}\quad	
	\includegraphics[width=0.315\textwidth]{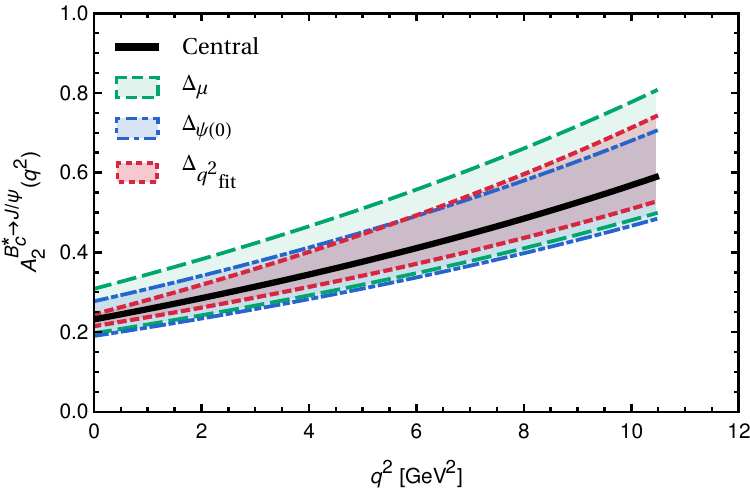}\quad
	\includegraphics[width=0.315\textwidth]{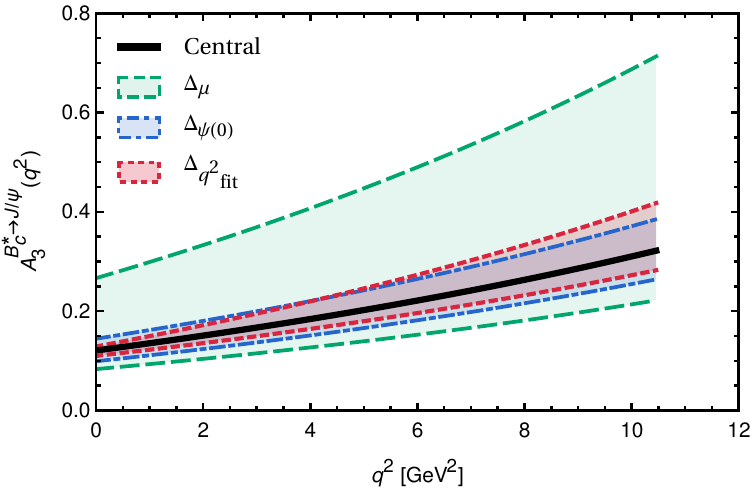}\quad
	\includegraphics[width=0.315\textwidth]{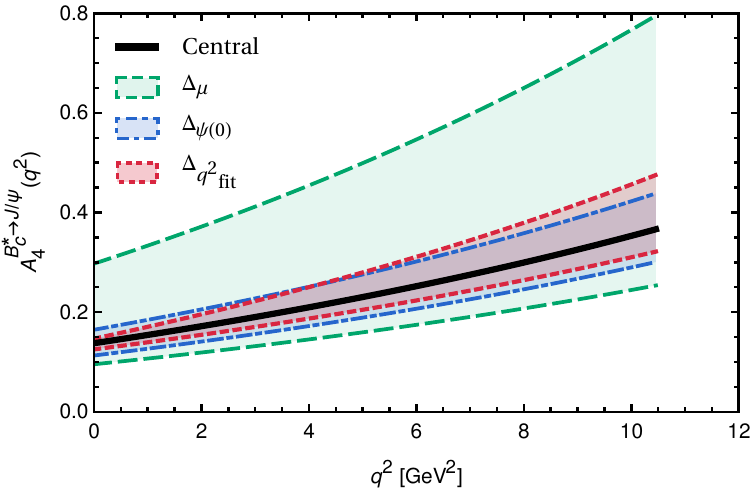}\quad		
	\includegraphics[width=0.315\textwidth]{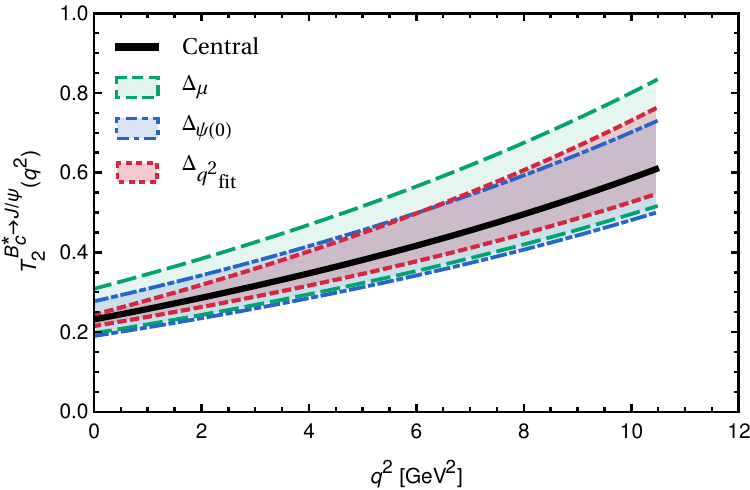}\quad
	\includegraphics[width=0.315\textwidth]{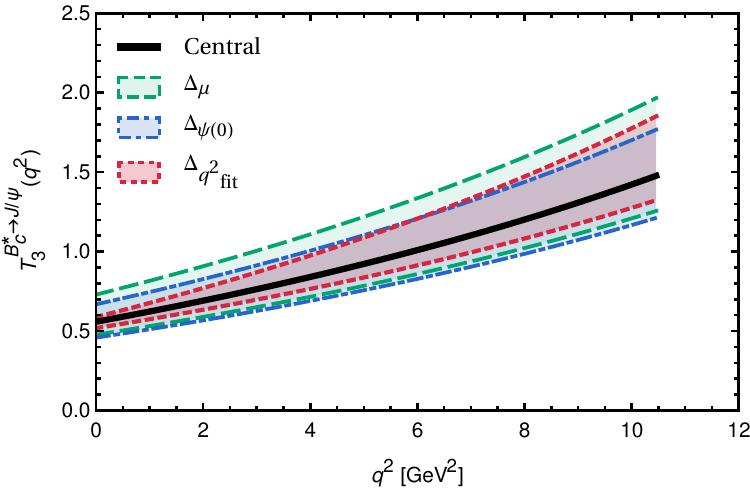}\quad
	\includegraphics[width=0.315\textwidth]{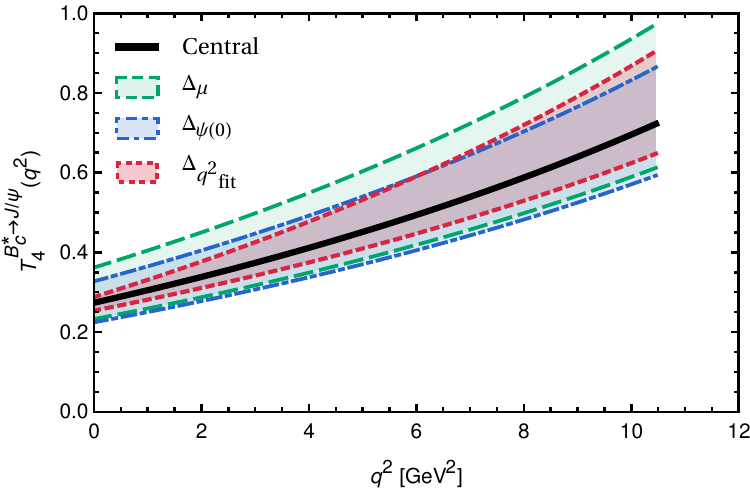}\quad
	\includegraphics[width=0.315\textwidth]{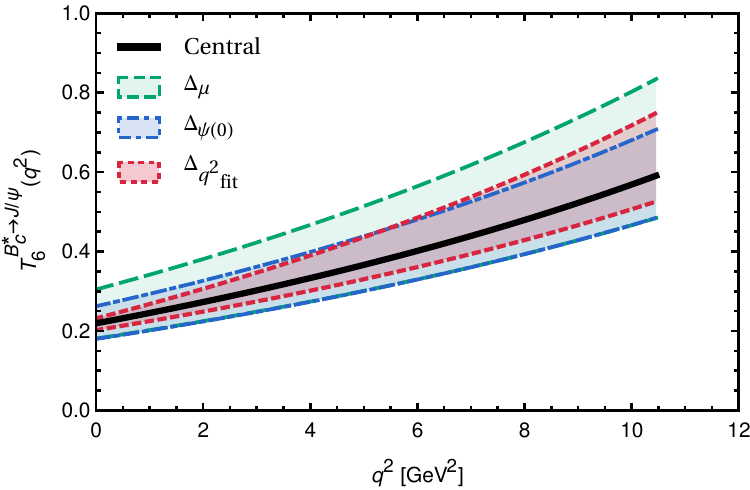}\quad
	\includegraphics[width=0.315\textwidth]{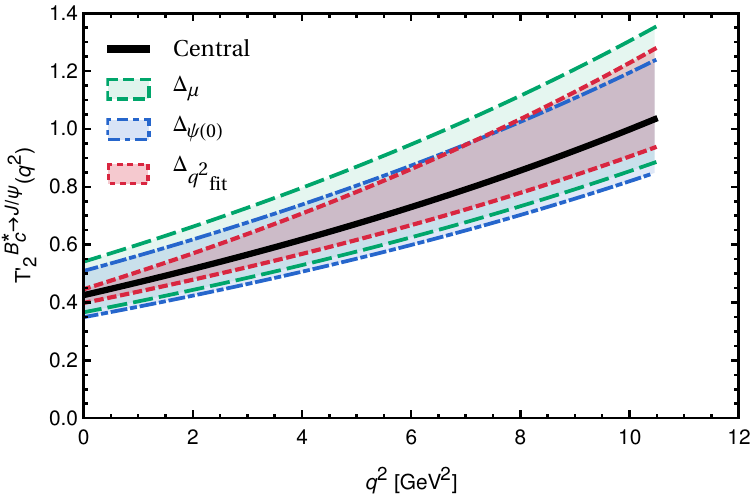}\quad
	\includegraphics[width=0.315\textwidth]{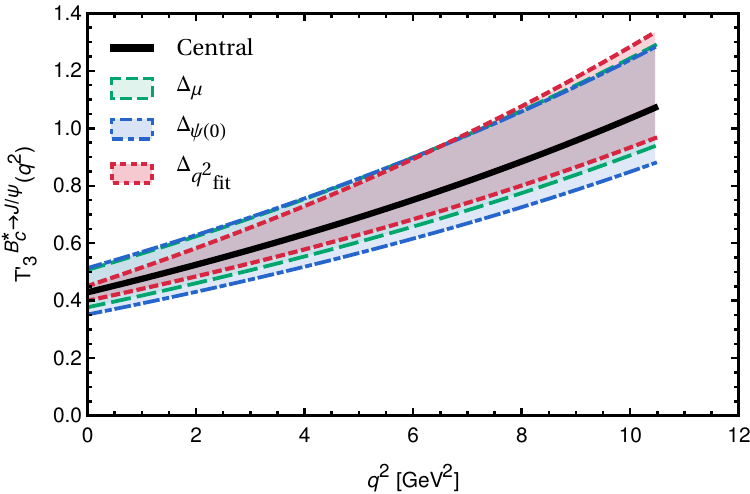}\quad
	\includegraphics[width=0.315\textwidth]{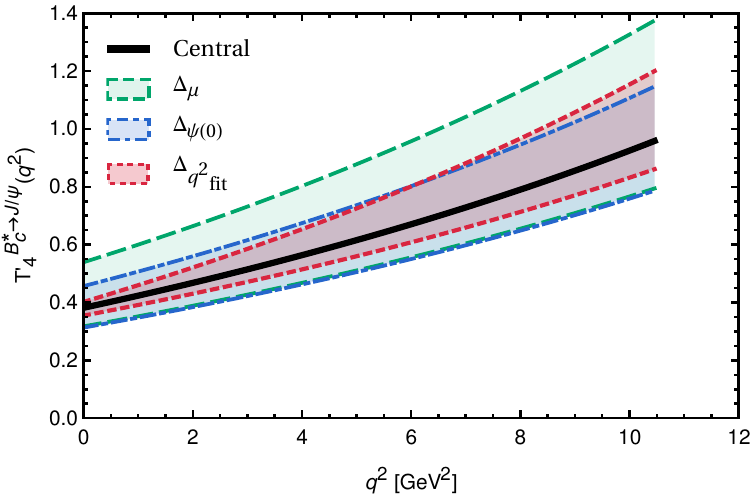}
	\caption{Same as figure~\ref{fig:ffq2nrzbcjpsi}, but for $B_c^*\to J/\psi$.}
	\label{fig:ffq2nrzbcvjpsi}
\end{figure}

\begin{figure}[htbp]%
	\centering
	\includegraphics[width=0.44\textwidth]{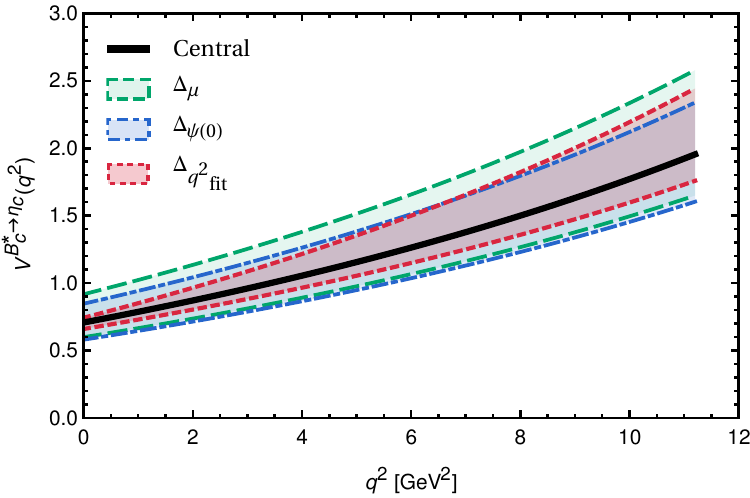}\qquad %
	\includegraphics[width=0.44\textwidth]{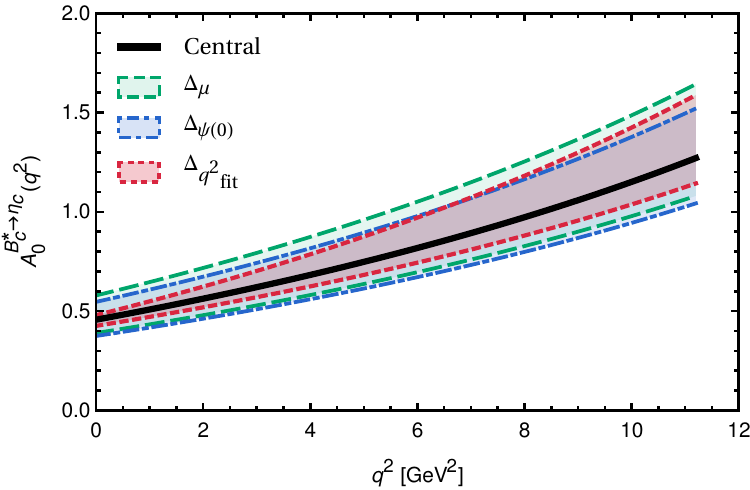}\qquad
	\includegraphics[width=0.44\textwidth]{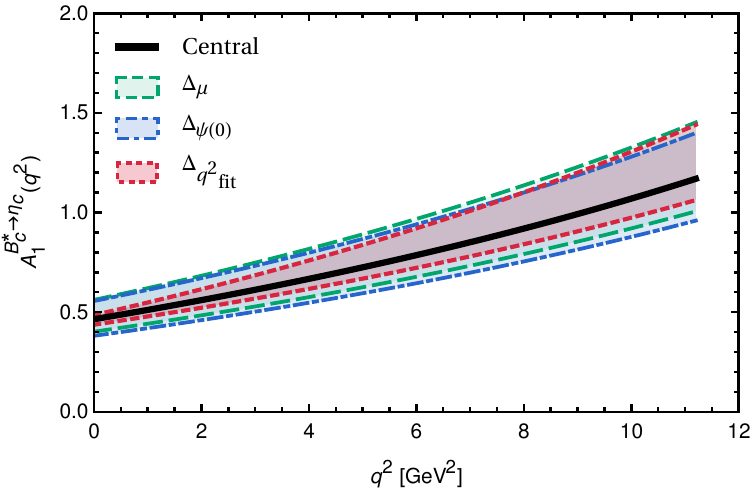}\qquad
	\includegraphics[width=0.44\textwidth]{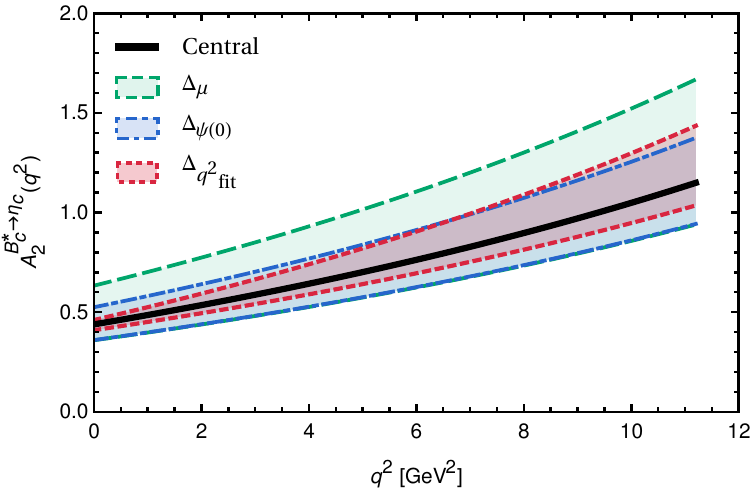}\qquad
	\includegraphics[width=0.44\textwidth]{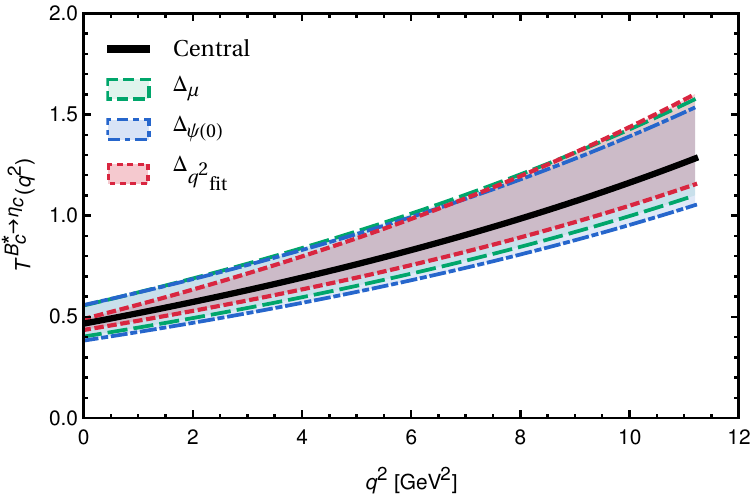}\qquad
	\includegraphics[width=0.44\textwidth]{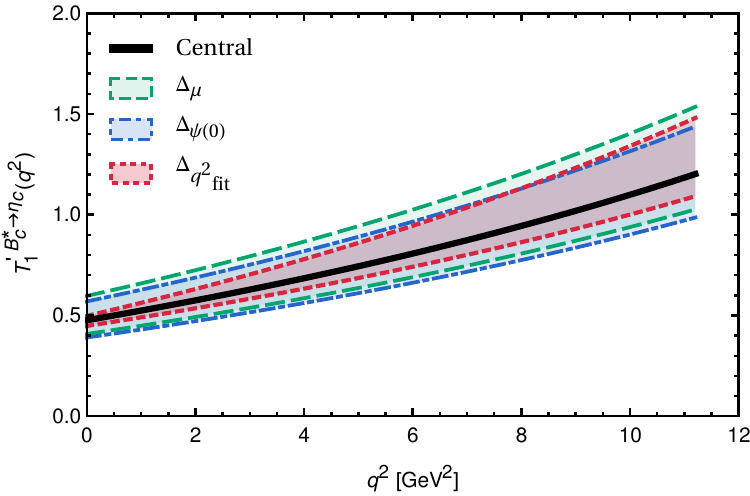}\qquad	
	\includegraphics[width=0.44\textwidth]{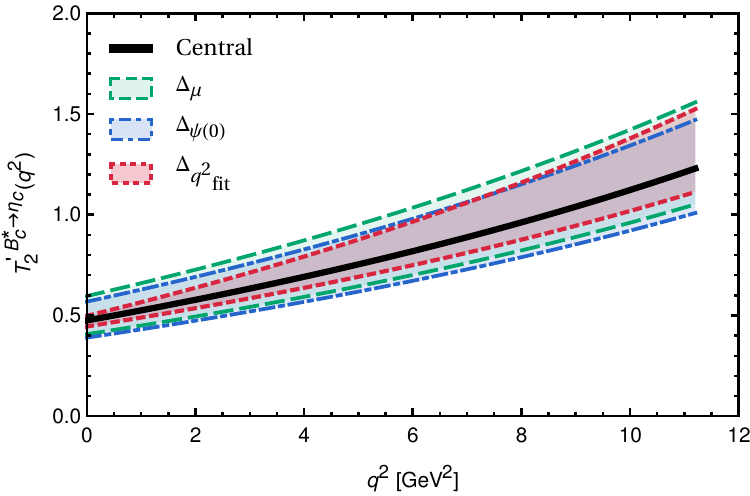}
	\caption{Same as figure~\ref{fig:ffq2nrzbcjpsi}, but for $B_c^*\to \eta_c$.}
	\label{fig:ffq2nrzbcvetac}
\end{figure}

In figures~\ref{fig:ffq2nrzbcjpsi}--\ref{fig:ffq2nrzbcvetac}, we present the resulting NNLO NRQCD+BCL $z$-series predictions for the $B_c^{(*)} \to J/\psi(\eta_c)$ form factors across the full physical range $0 \le q^2 \le q_{\max}^2$.
In each plot:
\begin{itemize}
	\item The thick solid black curve represents our central prediction, evaluated using the central inputs for the heavy-quark masses, default energy scales ($\mu = 4.75~\text{GeV}, \mu_f = 1.5~\text{GeV}$), and the midpoint wave-function parameters.
	
	\item The green dashed curves delimit the renormalization-scale uncertainty ($\Delta_\mu$), obtained by varying $\mu$ over the interval specified in eq.~\eqref{mbmcmumufval}.
	
	\item The blue dash-dotted curves delineate the nonperturbative parametric uncertainty ($\Delta_{\psi(0)}$), propagated from the lower and upper bounds of the wave functions at the origin extracted from decay constants and potential models.
	
	\item The red dotted curves enclosing the shaded band illustrate the fitting-point selection uncertainty ($\Delta_{q^2_{\text{fit}}}$), obtained by varying the set of NRQCD input points within the spacelike and near-zero $q^2$ region.
\end{itemize}

To assess the uncertainty arising from the choice of fitting points, we perform independent $z$-series fits using the following three different sets of $q^2$ points from our NRQCD dataset as input:
\begin{equation}
\begin{aligned}
q^2 &\in \{-10,-9,-8,-7,-6,-5,-4,-3,-2,-1,0,1\}\,\mathrm{GeV}^2,\\
q^2 &\in \{-5,-4,-3,-2,-1,0,1\}\,\mathrm{GeV}^2,\\
q^2 &\in \{-10,-9,-8,-7,-6,-5,-4,-3,-2,-1,0\}\,\mathrm{GeV}^2,
\end{aligned}
\label{eq:q2fitcenupdn}
\end{equation}
corresponding to the thick solid black central curve, the dotted upper edge, and the dotted lower edge of the red uncertainty band in each figure, respectively.  
As can be seen from the central curve and the upper and lower boundaries of the red uncertainty band, shifting the distribution of input fitting points toward the timelike or spacelike $q^2$ region influences the steepness of the extrapolated form factors: including timelike $q^2$ points in the fit leads to a steeper rise of the resulting NRQCD+$z$-series form-factor predictions, whereas shifting the input points deeper into the spacelike region results in a more gradual $q^2$ dependence over the physical region.

As shown in figures~\ref{fig:ffq2nrzbcjpsi}--\ref{fig:ffq2nrzbcvetac}, all NRQCD+$z$-series curves, including the central and boundary curves, exhibit a smooth and monotonically increasing $q^2$ dependence and remain stable and well behaved across the entire physical domain.
Overall, the three uncertainty bands broaden mildly as $q^2$ increases, while remaining reasonably well controlled up to $q_{\text{max}}^2$. 
A rough comparison among the three uncertainty bands indicates that the uncertainties arising from the renormalization scale ($\Delta_\mu$), the wave functions at the origin ($\Delta_{\psi(0)}$), and the choice of fitting points ($\Delta_{q^2_{\text{fit}}}$) are comparable in magnitude, with $\Delta_{q^2_{\text{fit}}}$ being relatively smaller, demonstrating the stability and robustness of our $z$-series fit.

Furthermore, a quantitative comparison with the latest lattice QCD results~\cite{Harrison:2025yan} shows broad agreement across the physical region. At $q^2=0$, our NNLO NRQCD+$z$-series central predictions $(V, A_0, A_1, A_2, T_1, T_2, T_3)^{B_c\to J/\psi} \approx (0.668, 0.434, 0.437, 0.438, 0.439, 0.446, 0.126)$ are close to the lattice central values $(0.751, 0.465, 0.456, 0.437, 0.498, 0.498, 0.250)$, except for $T_3^{B_c\to J/\psi}$, which is about half of the lattice value. At $q^2=10\,\mathrm{GeV}^2$, our central values $(1.680, 1.100, 1.016, 1.050, 1.102, 1.037, 0.287)$ remain broadly consistent with the lattice central values $(1.578, 1.053, 0.785, 0.708, 1.029, 0.817, 0.451)$, although ours rise somewhat more steeply toward $q^2_{\max}$, most notably for $A_2^{B_c\to J/\psi}$.

Overall, these results indicate that the NNLO NRQCD+$z$-series predictions provide a relatively reliable and theoretically controlled description of the $B_c^{(*)}\to J/\psi(\eta_c)$ form factors across the entire physical region.

\section{Semileptonic decay widths, branching fractions and LFU ratios~\label{sec:GmBrR}}

As a phenomenological application of our NRQCD+$z$-series predictions for the $B_c^{(*)} \to J/\psi (\eta_c)$ transition form factors, we now investigate the corresponding semileptonic decays $B_c^{(*)} \to J/\psi (\eta_c) \, \ell {\nu}_\ell$ with $\ell \in \{e, \mu, \tau\}$. 

In the Standard Model, the differential decay width $d\Gamma/dq^2$ for each semileptonic decay $H \to H' \, \ell {\nu}_\ell$ can be expressed in terms of the corresponding vector and axial-vector form factors via helicity amplitudes. 
The explicit expressions of $d\Gamma/dq^2$ formulated with the form factors defined in this work can be adapted from refs.~\cite{Harrison:2020gvo,Wang:2008xt,Chang:2020xvu,Chang:2016cdi} and are collected in appendix~\ref{app:dGmdq2ffs} for completeness.

Using our NRQCD+$z$-series results for the $B_c^{(*)} \to J/\psi(\eta_c)$ transition form factors as input, the semileptonic decay widths $\Gamma$ for $B_c^{(*)} \to J/\psi(\eta_c) \, \ell {\nu}_\ell$ are obtained by integrating the corresponding $d\Gamma/dq^2$ expressions over the full kinematically allowed $q^2$ range~\cite{Cohen:2018vhw}, from which the corresponding branching fractions $\mathcal{B}$ and lepton-flavor-universality (LFU) ratios $\mathcal{R}$ are defined as follows:
\begin{equation}
\Gamma(H\to H'\,\ell{\nu}_\ell) = \int_{m_\ell^2}^{(m_{H}-m_{H'})^2}\, \frac{{d}\Gamma(H\to H'\,\ell{\nu}_\ell)}{{d}q^2}\,{d}q^2 ,
\end{equation}
\begin{equation}
\mathcal{B}(H\to H'\,\ell{\nu}_\ell) = \frac{\Gamma(H\to H'\,\ell{\nu}_\ell)} {\Gamma_{\mathrm{tot}}(H)},
\end{equation}
\begin{equation}
\mathcal{R}(H\to H') = \frac{\mathcal{B}(H\to H'\,\tau{\nu}_\tau)} {\mathcal{B}(H\to H'\,\mu{\nu}_\mu)} = \frac{\Gamma(H\to H'\,\tau{\nu}_\tau)} {\Gamma(H\to H'\,\mu{\nu}_\mu)}, 
\end{equation}
where
\begin{equation}
H\in\{B_c,B_c^*\},\qquad H'\in\{J/\psi,\eta_c\},\qquad \ell\in\{e,\mu,\tau\},
\end{equation}
and the corresponding values of $\Gamma_{\mathrm{tot}}(H)$, $m_H$, $m_{H'}$, and $m_\ell$ are taken from table~\ref{tab:globalinputs}.

\begin{table}[htbp]
	\centering
	\caption{The LO, NLO, and NNLO NRQCD+BCL $z$-series predictions for the semileptonic decay widths %
				 $\Gamma(H \to H' \, \ell {\nu}_\ell)$ ($H \in \{B_c, B_c^*\}$, $H' \in \{J/\psi, \eta_c\}$, and $\ell \in \{e, \mu, \tau\}$).
					The LO, NLO, and NNLO central values of the semileptonic decay widths are calculated using the corresponding LO, NLO, and NNLO NRQCD+$z$-series form factors obtained with the default inputs:
		$m_b = 4.75~\mathrm{GeV}$, $m_c = 1.5~\mathrm{GeV}$, $\mu = 4.75~\mathrm{GeV}$, $\mu_f = 1.5~\mathrm{GeV}$, 
		$\psi_{B_c^{(*)}}(0)\psi_{J/\psi(\eta_c)}(0) = 0.325 \times 0.225~\mathrm{GeV}^3$, 
		and the fitting-point set $q^2_{\mathrm{fit}} = \{-10,-9,-8,-7,-6,-5,-4,-3,-2,-1,0,1\}~\mathrm{GeV}^2$. 
		The four uncertainty terms in the superscripts (subscripts) attached to each central value correspond to the shifts obtained by individually varying the following inputs to their upper (lower) values:
		the renormalization scale $\mu = 10.0\ (1.5)~\mathrm{GeV}$, 
		the factorization scale $\mu_f = 3.0\ (0.5)~\mathrm{GeV}$, 
		the wave-function product $\psi_{B_c^{(*)}}(0)\psi_{J/\psi(\eta_c)}(0) = 0.350 \times 0.250\ (0.300 \times 0.200)~\mathrm{GeV}^3$, 
		and the fitting-point set $q^2_{\mathrm{fit}} = \{-5,-4,-3,-2,-1,0,1\}\ (\{-10,-9,-8,-7,-6,-5,-4,-3,-2,-1,0\})~\mathrm{GeV}^2$, respectively.
			}
	\label{tab:decaywidths}
	\renewcommand{\arraystretch}{1.5} %
	\setlength{\tabcolsep}{6pt} 
	\resizebox{\textwidth}{!}
	{
		\begin{tabular}{l c c c}
			\toprule
				\makebox[0pt][l]{$\displaystyle\frac{\Gamma(H \to H'  \ell {\nu}_\ell)}{10^{-15}\,\text{GeV}}$} & LO & NLO & NNLO \\
			\midrule
			\multicolumn{4}{l}{\textbf{$B_c \to J/\psi \ell \nu_\ell$}} \\
			\quad $e\qquad$   & $5.8633^{-2.0843+0+2.5318+1.9728}_{+10.4703+0-1.9159-0.8467}$ & $16.5372^{-5.3368+0+7.1409+5.8483}_{+23.5655+0-5.4037-2.5121}$ & $24.7789^{-6.4258-4.1461+10.6997+9.0839}_{+14.0018+7.3488-8.0967-3.8934}$ \\
			\quad $\mu$ & $5.8421^{-2.0767+0+2.5227+1.9681}_{+10.4323+0-1.9090-0.8443}$ & $16.4776^{-5.3176+0+7.1151+5.8341}_{+23.4810+0-5.3842-2.5049}$ & $24.6899^{-6.4028-4.1312+10.6613+9.0616}_{+13.9531+7.3223-8.0677-3.8820}$ \\
			\quad $\tau$& $1.6558^{-0.5886+0+0.7150+0.6601}_{+2.9568+0-0.5410-0.2562}$ & $4.6678^{-1.5060+0+2.0156+1.9496}_{+6.6472+0-1.5253-0.7556}$ & $7.0011^{-1.8174-1.1712+3.0231+3.0233}_{+3.9887+2.0758-2.2877-1.1680}$ \\
			\addlinespace
			\multicolumn{4}{l}{\textbf{$B_c \to \eta_c \ell \nu_\ell$}} \\
			\quad $e$   & $1.9340^{-0.6875+0+0.8351+0.5380}_{+3.4536+0-0.6319-0.2751}$ & $5.9957^{-2.0158+0+2.5890+1.7500}_{+9.6206+0-1.9592-0.8925}$ & $9.1405^{-2.4553-1.7159+3.9470+2.7638}_{+5.0874+3.0848-2.9868-1.4037}$ \\
			\quad $\mu$ & $1.9318^{-0.6867+0+0.8342+0.5389}_{+3.4496+0-0.6312-0.2747}$ & $5.9892^{-2.0136+0+2.5862+1.7528}_{+9.6108+0-1.9570-0.8913}$ & $9.1311^{-2.4529-1.7141+3.9429+2.7681}_{+5.0838+3.0814-2.9837-1.4018}$ \\
			\quad $\tau$& $0.7817^{-0.2779+0+0.3376+0.3007}_{+1.3960+0-0.2554-0.1187}$ & $2.4385^{-0.8219+0+1.0530+0.9764}_{+3.9412+0-0.7968-0.3849}$ & $3.7407^{-1.0125-0.6981+1.6153+1.5406}_{+2.1615+1.2541-1.2223-0.6052}$ \\
			\addlinespace
			\multicolumn{4}{l}{\textbf{$B_c^* \to J/\psi \ell \nu_\ell$}} \\
			\quad $e$   & $4.9349^{-1.7543+0+2.1309+1.7341}_{+8.8123+0-1.6125-0.7842}$ & $13.1281^{-4.1229+0+5.6688+4.8860}_{+17.2645+0-4.2897-2.2136}$ & $19.3932^{-4.8956-2.9020+8.3741+7.5448}_{+10.8444+5.0822-6.3369-3.4141}$ \\
			\quad $\mu$ & $4.9200^{-1.7490+0+2.1245+1.7314}_{+8.7857+0-1.6077-0.7822}$ & $13.0887^{-4.1105+0+5.6518+4.8786}_{+17.2130+0-4.2768-2.2079}$ & $19.3352^{-4.8810-2.8933+8.3491+7.5335}_{+10.8127+5.0670-6.3180-3.4053}$ \\
			\quad $\tau$& $1.5340^{-0.5453+0+0.6624+0.6630}_{+2.7393+0-0.5013-0.2601}$ & $4.0678^{-1.2754+0+1.7565+1.8616}_{+5.3206+0-1.3292-0.7292}$ & $5.9974^{-1.5092-0.9000+2.5897+2.8673}_{+3.3173+1.5766-1.9597-1.1193}$ \\
			\addlinespace
			\multicolumn{4}{l}{\textbf{$B_c^* \to \eta_c \ell \nu_\ell$}} \\
			\quad $e$   & $2.8121^{-0.9996+0+1.2143+1.0123}_{+5.0215+0-0.9189-0.4165}$ & $8.2634^{-2.7161+0+3.5682+3.0887}_{+12.4283+0-2.7001-1.2725}$ & $12.4640^{-3.2802-2.1397+5.3820+4.7876}_{+6.9488+3.8041-4.0727-1.9687}$ \\
			\quad $\mu$ & $2.8033^{-0.9965+0+1.2105+1.0101}_{+5.0059+0-0.9160-0.4154}$ & $8.2376^{-2.7076+0+3.5571+3.0819}_{+12.3894+0-2.6917-1.2691}$ & $12.4251^{-3.2700-2.1330+5.3653+4.7770}_{+6.9274+3.7922-4.0600-1.9635}$ \\
			\quad $\tau$& $0.9489^{-0.3373+0+0.4098+0.3931}_{+1.6945+0-0.3101-0.1480}$ & $2.7775^{-0.9113+0+1.1993+1.1962}_{+4.1560+0-0.9076-0.4502}$ & $4.1854^{-1.0996-0.7202+1.8073+1.8528}_{+2.3310+1.2808-1.3676-0.6952}$ \\
			\bottomrule
		\end{tabular}
	}
\end{table}

\begin{table*}[htbp] %
	\centering
	\caption{Same as table~\ref{tab:decaywidths}, but for the semileptonic branching fractions %
	$\mathcal{B}(H\to H'\, \ell {\nu}_\ell)$. %
}
	\label{tab:branchingfractions}
	\setlength{\tabcolsep}{6pt}
	\renewcommand{\arraystretch}{1.5}
	\resizebox{\textwidth}{!}{
		\begin{tabular}{l c c c}
			\toprule
			& LO & NLO & NNLO \\
			\midrule
			\multicolumn{4}{l}{\textbf{$\mathcal{B}(B_c \to J/\psi \ell \nu_\ell)/(10^{-2})$}} \\
			$\quad e\qquad$    & $0.4543^{-0.1615+0+0.1962+0.1529}_{+0.8113+0-0.1484-0.0656}$ & $1.2814^{-0.4135+0+0.5533+0.4531}_{+1.8259+0-0.4187-0.1946}$ & $1.9199^{-0.4979-0.3213+0.8290+0.7038}_{+1.0849+0.5694-0.6274-0.3017}$ \\
			$\quad \mu$ & $0.4527^{-0.1609+0+0.1955+0.1525}_{+0.8083+0-0.1479-0.0654}$ & $1.2767^{-0.4120+0+0.5513+0.4520}_{+1.8194+0-0.4172-0.1941}$ & $1.9130^{-0.4961-0.3201+0.8261+0.7021}_{+1.0811+0.5673-0.6251-0.3008}$ \\
			$\quad \tau$& $0.1283^{-0.0456+0+0.0554+0.0511}_{+0.2291+0-0.0419-0.0199}$ & $0.3617^{-0.1167+0+0.1562+0.1511}_{+0.5150+0-0.1182-0.0585}$ & $0.5425^{-0.1408-0.0907+0.2342+0.2343}_{+0.3091+0.1608-0.1773-0.0905}$ \\
			\midrule
			\multicolumn{4}{l}{\textbf{$\mathcal{B}(B_c \to \eta_c \ell \nu_\ell)/(10^{-2})$}} \\
			$\quad e$     & $0.1498^{-0.0533+0+0.0647+0.0417}_{+0.2676+0-0.0490-0.0213}$ & $0.4646^{-0.1562+0+0.2006+0.1356}_{+0.7454+0-0.1518-0.0692}$ & $0.7082^{-0.1902-0.1330+0.3058+0.2141}_{+0.3942+0.2390-0.2314-0.1088}$ \\
			$\quad \mu$  & $0.1497^{-0.0532+0+0.0646+0.0418}_{+0.2673+0-0.0489-0.0213}$ & $0.4641^{-0.1560+0+0.2004+0.1358}_{+0.7447+0-0.1516-0.0691}$ & $0.7075^{-0.1901-0.1328+0.3055+0.2145}_{+0.3939+0.2388-0.2312-0.1086}$ \\
			$\quad \tau$ & $0.0606^{-0.0215+0+0.0262+0.0233}_{+0.1082+0-0.0198-0.0092}$ & $0.1889^{-0.0637+0+0.0816+0.0757}_{+0.3054+0-0.0617-0.0298}$ & $0.2898^{-0.0785-0.0541+0.1252+0.1194}_{+0.1675+0.0972-0.0947-0.0469}$ \\
			\midrule
			\multicolumn{4}{l}{\textbf{$\mathcal{B}(B_c^* \to J/\psi \ell \nu_\ell)/(10^{-7})$}} \\
			$\quad e$    & $0.8225^{-0.2924+0+0.3552+0.2890}_{+1.4687+0-0.2688-0.1307}$ & $2.1880^{-0.6871+0+0.9448+0.8143}_{+2.8774+0-0.7150-0.3689}$ & $3.2322^{-0.8159-0.4837+1.3957+1.2575}_{+1.8074+0.8470-1.0562-0.5690}$ \\
			$\quad \mu$ & $0.8200^{-0.2915+0+0.3541+0.2886}_{+1.4643+0-0.2679-0.1304}$ & $2.1815^{-0.6851+0+0.9420+0.8131}_{+2.8688+0-0.7128-0.3680}$ & $3.2225^{-0.8135-0.4822+1.3915+1.2556}_{+1.8021+0.8445-1.0530-0.5676}$ \\
			$\quad \tau$& $0.2557^{-0.0909+0+0.1104+0.1105}_{+0.4566+0-0.0835-0.0434}$ & $0.6780^{-0.2126+0+0.2928+0.3103}_{+0.8868+0-0.2215-0.1215}$ & $0.9996^{-0.2515-0.1500+0.4316+0.4779}_{+0.5529+0.2628-0.3266-0.1866}$ \\
			\midrule
			\multicolumn{4}{l}{\textbf{$\mathcal{B}(B_c^* \to \eta_c \ell \nu_\ell)/(10^{-7})$}} \\
			$\quad e$     & $0.4687^{-0.1666+0+0.2024+0.1687}_{+0.8369+0-0.1531-0.0694}$ & $1.3772^{-0.4527+0+0.5947+0.5148}_{+2.0714+0-0.4500-0.2121}$ & $2.0773^{-0.5467-0.3566+0.8970+0.7979}_{+1.1581+0.6340-0.6788-0.3281}$ \\
			$\quad \mu$  & $0.4672^{-0.1661+0+0.2017+0.1683}_{+0.8343+0-0.1527-0.0692}$ & $1.3729^{-0.4513+0+0.5928+0.5136}_{+2.0649+0-0.4486-0.2115}$ & $2.0709^{-0.5450-0.3555+0.8942+0.7962}_{+1.1546+0.6320-0.6767-0.3272}$ \\
			$\quad \tau$ & $0.1582^{-0.0562+0+0.0683+0.0655}_{+0.2824+0-0.0517-0.0247}$ & $0.4629^{-0.1519+0+0.1999+0.1994}_{+0.6927+0-0.1513-0.0750}$ & $0.6976^{-0.1833-0.1200+0.3012+0.3088}_{+0.3885+0.2135-0.2279-0.1159}$ \\
			\bottomrule
		\end{tabular}
	}
\end{table*}

\begin{table}[htbp]
	\centering
	\caption{
Same as table~\ref{tab:decaywidths}, but for the semileptonic LFU ratios %
$\mathcal{R}(H\to H')$. %
	}
	\label{tab:Rtaumu}
	\renewcommand{\arraystretch}{1.5} %
	\setlength{\tabcolsep}{6pt} 
	\resizebox{\textwidth}{!}
	{
		\begin{tabular}{lccc}
			\toprule
			& LO & NLO & NNLO \\
			\midrule
			$\mathcal{R}(B_c\to J/\psi)$
			& $0.2834^{-0+0+0+0.0131}_{+0+0+0-0.0034}$ & $0.2833^{+0.0000+0+0+0.0133}_{-0.0001+0+0-0.0033}$ & $0.2836^{-0.0001+0.0000+0+0.0134}_{+0.0008-0.0000+0-0.0032}$ \\
			$\mathcal{R}(B_c\to\eta_c)$
			& $0.4047^{-0+0+0+0.0335}_{+0+0+0-0.0046}$ & $0.4071^{-0.0005+0+0+0.0339}_{+0.0018+0+0-0.0043}$ & $0.4097^{-0.0011+0.0005+0+0.0342}_{+0.0055-0.0007+0-0.0040}$ \\
			$\mathcal{R}(B_c^*\to J/\psi)$
			& $0.3118^{-0+0+0+0.0185}_{+0+0+0-0.0039}$ & $0.3108^{+0.0002+0+0+0.0192}_{-0.0010+0+0-0.0040}$ & $0.3102^{+0.0003-0.0002+0+0.0197}_{-0.0012+0.0002+0-0.0040}$ \\
			$\mathcal{R}(B_c^*\to\eta_c)$
			& $0.3385^{-0+0+0+0.0134}_{+0+0+0-0.0031}$ & $0.3372^{+0.0003+0+0+0.0139}_{-0.0010+0+0-0.0032}$ & $0.3369^{+0.0002-0.0002+0+0.0142}_{-0.0001+0.0002+0-0.0032}$ \\
			\bottomrule
		\end{tabular}
	}
\end{table}

Based on the above definitions and inputs, the resulting NRQCD+$z$-series predictions at LO, NLO, and NNLO QCD accuracy for the decay widths $\Gamma$, branching fractions $\mathcal{B}$, and LFU ratios $\mathcal{R}$ for the semileptonic decays $B_c^{(*)}\to J/\psi(\eta_c)\,\ell {\nu}_\ell$ are summarized in tables~\ref{tab:decaywidths}, \ref{tab:branchingfractions}, and \ref{tab:Rtaumu}, respectively.
The central values at LO, NLO, and NNLO QCD accuracy are obtained from the NRQCD+$z$-series form factors calculated at the corresponding perturbative orders, with the input parameters fixed at their default (central) values. The four superscript (subscript) uncertainties attached to each central value correspond, in order, to the shifts induced by individually varying the renormalization scale $\mu$, the factorization scale $\mu_f$, the wave-function product $\psi_H(0)\psi_{H'}(0)$, and the fitting-point set $q^2_{\rm fit}$ to their upper (lower) values, respectively.

From the numerical results for $\Gamma$ in table~\ref{tab:decaywidths} and $\mathcal{B}$ in table~\ref{tab:branchingfractions}, we observe that the NLO QCD corrections to the semileptonic decay widths and branching fractions are large and positive in all decay channels, 
amounting to roughly $170\%$--$210\%$ of the corresponding LO predictions.
This substantial enhancement is partly due to the fact that the decay widths are proportional to the modulus squared of the form factors, which amplifies the QCD radiative corrections. The NNLO corrections remain sizable but are systematically smaller than the NLO corrections, amounting to about $47\%$--$53\%$ of the NLO predictions. Thus, although the perturbative corrections are substantial, their reduction in size at successive orders indicates improved perturbative convergence at NNLO.
The uncertainties propagated from the renormalization scale and the wave functions at the origin are relatively large, while the uncertainty arising from the factorization scale is comparatively the smallest.
The relative renormalization-scale uncertainty decreases only mildly from LO to NLO, but is substantially reduced at NNLO compared with NLO, indicating that the inclusion of higher-order QCD corrections---especially at NNLO---significantly suppresses the scale dependence and enhances the stability and reliability of our theoretical predictions.

In contrast to the absolute decay widths and branching fractions, the LFU ratios $\mathcal{R}(H\to H')$ in table~\ref{tab:Rtaumu} exhibit remarkably weak dependence on both perturbative and nonperturbative inputs in NRQCD, with the uncertainties from the renormalization scale, factorization scale, and wave functions at the origin largely canceling between the $\tau$ and muon modes. The central predictions remain remarkably stable from LO to NLO and NNLO, further demonstrating the strong cancellation of common higher-order QCD correction effects in the ratios. Consequently, the residual uncertainty in $\mathcal{R}$ is dominated by the choice of the fitting-point set $q^2_{\mathrm{fit}}$ used in the $z$-series parametrization.

As shown in table~\ref{tab:branchingfractions}, the semileptonic branching fractions of $B_c^*$ are suppressed by about $4$--$5$ orders of magnitude relative to those of $B_c$. This is mainly due to the vastly different total widths: $B_c$ decays only through the weak interaction, whereas the total width of $B_c^*$ is dominated by the electromagnetic transition $B_c^*\to B_c\gamma$, thereby strongly suppressing its semileptonic weak branching fractions.
For the ground-state $B_c$, which decays only through the weak interaction, the total width is of order $10^{-3}$ eV, leading to semileptonic branching fractions at the percent level. In contrast, the total width of $B_c^*$ is dominated by the electromagnetic M1 transition $B_c^*\to B_c\gamma$ and is of order tens of eV, about $10^4$--$10^5$ times larger than that of $B_c$, thereby suppressing its semileptonic branching fractions to the $\mathcal O(10^{-7})$ level.

For $B_c^{(*)}\to J/\psi(\eta_c)\,\ell\nu_\ell$, our NNLO NRQCD+$z$-series predictions for the semileptonic decay widths, branching fractions, and LFU ratios are broadly comparable in magnitude with other Standard Model calculations based on QCD sum rules~\cite{Kiselev:2002vz,Wang:2012hu}, quark potential models~\cite{Hernandez:2006gt,Ivanov:2006ni,Ebert:2003cn,Chang:2026ctr}, and state-of-the-art lattice QCD determinations~\cite{Harrison:2020gvo,Harrison:2020nrv,Harrison:2025yan}. In particular, our baseline predictions, $\mathcal{B}(B_c\to J/\psi\,\mu\nu_\mu)=0.0191$ and $\mathcal{R}(B_c\to J/\psi)=0.2836$, are somewhat larger than, but broadly comparable to, the latest HPQCD results~\cite{Harrison:2025yan} for the corresponding observables, $0.0153\pm0.0004\pm0.0006\pm0.0003$ and $0.2597\pm0.0027$, respectively. 
The experimental determinations of $\mathcal{R}(B_c\to J/\psi)$ have evolved from the earlier LHCb Run 1 result ($0.71\pm0.17\pm0.18$)~\cite{LHCb:2017vlu} to the CMS measurement ($0.17\pm0.33$)~\cite{CMS:2024seh}, the updated CMS combination ($0.49\pm0.26$)~\cite{CMS:2025jfx}, and most recently the LHCb Run 2 measurement ($0.51\pm0.12\pm0.08$)~\cite{LHCb:2026dev}. 
Notably, the latest LHCb central value has moved downward relative to Run 1, although it remains somewhat above both our baseline prediction and the latest HPQCD result.

A closer look at the $q^2_{\rm fit}$ dependence of our predictions in table~\ref{tab:Rtaumu} reveals that $\mathcal{R}$ increases as the fitting-point set $q^2_{\rm fit}$ shifts rightward (toward larger $q^2$). Combined with the form-factor extrapolation behavior observed in figures~\ref{fig:ffq2nrzbcjpsi}--\ref{fig:ffq2nrzbcvetac}, 
this indicates that shifting $q^2_{\rm fit}$ rightward results in a steeper $q^2$ dependence of the form factors, which in turn leads to a larger $\mathcal{R}$.
This can be roughly interpreted as $\mathcal{R} \propto b/(a+b)$ increasing as the steepening of the form factors elevates $b$ relative to $a$, where $b$ denotes the high-$q^2$ integrated contribution of squared form factors accessible to both $\tau$ and the light lepton, while $a$ represents the low-$q^2$ contribution exclusive to the light lepton.
As an illustrative test, if we instead took a more aggressive physical fitting-point set, e.g., $q^2_{\rm fit}=\{0,1,2,3,4,5\}\,\mathrm{GeV}^2$, our predictions would reach $\mathcal{B}(B_c\to J/\psi\,\mu\nu_\mu)\approx0.10$ and $\mathcal{R}(B_c\to J/\psi)\approx0.36$.
Such a fitting-point choice is not adopted in our baseline analysis, but it illustrates the possibility of further alleviating the remaining theory--experiment tension.

\section{Summary~\label{Summary}}

We have calculated the QCD corrections through NNLO, $\mathcal{O}(\alpha_s^3)$, to all 34 vector, axial-vector, tensor, and axial-tensor form factors for $B_c^{(*)}\to J/\psi(\eta_c)$ within NRQCD factorization, at lowest order in the velocity expansion. The complete results, covering 35 points in the $(m_b,m_c,q^2)$ parameter space, are provided in the ancillary files accompanying this paper.

Throughout the numerical analysis, the wave functions at the origin are treated as independent nonperturbative inputs (with ranges informed by decay constants and potential models), and lattice transition form factors are used for comparison rather than as inputs. 
At $q^2=0$, the NLO corrections typically enhance the nonvanishing LO form factors by about $70\%$, while the NNLO corrections amount to about $20\%$ of the NLO predictions, indicating improved perturbative convergence. The NNLO predictions also show a substantially reduced renormalization-scale dependence compared with lower orders and only a mild factorization-scale dependence.
Extending the analysis across $q^2\in[-10,10]~\mathrm{GeV}^2$ shows that the perturbative behavior further improves toward the spacelike region ($q^2<0$), whereas fixed-order NRQCD predictions become less reliable at large timelike $q^2$ toward zero recoil, with the form factors exhibiting increasingly steep growth and deteriorating perturbative stability.

We have therefore used the NRQCD form factors at spacelike and near-zero $q^2$ directly as inputs to the $z$-series fits to extrapolate all 34 form factors over the full physical region. %
The resulting NNLO NRQCD+$z$-series form factors increase smoothly and monotonically with $q^2$, with comparable and reasonably well-controlled uncertainties from the renormalization scale, the wave functions at the origin, and the choice of fitting points.
For $B_c\to J/\psi$, most of our form factors are broadly consistent with the latest HPQCD lattice results~\cite{Harrison:2025yan} across the physical region, although ours rise somewhat more steeply toward $q_{\max}^2$.

Using the vector and axial-vector NRQCD+$z$-series form factors at LO, NLO, and NNLO, we obtained the corresponding LO, NLO, and NNLO predictions for the decay widths, branching fractions, and $\tau$-to-muon LFU ratios of $B_c^{(*)}\to J/\psi(\eta_c)\,\ell\nu_\ell$ ($\ell=e,\mu,\tau$).
The NLO corrections to the decay widths and branching fractions are roughly twice as large as the LO results, whereas the NNLO corrections amount to about half of the NLO predictions. For these absolute decay rates, the relative renormalization-scale uncertainty is substantially reduced at NNLO compared with the lower orders, indicating improved perturbative stability of our predictions, 
although the remaining uncertainties from the renormalization scale, the wave functions at the origin, and the choice of fitting points are still sizable. 
With the default inputs at NNLO, we obtain $\mathcal{B}(B_c\to J/\psi\,\mu\nu_\mu)=1.9130\times 10^{-2}$ and $\mathcal{B}(B_c\to\eta_c\,\mu\nu_\mu)=0.7075\times 10^{-2}$. For the adopted total width $\Gamma_{\mathrm{tot}}(B_c^*)=6.0\times 10^{-8}~\mathrm{GeV}$, the corresponding $B_c^*$ branching fractions, $\mathcal{B}(B_c^*\to J/\psi\,\mu\nu_\mu)=3.2225\times 10^{-7}$ and $\mathcal{B}(B_c^*\to\eta_c\,\mu\nu_\mu)=2.0709\times 10^{-7}$, are suppressed by $4$--$5$ orders of magnitude relative to those of $B_c$, reflecting the dominance of the electromagnetic M1 transition $B_c^*\to B_c\gamma$ in the total width of $B_c^*$ rather than a comparable suppression of the semileptonic decay widths.

Higher-order QCD corrections substantially enhance the absolute decay rates but largely cancel between the $\tau$ and muon modes in the LFU ratios $\mathcal{R}$, which remain remarkably stable from LO to NNLO and are largely insensitive to the scales and the wave functions at the origin. At NNLO, our central predictions are $\mathcal{R}(B_c\to J/\psi)=0.2836$, $\mathcal{R}(B_c\to\eta_c)=0.4097$, $\mathcal{R}(B_c^*\to J/\psi)=0.3102$, and $\mathcal{R}(B_c^*\to\eta_c)=0.3369$, with residual uncertainties dominated by the choice of fitting points. Shifting the fitting points toward larger $q^2$ produces steeper extrapolated form factors and larger LFU ratios.

Our NNLO prediction for $\mathcal{B}(B_c\to J/\psi\,\mu\nu_\mu)$ is somewhat larger than, but broadly comparable to, the latest HPQCD result~\cite{Harrison:2025yan}. For $\mathcal{R}(B_c\to J/\psi)$, our central value lies above the corresponding HPQCD determination and below the central values of the updated CMS combination~\cite{CMS:2025jfx} and the LHCb Run~2 measurement~\cite{LHCb:2026dev}. The downward shift of the LHCb central value from Run~1~\cite{LHCb:2017vlu} to Run~2 narrows the gap with both theoretical predictions, although the experimental uncertainties remain sizable.

Overall, our NNLO form-factor results and semileptonic predictions
provide useful theoretical input for precision tests of the
Standard Model, investigations of possible new physics in
$b\to c$ transitions, and future experimental studies of
$B_c^{(*)}$ decays.

\acknowledgments
{\emergencystretch=1em
We thank Junfeng Sun, Jian-Ping Ma, Jinmin Yang, Li-Ting Wang,
Huijing Li, Neng-Chang Wei, Wenlong Sang, Vladyslav Shtabovenko, Guido Bell,
and Matthias Steinhauser for helpful discussions.
This work is supported by the Postdoctoral Research Funding Project
of Henan Province (Grant No.~HN2026009),
the National Natural Science Foundation of China
(NSFC, Grant Nos.~1267050014 and 12275067),
the Science and Technology R\&D Program Joint Fund Project
of Henan Province (Grant No.~225200810030),
the Science and Technology Innovation Leading Talent Support Program
of Henan Province (Grant No.~254000510039),
the National Key R\&D Program of China (Grant No.~2023YFA1606000),
and by the High Performance Computing Center of Henan Normal University.
This work is partially supported by the NSFC
(Grant Nos.~12322503 and 12335003).
We also thank the National Supercomputing Center in Zhengzhou
for making its computing resources available to us.\par}

\appendix
\section{Differential decay widths for $B_c^{(*)} \to J/\psi(\eta_c) \, \ell {\nu}_\ell$ in terms of transition form factors}
\label{app:dGmdq2ffs}
In this appendix, we collect the analytical expressions for the kinematics, the helicity amplitudes, and the differential decay width $d\Gamma/dq^2$ for each semileptonic decay $B_c^{(*)} \to J/\psi (\eta_c) \, \ell {\nu}_\ell$, written in terms of the corresponding vector and axial-vector transition form factors defined in section~\ref{sec:ffdef}.

The kinematic K\"all\'en function $\lambda$ is defined as:
\begin{equation}
\lambda = \left[ (M - M')^2 - q^2 \right] \left[ (M + M')^2 - q^2 \right],
\end{equation}
where $M \in \{m_{B_c}, m_{B_c^*}\}$ and $M' \in \{m_{J/\psi}, m_{\eta_c}\}$ represent the masses of the initial and final-state mesons, respectively. 

\subsection{$B_c \to J/\psi \, \ell {\nu}_\ell$}
The helicity amplitudes in terms of the form factors are given by~\cite{Harrison:2020gvo}
\begin{align}
H_+ &=  - \frac{\sqrt{\lambda}}{M + M'} V^{B_c \to J/\psi}+(M + M') A_1^{B_c \to J/\psi} , \\[0.6em]
H_- &=  \frac{\sqrt{\lambda}}{M + M'} V^{B_c \to J/\psi}+(M + M') A_1^{B_c \to J/\psi}  , \\[0.6em]
H_0 &= \frac{1}{2 M' \sqrt{q^2}} \left[ (M + M')(M^2 - M'^2 - q^2) A_1^{B_c \to J/\psi} - \frac{\lambda}{M + M'} A_2^{B_c \to J/\psi} \right] , \\[0.6em]
H_t &= \sqrt{\frac{\lambda}{q^2}} A_0^{B_c \to J/\psi} .
\end{align}
The corresponding differential decay width reads
\begin{equation}
\begin{aligned}
\frac{d\Gamma}{dq^2} = \frac{G_F^2 |V_{cb}|^2 \lambda^{1/2} }{192 \pi^3 M^3}  q^2 \left( 1 - \frac{m_\ell^2}{q^2} \right)^2 \,  \bigg[ & H_+^2 + H_-^2 + H_0^2 \\
& + \frac{m_\ell^2}{2 q^2} \left( H_+^2 + H_-^2 + H_0^2 + 3 H_t^2 \right) \bigg] ,
\end{aligned}
\end{equation}
where $m_\ell$ denotes the mass of the charged lepton ($\ell \in\{ e, \mu, \tau\}$).

\subsection{$B_c \to \eta_c \, \ell {\nu}_\ell$}
The differential decay width expressed in terms of the form factors reads~\cite{Wang:2008xt,Richman:1995wm,Qiao:2012vt,Berns:2018vpl,Wang:2018duy,Wang:2012lrc,Ebert:2003cn}
\begin{equation}
\begin{aligned}
\frac{d\Gamma}{dq^2} = \frac{G_F^2 |V_{cb}|^2 \lambda^{3/2}}{384 \pi^3 M^3}    \left( 1 - \frac{m_\ell^2}{q^2} \right)^2 \, \bigg[ &   (2  +\frac{m_\ell^2}{q^2} ) \left(F_p^{B_c \to \eta_c}\right)^2 \\
& +  \frac{ 3 m_\ell^2}{q^2} \frac{(M^2 - M'^2)^2}{\lambda} \left(F_0^{B_c \to \eta_c}\right)^2 \bigg].
\end{aligned}
\end{equation}

\subsection{$B_c^* \to J/\psi\,\ell\nu_\ell$}
The helicity amplitudes in terms of the form factors are given by~\cite{Chang:2020xvu,Sheng:2022peg,Sheng:2021iss}
\begin{align}
H_{t++} &=  \frac{M^2 - M'^2}{\sqrt{q^2}} V_1^{B_c^* \to J/\psi} - \sqrt{q^2} V_2^{B_c^* \to J/\psi} -\frac{\sqrt{\lambda}}{\sqrt{q^2}} A_1^{B_c^* \to J/\psi} , \\[0.6em]
H_{t--} &=  \frac{M^2 - M'^2}{\sqrt{q^2}} V_1^{B_c^* \to J/\psi} - \sqrt{q^2} V_2^{B_c^* \to J/\psi} +\frac{\sqrt{\lambda}}{\sqrt{q^2}} A_1^{B_c^* \to J/\psi} , \\[0.6em]
H_{t00} &= \frac{(M^2 - M'^2)(M^2 + M'^2 - q^2)}{2 M M' \sqrt{q^2}} V_1^{B_c^* \to J/\psi} - \frac{(M^2 + M'^2 - q^2)\sqrt{q^2}}{2 M M'} V_2^{B_c^* \to J/\psi} \nonumber \\
&\quad + \frac{\lambda}{4 M M' \sqrt{q^2}} \left( -V_3^{B_c^* \to J/\psi} + V_4^{B_c^* \to J/\psi} + V_6^{B_c^* \to J/\psi} \right) , \\[0.6em]
H_{0++} &= \frac{\sqrt{\lambda}}{\sqrt{q^2}} V_1^{B_c^* \to J/\psi} - \frac{M^2 - M'^2}{\sqrt{q^2}} A_1^{B_c^* \to J/\psi} + \sqrt{q^2} A_2^{B_c^* \to J/\psi} , \\[0.6em]
H_{0--} &= \frac{\sqrt{\lambda}}{\sqrt{q^2}} V_1^{B_c^* \to J/\psi} + \frac{M^2 - M'^2}{\sqrt{q^2}} A_1^{B_c^* \to J/\psi} - \sqrt{q^2} A_2^{B_c^* \to J/\psi} , \\[0.6em]
H_{-+0} &= -\frac{\sqrt{\lambda}}{2 M'} V_4^{B_c^* \to J/\psi}  - \frac{M^2 + 3 M'^2 - q^2}{2 M'} A_1^{B_c^* \to J/\psi}  \nonumber \\
&\quad + \frac{M^2 - M'^2 - q^2}{2 M'} A_2^{B_c^* \to J/\psi}- \frac{\lambda}{2 M'(M^2 - M'^2)} A_3^{B_c^* \to J/\psi}, \\[0.6em]
H_{+-0} &= -\frac{\sqrt{\lambda}}{2 M'} V_4^{B_c^* \to J/\psi}+ \frac{M^2 + 3 M'^2 - q^2}{2 M'} A_1^{B_c^* \to J/\psi}  \nonumber \\
&\quad - \frac{M^2 - M'^2 - q^2}{2 M'} A_2^{B_c^* \to J/\psi}  + \frac{\lambda}{2 M'(M^2 - M'^2)} A_3^{B_c^* \to J/\psi}, \\[0.6em]
H_{+0+} &= -\frac{\sqrt{\lambda}}{2 M} V_3^{B_c^* \to J/\psi} + \frac{3 M^2 + M'^2 - q^2}{2 M} A_1^{B_c^* \to J/\psi} \nonumber \\
&\quad  - \frac{M^2 - M'^2 + q^2}{2 M} A_2^{B_c^* \to J/\psi}+ \frac{\lambda}{2 M(M^2 - M'^2)} A_4^{B_c^* \to J/\psi} , \\[0.6em]
H_{-0-} &= -\frac{\sqrt{\lambda}}{2 M} V_3^{B_c^* \to J/\psi} - \frac{3 M^2 + M'^2 - q^2}{2 M} A_1^{B_c^* \to J/\psi} \nonumber \\
&\quad  + \frac{M^2 - M'^2 + q^2}{2 M} A_2^{B_c^* \to J/\psi} - \frac{\lambda}{2 M(M^2 - M'^2)} A_4^{B_c^* \to J/\psi}, \\[0.6em]
H_{000} &=  \frac{(M^2 + M'^2 - q^2)\sqrt{\lambda}}{2 M M'\sqrt{q^2}} V_1^{B_c^* \to J/\psi} - \frac{(M^2 - M'^2 - q^2)\sqrt{\lambda}}{4 M M'\sqrt{q^2}} V_3^{B_c^* \to J/\psi} \nonumber \\
&\quad  + \frac{(M^2 - M'^2 + q^2)\sqrt{\lambda}}{4 M M'\sqrt{q^2}} V_4^{B_c^* \to J/\psi} +\frac{\lambda^{3/2}}{4 M M'(M^2 - M'^2)\sqrt{q^2}} V_5^{B_c^* \to J/\psi}.
\end{align}
The corresponding differential decay width reads
\begin{equation}
\begin{aligned}
\frac{d\Gamma}{dq^2} &= \frac{G_F^2 |V_{cb}|^2  \lambda^{1/2}}{576 \pi^3 M^3}   q^2\left( 1 - \frac{m_\ell^2}{q^2} \right)^2\,  \bigg[  \frac{3 m_\ell^2}{2 q^2} \left( H_{t00}^2 + H_{t--}^2 + H_{t++}^2 \right) \\
&\quad + \left( 1 + \frac{m_\ell^2}{2 q^2} \right)\left( H_{000}^2 + H_{0--}^2 + H_{0++}^2 + H_{-0-}^2 + H_{-+0}^2 + H_{+0+}^2 + H_{+-0}^2 \right)  \bigg] .
\end{aligned}
\end{equation}

\subsection{$B_c^* \to \eta_c\,\ell\nu_\ell$}
The helicity amplitudes in terms of the form factors are given by~\cite{Wang:2024cyi,Wang:2018ryc,Cohen:2018vhw,Chang:2016cdi,Zhang:2019hth,Ray:2019gkv,Chang:2026ctr}
\begin{align}
H_{0t} &= \frac{\sqrt{\lambda}}{\sqrt{q^2}} A_0^{B_c^* \to \eta_c} , \\[0.6em]
H_{00} &= \frac{1}{2 M \sqrt{q^2}} \bigg[ (M + M') (M^2 - M'^2 + q^2) A_1^{B_c^* \to \eta_c}+\frac{\lambda}{M + M'} A_2^{B_c^* \to \eta_c}  \bigg], \\[0.6em]
H_{+-} &= - \frac{\sqrt{\lambda}}{M + M'}  V^{B_c^* \to \eta_c}-(M + M') A_1^{B_c^* \to \eta_c} , \\[0.6em]
H_{-+} &=  \frac{\sqrt{\lambda}}{M + M'} V^{B_c^* \to \eta_c}  -(M + M') A_1^{B_c^* \to \eta_c}.
\end{align}
The corresponding differential decay width reads
\begin{equation}
\begin{aligned}
\frac{d\Gamma}{dq^2} = \frac{G_F^2 |V_{cb}|^2  \lambda^{1/2}}{576 \pi^3 M^3} q^2  \left( 1 - \frac{m_\ell^2}{q^2} \right)^2\,  \bigg[  \left( 1 + \frac{m_\ell^2}{2 q^2} \right)\left( H_{00}^2 + H_{-+}^2 + H_{+-}^2 \right)  
+ \frac{3  m_\ell^2}{2 q^2} H_{0t}^2\bigg] .
\end{aligned}
\end{equation}

\bibliographystyle{JHEP}
\bibliography{refs}

\end{document}